\documentclass[11pt]{article}

\usepackage{amsmath}
\usepackage{amssymb}
\usepackage{amsthm}
\usepackage{natbib}
\usepackage{graphicx}
\usepackage{threeparttable}
\usepackage{bm}
\usepackage{bbm}
\usepackage{comment}
\usepackage{enumitem}   
\usepackage{xcolor}
\usepackage{setspace}
\usepackage{caption}
\usepackage{subcaption}
\usepackage{soul}
\usepackage{hyperref}
\hypersetup{hidelinks}
\hypersetup{
  colorlinks   = true, %Colours links instead of ugly boxes
  urlcolor     = blue, %Colour for external hyperlinks
  linkcolor    = blue, %Colour of internal links
  citecolor   = blue %Colour of citations
}

\usepackage{multibib}
\newcites{sm}{Additional References}

\theoremstyle{plain}
\newtheorem{prop}{Proposition}

\newtheorem{coro}{Corollary}

\theoremstyle{remark}

\newtheorem{ex}{Example}

\graphicspath{{figs/}}

\newcommand{\x}{\bm{x}}
\newcommand{\X}{\bm{X}}

\newcommand{\y}{\bm{y}}
\newcommand{\bV}{\bm{V}}
\newcommand{\bD}{\bm{D}}

\newcommand{\tz}{\tilde{\bm{z}}}

\newcommand{\trho}{\tilde{\rho}}
\newcommand{\bs}{\bm{s}}
\newcommand{\bS}{\bm{S}}
\newcommand{\bx}{\bm{x}}
\newcommand{\by}{\bm{y}}

\newcommand{\bz}{\bm{z}}
\newcommand{\bZ}{\bm{Z}}
\newcommand{\bbeta}{\bm{\beta}}
\newcommand{\bga}{\bm{\gamma}}
\newcommand{\bzeta}{\bm{\zeta}}

\newcommand{\btheta}{\bm{\theta}}

\newcommand{\la}{\lambda}
\newcommand{\ga}{\gamma}

\newcommand{\D}{\mathcal{D}}
\newcommand{\R}{\mathbb{R}}
\DeclareMathOperator*{\BS}{\mathcal{S}}
\DeclareMathOperator*{\V}{\mathcal{V}}
\newcommand{\sv}{\bm{v}}
\newcommand{\su}{\bm{u}}
\DeclareMathOperator*{\U}{\mathcal{U}}

\newcommand{\tNe}{\text{Ne}}
\newcommand{\cov}{\mathrm{Cov}} 
\newcommand{\bmu}{\bm{\mu}}
\newcommand{\bxi}{\bm{\xi}}
\newcommand{\bla}{\bm{\la}}
\newcommand{\iL}{\mathcal{L}}
\newcommand{\iH}{\mathcal{H}}

\newcommand{\blind}{0}

\begin{document}

\def\spacingset#1{\renewcommand{\baselinestretch}
		{#1}\small\normalsize} \spacingset{1}

	%%%%%%%%%%%%%%%%%%%%%%%%%%%%%%%%%%%%%%%%%%%%%%%%%%%%%%%%
	
	\if0\blind
	{
		\title{\bf Nearest-Neighbor Mixture Models for Non-Gaussian Spatial Processes}
		\author{
		Xiaotian Zheng,
		%\thanks{X. Zheng (xiaotian@ucsc.edu) is doctoral student, Department of Statistics, University of California, Santa Cruz.}
        Athanasios Kottas,
        %\thanks{A. Kottas (thanos@soe.ucsc.edu) is Professor, Department of Statistics, University of California, Santa Cruz.} 
        and
        Bruno Sans\'o
        %\thanks{B. Sans\'o (bruno@soe.ucsc.edu) is Professor, Department of Statistics, University of California, Santa Cruz.}
        \\
        Department of Statistics, University of California, Santa Cruz %California, USA
        }
		\maketitle
	} \fi
	
	\if1\blind
	{
		\bigskip
		\bigskip
		\bigskip
		\begin{center}
			{\bf Nearest-Neighbor Mixture Models for Non-Gaussian Spatial Processes}
		\end{center}
		\medskip
	} \fi
	
	\bigskip
\begin{abstract}
We develop a class of nearest-neighbor mixture models
that provide direct, computationally efficient, probabilistic modeling for 
non-Gaussian geospatial data. The class is defined over 
a directed acyclic graph, which implies conditional independence
in representing a multivariate distribution through factorization
into a product of univariate conditionals, and is extended to a full spatial process. 
We model each conditional as a mixture of spatially varying transition kernels, 
with locally adaptive weights, for each one of a given number of nearest neighbors.
The modeling framework emphasizes the description of non-Gaussian 
dependence at the data level, in contrast with approaches that introduce a 
spatial process for transformed data, or for functionals of the data 
probability distribution. Thus, it facilitates efficient, full 
simulation-based inference. We study model construction and properties 
analytically through specification of bivariate distributions that define 
the local transition kernels, providing a general strategy for modeling 
general types of non-Gaussian data. Regarding computation, the framework 
lays out a new approach to handling spatial data sets, leveraging a 
mixture model structure to avoid computational issues that arise from large 
matrix operations. We illustrate the methodology using synthetic data examples 
and an analysis of Mediterranean Sea surface temperature observations.
\end{abstract}
	
	
	\vfill
	
	\newpage

	\spacingset{1.2} % DON'T change the spacing!
	
\section{Introduction}

Gaussian processes have been widely used as an underlying structure in 
model-based analysis of irregularly located spatial data in order to capture 
short range variability. The fruitfulness of these spatial models owes to the 
simple characterization of the Gaussian process by a mean and a covariance 
function, and the optimal prediction it provides that justifies kriging.
However, the assumption of Gaussianity is restrictive in many 
fields where the data exhibits non-Gaussian features, for example, vegetation 
abundance \citep{eskelson2011estimating}, precipitation data \citep{sun2015stochastic}, 
temperature data \citep{north2011correlation}, and wind speed data 
\citep{bevilacqua2020modeling}. This article aims at developing a 
flexible class of geostatistical models that is customizable to general 
non-Gaussian distributions, with particular focus on continuous data.

Several approaches have been developed for non-Gaussian geostatistical modeling. 
A straightforward approach consists of fitting 
a Gaussian process after transformation of the original data.
Possible transformations include Box-Cox \citep{de1997bayesian}, 
power \citep{allcroft2003latent}, square-root 
\citep{johns2003infilling}, and Tukey g-and-h 
\citep{xu2017tukey} transforms, to name a few. 
An alternative approach is to represent a non-Gaussian distribution as
a location-scale  mixture of Gaussian distributions.
This yields Gaussian process extensions
that are able to capture skewness and long tails
\citep{kim2004bayesian,palacios2006non, zhang2010spatial, mahmoudian2017skewed,
morris2017space, zareifard2018modeling, bevilacqua2021non}. 
Beyond methods based on continuous mixtures of Gaussian distributions,
Bayesian nonparametric methods  have been explored for
geostatistical data modeling, starting with the approach in 
\cite{gelfand2005bayesian} which extends the 
Dirichlet process \citep{ferguson1973bayesian} to a prior model for 
random  spatial surfaces. We refer to \cite{mueller2018nonparametric} for a review.
From a different perspective, \cite{bolin2014spatial} formulates
stochastic partial differential equations driven by non-Gaussian noise,
resulting in a class of non-Gaussian Mat\'ern fields.

An alternative popular approach involves 
a hierarchical model structure that assumes conditionally independent non-Gaussian
marginals, combined with a latent spatial process that is associated with some
functional or link function of the first-stage marginals. Hereafter, we refer to these models 
as hierarchical first-stage non-Gaussian models. If the latent process is linked
through a function of some parameter(s) of the first-stage marginal which belongs to
the exponential dispersion family, the approach is known as
the spatial generalized linear mixed model and its extensions \citep{diggle1998model}. 
Non-Gaussian spatial models that build
from copulas \citep{joe2014dependence} can also be classified into this category.
Copula models assume pre-specified families of marginals for observations, with 
a multivariate distribution underlying the copula for a vector of latent variables
that are probability integral transformations of the observations \citep{danaher2011modeling}. 
Spatial copula models replace the multivariate distribution with one 
that corresponds to a spatial process, thus introducing spatial dependence \citep{bardossy2006copula,ghosh2011hierarchical,krupskii2018factor,beck2020predicting}.

The non-Gaussian modeling framework proposed in this article is 
distinctly different from the previously mentioned approaches. Our 
methodology builds on the class of nearest-neighbor processes obtained by 
extending a joint density for a reference set of locations to the entire 
spatial domain. The original joint density is factorized into a product of 
conditionals with respect to a sparse directed acyclic graph (DAG).
Deriving each conditional 
from a Gaussian process results in the
nearest-neighbor Gaussian process \citep{datta2016hierarchical}.
Models defined over DAGs have received substantial attention; see, e.g., 
\cite{datta2016nonseparable, finley2019efficient, Peruzzi2020mesh,peruzzi2022spatial}.
The class of DAG-based models originates from 
the Vecchia approximation framework \citep{vecchia1988estimation}.
\cite{katzfuss2021general} provide further generalization.
Considerably less attention, however, has been devoted to defining models
over a sparse DAG with non-Gaussian distributions  for the conditionals of the 
joint density. This is in general a difficult
problem, as each conditional involves, say, a $p$-dimensional conditioning set,
which requires a coherent model for a $(p+1)$-dimensional non-Gaussian 
distribution, with $p$ potentially large. In this article, we take on the challenging 
task of developing a computationally efficient, interpretable framework that provides
generality for modeling different types of non-Gaussian data and
flexibility for complex spatial dependence.

We overcome the aforementioned challenge by modeling each 
conditional of the joint density as a weighted combination of 
spatially varying transition kernels, each of which depends on a specific neighbor. 
This approach produces multivariate non-Gaussian distributions by specification of 
the bivariate distributions that define the local transition kernels. 
Thus, it provides generality for modeling different non-Gaussian behaviors, 
since, relative to the multivariate analogue, constructing bivariate distributions 
is substantially easier, for instance, using bivariate copulas. Moreover, 
such a model structure offers the convenience of quantifying 
multivariate dependence through the collection of bivariate distributions. 
As an illustration, we study tail dependence properties under appropriate 
families of bivariate distributions, and provide results that guide modeling 
choices. The modeling framework 
achieves flexibility by letting both the weights and transition kernels 
be spatially dependent, inducing sufficient local dependence to 
describe a wide range of spatial variability. 
We refer to the resulting geospatial process as the
nearest-neighbor mixture process (NNMP).

An important feature of the 
model structure is that it facilitates the study of conditions 
for constructing NNMPs with pre-specified families of marginal distributions.
Such conditions are easily implemented without parameter constraints, thus 
resulting in a general modeling tool to describe spatial data distributions
that are skewed, heavy-tailed, positive-valued, or have bounded support, as
illustrated through several examples in Section \ref{sec:data_examples} and in
the supplementary material. The NNMP framework emphasizes direct modeling
by introducing spatial dependence at the data level. It avoids the use of
transformations that may distort the Gaussian process properties 
\citep{wallin2015geostatistical}. It is fundamentally different from the class of
hierarchical first-stage non-Gaussian models that introduce spatial dependence
through functionals of the data probability distribution, such as the transformed 
mean. Regarding computation, NNMP models do not require estimation of 
potentially large vectors of spatially correlated latent variables, 
something unavoidable with hierarchical first-stage non-Gaussian 
models. In fact, approaches for such models typically resort to approximate inference,
either directly or combined with a scalable model \citep{zilber2021vecchia}.
Estimation of NNMPs is analogous to that of a finite mixture model, 
thus avoiding the need to perform costly matrix operations for large data sets, 
and allowing for computationally efficient, full simulation-based inference. 
Overall, the NNMP framework offers a flexible class of models that is able to
describe complex spatial dependence, coupled with an efficient
computational approach, leveraged from the mixture structure of the model.

The rest of the article is organized as follows. In Section \ref{sec:framework}, 
we formulate the NNMP framework and study model properties. Specific 
examples of NNMP models illustrate different components of the methodology.
Section \ref{sec:inference} develops the general approach to Bayesian estimation 
and prediction under NNMP models. In Section \ref{sec:data_examples}, we 
demonstrate different NNMP models with a synthetic data example and 
with the analysis of Mediterranean Sea surface temperature data, respectively. 
Finally, Section \ref{sec:sum} concludes with a summary and discussion of future work.

\section{Nearest-Neighbor Mixture Processes for Spatial Data}
\label{sec:framework}

\subsection{Modeling Framework}

Consider a univariate spatial process $\{ Z(\sv): \sv \in \D \}$, 
where $\D\subset\R^p$, for $p\geq 1$.
Let $\BS = \{\bs_1,\dots,\bs_n\}$ be a finite collection of locations in $\D$, 
referred to as the reference set. We write the joint density of the random vector 
$\bz_{\BS} = (Z(\bs_1), \dots, Z(\bs_n))^\top$ as
$p(\bz_{\BS}) = p(z(\bs_1))\prod_{i=2}^n p(z(\bs_i)\mid z(\bs_{i-1}),\dots,z(\bs_1))$.
If we regard the conditioning set of $z(\bs_i)$ as the parents of $z(\bs_i)$,
the joint density $p(\bz_{\bS})$ 
is a factorization 
according to a DAG whose vertices are $z(\bs_i)$. We obtain a sparse DAG by 
reducing the conditioning set of $z(\bs_i)$ to a smaller subset, denoted as $\bz_{\tNe(\bs_i)}$,
with $\tNe(\bs_i)\subset\BS_i = \{\bs_1,\dots,\bs_{i-1}\}$.
We refer to $\tNe(\bs_i)$ as the neighbor set for $\bs_i$,
having at most $L$ elements with $L\ll n$.
The resulting density for the sparse DAG is
\begin{equation}\label{eq:dag}
\tilde{p}(\bz_{\BS}) = p(z(\bs_1))\prod_{i=2}^{n} p(z(\bs_i)\mid \bz_{\tNe(\bs_i)}),
\end{equation}
which is a proper density \citep{lauritzen1996graphical}.
Choosing the neighbor sets $\tNe(\bs_i)$ creates different sparse DAGs.
There are different ways to select members from $\BS_i$ for $\tNe(\bs_i)$;
see, e.g., \cite{vecchia1988estimation}, \cite{stein2004approximating},
and \cite{gramacy2015local}. Our selection is based on the geostatistical distance 
between $\bs_i$ and $\bs_j \in \BS_i$. 
The selected locations $\bs_j$ are placed in ascending order 
according to the distance, denoted as $\bs_{(i1)}, \dots, \bs_{(i,i_{L})}$,
where $i_L := (i-1)\wedge L$. We note that the development of the proposed framework
holds true for any choice of the neighbor sets.

Constructing a nearest-neighbor process involves specification 
of the conditional densities $p(z(\bs_i)\mid \bz_{\tNe(\bs_i)})$ in \eqref{eq:dag}, 
and extension to an arbitrary finite set in $\D$ that is not overlapping with $\BS$.
We approach the problem of constructing a nearest-neighbor non-Gaussian process 
following this idea. A notable difference, however, from the nearest-neighbor 
Gaussian process approach in \cite{datta2016hierarchical} is that we do not posit 
a parent process for $p(\bz_{\BS})$ when deriving $\tilde{p}(\bz_{\BS})$.
\cite{datta2016hierarchical} assume a Gaussian process for $p(\bz_{\BS})$, and use
it to model the conditional densities $p(z(\bs_i)\mid \bz_{\tNe(\bs_i)})$.
Similar ideas underlie the Vecchia approximation framework which considers 
\eqref{eq:dag} as an approximation for the density of a Gaussian process realization.
We instead utilize the structure in \eqref{eq:dag} to develop nearest-neighbor 
models for non-Gaussian spatial processes. 
To this end, we define the conditional density in the product in \eqref{eq:dag} as
\begin{equation}\label{eq:nnmp1}
    p(z(\bs_i)\mid\bz_{\tNe(\bs_i)}) = \sum_{l=1}^{i_L} w_l(\bs_i) \,
    f_{\bs_i,l}(z(\bs_i)\mid z(\bs_{(il)})),
\end{equation}
where $f_{\bs_i,l}$ is the $l$th component of the mixture
density $p$, and the weights satisfy $w_l(\bs_i) \geq 0$, for all $l$, 
and $\sum_{l=1}^{i_L}w_l(\bs_i) = 1$, for every $\bs_i\in\BS$.
In a sparse DAG, nearest neighbors in set $\tNe(\bs_i)$
are nodes that have directed edges pointing to $\bs_i$. Thus, it is appealing 
to consider a high-order Markov model in which temporal lags have a similar 
notion of direction. Our approach to formulate \eqref{eq:nnmp1} is motivated 
from a class of mixture transition distribution models \citep{le1996modeling}, 
which consists of a mixture of first-order transition densities with a vector of 
common weights. A key feature of the formulation in \eqref{eq:nnmp1} is the 
decomposition of a non-Gaussian conditional density, with a potentially large 
conditioning set, into a weighted sum of local conditional densities. This 
provides flexible, parsimonious modeling of $p(z(\bs_i)\mid\bz_{\tNe(\bs_i)})$
through specifying bivariate distributions that define the local
conditionals $f_{\bs_i,l}(z(\bs_i)\mid z(\bs_{(il)})$. We provide further
discussion on this feature for model construction and relevant properties 
in the following sections.

Spatial dependence characterized by \eqref{eq:nnmp1} is twofold. 
First, each component $f_{\bs_i,l}$ is associated 
with spatially varying parameters indexed at $\bs_i\in\BS$, 
defined by a probability model or a link function. 
Secondly, the weights $w_l(\bs_i)$ are spatially varying. 
As each component density $f_{\bs_i,l}$ depends on a specific neighbor, the weights
indicate the contribution of each neighbor of $\bs_i$. Besides, 
the weights adapt to the change of locations. For two different
$\bs_i,\bs_j$ in $\BS$, the relative locations of the nearest neighbors
$\tNe(\bs_i)$ to $\bs_i$ are different from that of $\tNe(\bs_j)$ to $\bs_j$. 
If all elements of $\tNe(\bs_i)$ are very close to $\bs_i$, then 
values of $(w_1(\bs_i),\dots,w_{i_{L}}(\bs_i))^\top$ should be quite even. 
On the other hand, if, for $\bs_j$, only a subset of its neighbors in $\tNe(\bs_j)$ are
close to $\bs_j$, then the weights corresponding to this subset should receive larger
values. We remark that in general, probability models or link functions for the spatially 
varying parameters should be considered case by case, given different specifications on the 
components $f_{\bs_i,l}$. Details of the construction for the component densities
and the weights are deferred to later sections.

We obtain the NNMP, a legitimate spatial process, by extending
\eqref{eq:nnmp1} to an arbitrary set of non-reference locations 
$\;\U = \{\su_1,\dots,\su_r\}$ where $\U\subset\D\setminus\BS$. In particular, 
we define the conditional density of $\bz_{\U}$ given $\bz_{\BS}$ as
\begin{equation}\label{eq:nnmp2}
    \tilde{p}(\bz_{\U}\mid \bz_{\BS}) = \prod_{i=1}^r
    p(z(\su_i)\mid \bz_{\tNe(\su_i)}) = 
    \prod_{i=1}^r\sum_{l=1}^L w_l(\su_i) \, f_{\su_i,l}(z(\su_i)\mid z(\su_{(il)})),
\end{equation}
where the specification on $w_l(\su_i)$ and $f_{\su_i,l}$ for all $i$ and all $l$ is 
analogous to that for \eqref{eq:nnmp1}, except that 
$\tNe(\su_i) = \{\su_{(i1)},\dots,\su_{(iL)}\}$ are the first $L$ locations
in $\BS$ that are closest to $\su_i$ in terms of geostatistical distance. 
Building the construction of the neighbor sets $\tNe(\su_i)$ on the reference set  
ensures that $\tilde{p}(\bz_{\U}\,|\,\bz_{\BS})$ is a proper density.

Given \eqref{eq:nnmp1} and \eqref{eq:nnmp2}, 
we can obtain the joint density $\tilde{p}(\bz_{\V})$ of a realization $\bz_{\V}$ 
over any finite set of locations $\V\subset\D$. When $\V\subset\BS$,
the joint density $\tilde{p}(\bz_{\V})$ is directly available as the 
appropriate marginal of $\tilde{p}(\bz_{\BS})$. Otherwise, we have 
$\tilde{p}(\bz_{\V}) = \int \tilde{p}(\bz_{\U}\,|\,\bz_{\BS})\tilde{p}(\bz_{\BS})
\prod_{\{\bs_i\in\BS\setminus\V\}}d z(\bs_i)$,
where $\U = \V\setminus\BS$. If $\BS\setminus\V$ is empty,
$\tilde{p}(\bz_{\V})$ is simply $\tilde{p}(\bz_{\U}\,|\,\bz_{\BS})\tilde{p}(\bz_{\BS})$.
In general, the joint density $\tilde{p}(\bz_{\V})$ of an NNMP is intractable. 
However, since both $\tilde{p}(\bz_{\U}\,|\,\bz_{\BS})$ and 
$\tilde{p}(\bz_{\BS})$ are products of mixtures, we can recognize that
$\tilde{p}(\bz_{\V})$ is a finite mixture, which suggests flexibility of the model 
to capture complex non-Gaussian dependence over the domain $\D$. Moreover, we show 
in Section \ref{sec:construction} that for some NNMPs, the joint density 
$\tilde{p}(\bz_{\V})$ has a closed-form expression.
In the subsequent development of model properties, 
we will use the conditional density
\begin{equation}\label{eq:nnmp3}
p(z(\sv)\mid \bz_{\tNe(\sv)}) = \sum_{l=1}^L w_l(\sv) \,
f_{\sv,l}(z(\sv)\mid z(\sv_{(l)})),\;\;\sv\in\D,
\end{equation}
to characterize an NNMP, where $\tNe(\sv)$ contains the first $L$
locations that are closest to $\sv$, selected from locations 
in $\BS$. These locations in $\tNe(\sv)$ are placed in ascending order according to 
distance, denoted as $\sv_{(1)},\dots, \sv_{(L)}$.

Before closing this section, we note that spatial locations are not naturally ordered. 
Given a distance function,
a different ordering on the locations results in different neighbor 
sets. Therefore, a different sparse DAG with density $\tilde{p}(\bz_{\BS})$
is created accordingly for model inference.
For the NNMP models illustrated in the data examples, 
we found through simulation experiments that there were no discernible differences
between the inferences based on $\tilde{p}(\bz_{\BS})$, 
given two different orderings. This observation is coherent with that 
from the literature that considers nearest-neighbor likelihood approximations.
Since the approximation of $\tilde{p}(\bz_{\BS})$ to $p(\bz_{\BS})$ depends on the 
information borrowed from the neighbors, as outlined in \cite{datta2016hierarchical},
the effectiveness is determined by the size of $\tNe(\bs_i)$ rather than the ordering.
A further remark is that, the ordering of the reference set $\BS$ is typically 
reserved for observed data. Thus, the ordering effect lies only in the model 
estimation based on \eqref{eq:nnmp1} with realization $\bz_{\BS}$. Spatial prediction 
typically rests on locations outside $\BS$ using \eqref{eq:nnmp2}, 
where the ordering effect disappears.

\subsection{NNMPs with Stationary Marginal Distributions}

We develop a sufficient condition to construct NNMPs with general stationary marginal 
distributions. The key feature of this result is that the condition relies on the 
bivariate distributions that define the first order transition kernels in \eqref{eq:nnmp3} 
without the need to impose restrictions on the parameter space. The supplementary material 
includes the proof of Proposition 1, as well as of Propositions 2, 3 and 4, and of 
Corollary 1, formulated later in Sections \ref{sec:construction} and \ref{sec:tail}.

\begin{prop}\label{prop:stationary}
Consider an NNMP for which the component density $f_{\sv,l}$ is specified by 
the conditional density of $U_{\sv,l}$ given $V_{\sv,l}$, where the random 
vector $(U_{\sv,l},V_{\sv,l})$ follows a bivariate distribution with marginal
densities $f_{U_{\sv,l}}$ and $f_{V_{\sv,l}}$, for $l = 1,\dots, L$.
The NNMP has stationary marginal density $f_Z$ if it satisfies the invariant 
condition: $Z(\bs_1)\sim f_Z$, $\bs_1\in\BS$, and for every $\sv\in\D$, 
$f_Z(z) =$ $f_{U_{\sv,l}}(z) =$ $f_{V_{\sv,l}}(z)$, for all $z$ and for all $l$.
\end{prop}

This result builds from the one in \cite{zheng2021construction} where 
mixture transition distribution models with stationary marginal distributions 
were constructed. It applies regardless of $Z(\sv)$ being a continuous, discrete 
or mixed random variable, thus allowing for a wide range of non-Gaussian marginal 
distributions and a general functional form, either linear or non-linear, for the 
expectation with respect to the conditional density $p$ in \eqref{eq:nnmp3}.

As previously discussed, the mixture model formulation for the conditional density 
in \eqref{eq:nnmp3} induces a finite mixture for the NNMP finite-dimensional distributions. 
On the other hand, due to the mixture form, an explicit expression for the covariance 
function is difficult to derive. A recursive equation can be obtained for a class of NNMP models 
for which the conditional expectation with respect to $(U_{\sv,l},V_{\sv,l})$ is linear, 
that is, $E(U_{\sv,l}\,|\,V_{\sv,l} = z) = a_l(\sv) + b_l(\sv) \, z$ for 
some $a_l(\sv), b_l(\sv)\in\mathbb{R},\;l = 1,\dots, L$, and for all $\sv\in\D$.
Suppose the NNMP has a stationary marginal distribution with finite first
and second moments. Without loss of generality, we assume the first moment is
zero. Then the covariance over any two locations
$\sv_1,\sv_2\in\D$ is
\begin{equation}\label{eq:cov}
\begin{aligned}
\cov(Z(\sv_1), Z(\sv_2)) = \begin{cases}
\sum_{l=1}^L w_l(\bs_i) \, b_l(\bs_i) \, E(Z(\bs_j)Z(\bs_{(il)})),
\;\;\sv_1\equiv\bs_i\in\BS,\sv_2\equiv\bs_j\in\BS,\\
\sum_{l=1}^L w_l(\sv_1) \, b_l(\sv_1) \, E(Z(\bs_j)Z(\sv_{(1l)})),\;\;
\sv_1\notin\BS, \sv_2\equiv\bs_j\in\BS,\\
\sum_{l=1}^L\sum_{l'=1}^Lw_{ll'} \, 
\{ a_{ll'}+ b_{ll'} E(Z(\sv_{(1l)})Z(\sv_{(2l')})) \}, 
\;\; \sv_1,\sv_2\notin\BS,
\end{cases}
\end{aligned}
\end{equation}
where $w_{ll'}\equiv w_l(\sv_1)w_{l'}(\sv_2)$,
$a_{ll'}\equiv a_l(\sv_1)a_{l'}(\sv_2)$,
$b_{ll'}\equiv b_l(\sv_1)b_{l'}(\sv_2)$, and
without loss of generality, we assume $i > j$.
The covariance in \eqref{eq:cov} implies that, even though the NNMP has a 
stationary marginal distribution, it is second-order non-stationary.

\subsection{Construction of NNMP Models}
\label{sec:construction}

The spatially varying conditional densities $f_{\sv,l}$ in \eqref{eq:nnmp3} correspond to 
a sequence of bivariate distributions indexed at $\sv$, namely, the distributions 
of $(U_{\sv,l},V_{\sv,l})$, for $l=1,...,L$.
To balance model flexibility and scalability, we build spatially varying 
distributions by considering the distribution of 
random vector $(U_l,V_l)$, for $l = 1,\dots, L$, and extending
some of its parameters to be spatially varying, that is, indexed in $\sv$. 
To this end, we use a probability model or a link function.
We refer to the random vectors $(U_l,V_l)$ as 
the set of base random vectors. With a careful choice of the model/function
for the spatially varying parameter(s), this construction method
reduces significantly the dimension of the parameter space, while preserving
the capability of the NNMP model structure to capture spatial dependence.

We illustrate the method with several examples below, starting 
with a bivariate Gaussian distribution and its continuous mixtures for real-valued
data, followed by a general strategy using bivariate copulas that can model data 
with general support. Before proceeding to the examples, we emphasize that our 
method allows for general bivariate distributions. One can also consider using 
a pair of compatible conditionals to specify bivariate distributions 
\citep{arnold1999conditional}, for instance, a pair of Lomax conditionals.
This is illustrated in Example \ref{ex:lomax-nnmp} in Section \ref{sec:tail}.

\begin{ex}\label{ex:gnnmp}

\normalfont \textit{Gaussian and continuous mixture of Gaussian NNMP models.}

\vspace{5pt}

For $l = 1,\dots,L$, take $(U_l,V_l)$ to be a bivariate Gaussian random 
vector with mean $\mu_l\bm 1_2$ and covariance matrix 
$\Sigma_l = \sigma_l^2\left(\begin{smallmatrix}1 & \rho_l\\ 
\rho_l & 1\end{smallmatrix}\right)$, where $\bm 1_2$ is 
the two-dimensional column vector of ones,
resulting in a Gaussian conditional density 
$f_{U_l|V_l}(u_l\,|\, v_l) = N(u_l\,|\,(1-\rho_l)\mu_l +
\rho_lv_l,\sigma_l^2(1-\rho_l^2))$. If we extend the correlation parameter
to be spatially varying, $\rho_l(\sv) =$ $k_l(\sv,\sv_{(l)})$, for a correlation 
function $k_l$, we obtain the spatially varying conditional density,
\begin{equation}\label{eq:GNNMP}
p(z(\sv) \mid \bz_{\tNe(\sv)}) = \sum_{l=1}^{L} w_l(\sv) \, 
N(z(\sv) \mid (1-\rho_l(\sv))\mu_l + 
\rho_l(\sv)z(\sv_{(l)}), \sigma_l^2(1-(\rho_l(\sv))^2)).
\end{equation}
This NNMP is referred to as the Gaussian NNMP. If we take $Z(\bs_1) \sim N(z\,|\,\mu,\sigma^2)$, 
and set $\mu_l = \mu$ and $\sigma^2_l = \sigma^2$, for all $l$, the resulting model
satisfies the invariant condition of Proposition \ref{prop:stationary} with 
stationary marginal given by the $N(\mu,\sigma^2)$ distribution. 
The finite-dimensional distribution of the Gaussian NNMP is characterized 
by the following proposition.
\end{ex}

\begin{prop}\label{prop:gausfdd}
Consider the Gaussian NNMP in \eqref{eq:GNNMP} 
with $\mu_l = \mu$ and $\sigma^2_l = \sigma^2$, for all $l$. 
If $Z(\bs_1)\sim N(z\,|\,\mu,\sigma^2)$, the Gaussian NNMP has the $N(\mu,\sigma^2)$ 
stationary marginal distribution, and its finite-dimensional distributions are  
mixtures of multivariate Gaussian distributions.
\end{prop}

We refer to the model in Proposition \ref{prop:gausfdd} as the stationary Gaussian NNMP. Based on the 
Gaussian NNMP, various NNMP models with different families for $(U_l,V_l)$ can be 
constructed by exploiting location-scale mixtures of Gaussian distributions. 
We illustrate the approach with the skew-Gaussian NNMP model. 
Denote by $\mathrm{TN}(\mu,\sigma^2; a, b)$ the Gaussian distribution
with mean $\mu$ and variance $\sigma^2$, truncated at the interval $(a, b)$.
Building from the Gaussian NNMP, we start with a
conditional bivariate Gaussian distribution for $(U_l,V_l)$, given
$z_0\sim\mathrm{TN}(0,1;0,\infty)$, where $\mu_l$ is replaced
with $\mu_l + \la_lz_0$. Marginalizing out $z_{0}$ yields
the bivariate skew-Gaussian distribution for $(U_l,V_l)$ \citep{azzalini2013skew}. 
Extending again $\rho_l$ to $\rho_l(\sv)$, for all $l$, we can express the conditional 
density $p(z(\sv)\,|\,\bz_{\tNe(\sv)})$ for the skew-Gaussian NNMP model as
$\sum_{l=1}^{L} w_l(\sv)\int_0^{\infty}N(z(\sv)\,|\,\mu_l(\sv),\sigma_l^2(\sv))
\, \mathrm{TN}(z_0(\sv)\,|\,\mu_{0l}(\sv_{(l)}),\sigma_{0l}^2;0,\infty)dz_0(\sv),$
where we have: $\mu_l(\sv) =$ $\{ 1-\rho_l(\sv) \} 
\{ \mu_l+\la_lz_0(\sv) \} + \rho_l(\sv)z(\sv_{(l)})$;
$\sigma_l^2(\sv) =$ $\sigma_l^2 \{ 1-(\rho_l(\sv))^2 \}$;
$\mu_{0l}(\sv_{(l)}) =$ $\{ z(\sv_{(l)})-\mu_l \} \la_l/(\sigma^2_l + \la_l^2)$;
and $\sigma_{0l}^2 =$ $\sigma^2_l/(\sigma^2_l + \la_l^2)$.
Setting $\la_l = \la$, $\mu_l = \mu$, and $\sigma^2_l = \sigma^2$, for all $l$, 
we obtain the stationary skew-Gaussian NNMP model, with skew-Gaussian marginal 
$f_Z(z) = 2 \, N(z\,|\, \mu,\la^2+\sigma^2) \, \Phi((z-\mu)\la/(\sigma\sqrt{\la^2 + \sigma^2}))$,
denoted as $\mathrm{SN}(\mu, \la^2+\sigma^2, \la/\sigma)$.

The skew-Gaussian NNMP model is an example of a location mixture of Gaussian distributions.
Scale mixtures can also be considered to obtain, for example, the Student-t
model. In that case, we replace the covariance matrix $\Sigma_l$ 
with $c\Sigma_l$, taking $c$ as a random variable with an appropriate 
inverse-gamma distribution. Important families that admit a location and/or scale 
mixture of Gaussians representation include the skew-t, Laplace, and asymmetric 
Laplace distributions. Using a similar approach to the one for the skew-Gaussian 
NNMP example, we can construct the corresponding NNMP models.

\begin{ex}\label{ex:cop-nnmp}

\normalfont \textit{Copula NNMP models.}

\vspace{5pt}

A copula function $C: [0,1]^p \rightarrow [0,1]$ is a
function such that, for any multivariate distribution $F(z_1,\dots,z_p)$, 
there exists a copula $C$ for which $F(z_1,\dots,z_p) = C(F_1(z_1),\dots,F_p(z_p))$, 
where $F_j$ is the marginal distribution function of $Z_j$, $j = 1,\dots,p$
\citep{sklar1959fonctions}.
If $F_j$ is continuous for all $j$, $C$ is unique. A copula enables us to separate
the modeling of the marginal distributions from the dependence. Thus, the invariant
condition in Proposition \ref{prop:stationary} can be attained by specifying the
stationary distribution $F_Z$ as the marginal distribution of $(U_l,V_l)$ for all
$l$. The copula parameter that determines the dependence  of $(U_l,V_l)$ can be
modeled as spatially varying to create a sequence of spatially dependent bivariate
vectors $(U_{\sv,l},V_{\sv,l})$. Here, we focus on continuous distributions,
although this strategy can be applied for any family of distributions for $F_Z$. 
We consider bivariate copulas with a single copula parameter, and illustrate next 
the construction of a copula NNMP given a stationary marginal density $f_Z$.

For the bivariate distribution of each $(U_l,V_l)$ with marginals $f_{U_l}$
and $f_{V_l}$, we consider a copula $C_l$ with parameter $\eta_l$, for 
$l = 1,\dots, L$. We obtain a spatially varying copula $C_{\sv,l}$
for $(U_{\sv,l},V_{\sv,l})$ by extending $\eta_l$ to $\eta_l(\sv)$.
The joint density of $(U_{\sv,l},V_{\sv,l})$ is given by
$c_{\sv,l}(z(\sv),z(\sv_{(l)}))f_{U_{\sv,l}}(z(\sv))f_{V_{\sv,l}}(z(\sv_{(l)}))$,
where $c_{\sv,l}$ is the copula density of $C_{\sv,l}$, and
$f_{U_{\sv,l}} = f_{U_l}$ and $f_{V_{\sv,l}} = f_{V_l}$ are the marginal 
densities of $U_{\sv,l}$ and $V_{\sv,l}$, respectively.
Given a pre-specified stationary marginal $f_Z$, 
we replace both $f_{U_{\sv,l}}$ and $f_{V_{\sv,l}}$ with $f_Z$,
for every $\sv$ and for all $l$.
We then obtain the conditional density
\begin{equation}\label{eq:copula-nnmp}
p(z(\sv)\mid\bz_{\tNe(\sv)}) = \sum_{l=1}^L w_l(\sv) \,
c_{\sv,l}(z(\sv),z(\sv_{(l)}))\,f_Z(z(\sv))
\end{equation}
that characterizes the stationary copula NNMP.

Under the copula framework, one strategy to specify the spatially varying
parameter is through the Kendall's $\tau$ coefficient.
The Kendall's $\tau$, taking values in $[-1,1]$, is a bivariate concordance 
measure with properties useful for non-Gaussian modeling.
In particular, its existence does not 
require finite second moment and it is invariant under strictly increasing 
transformations. If $(U_l,V_l)$ is continuous with a copula $C_l$, its
Kendall's $\tau$ is 
$\rho_{\tau,l} = 4\int_{[0,1]^2}C_ldC_l - 1$.
Taking $A_l \subset [-1,1]$ as the range of $\rho_{\tau,l}$, we
can construct a composition function $h_l:= g_l \circ k_l$ for some link function
$g_l: A_l\rightarrow H_l$  and kernel function $k_l: \D\times \D \rightarrow A_l$,
where $H_l$ is the parameter space associated with $C_l$.
The kernel $k_l$ should be specified with caution; $k_l$ must satisfy axioms in the 
definition of a bivariate concordance measure (\citealt{joe2014dependence}, Section 2.12).
We illustrate the strategy with the following example.

\begin{ex}\label{ex:gumbel}
\normalfont
The bivariate Gumbel copula is an asymmetric copula useful for 
modeling dependence when the marginals are positive and 
heavy-tailed. The spatial Gumbel copula can be defined as
$C_{\sv,l} = \exp\big(-[\{-\log F_{U_{\sv,l}}(z(\sv))\}^{\eta_l(\sv)} + 
\{-\log F_{V_{\sv,l}}(z(\sv_{(l)}))\}^{\eta_l(\sv)}]^{1/\eta_l(\sv)}\big)$,
where $\eta_l(\sv)\in [1,\infty)$ and perfect dependence is obtained if 
$\eta_l(\sv)\rightarrow\infty$. The Kendall's $\tau$ is
$\rho_{\tau,l}(\sv) = 1 - \eta_{l}^{-1}(\sv)$, taking values in $[0,1]$. 
We define $\rho_{\tau,l}(\sv) := k_l(||\sv - \sv_{(l)}||)$, an isotropic
correlation function. Let $g_l(x) = (1 - x)^{-1}$. Then, the function
$h_l(||\sv-\sv_{(l)}||) = g_l\circ k_l(||\sv-\sv_{(l)}||) = (1 - k_l(||\sv -\sv_{(l)}||))^{-1}$.
Thus, the parameter $\eta_l(\sv)\equiv \eta(||\sv-\sv_{(l)}||)$ is given by 
$h_l(||\sv-\sv_{(l)}||)$, and 
$\eta_{l}(\sv)\rightarrow\infty$ as $||\sv - \sv_{(l)}||\rightarrow0$.
\end{ex}

After we define a spatially varying copula, we obtain a family of copula NNMPs 
by choosing a desired family of marginal distributions. Section \ref{sec:sim}
illustrates Gaussian and Gumbel copula NNMP models with gamma marginals,
and the supplementary material provides an additional example of a Gaussian 
copula NNMP with beta marginals.

Copula NNMP models offer avenues to capture complex dependence using 
general bivariate copulas. Traditional spatial copula models specify the finite 
dimensional distributions of the underlying spatial process with a multivariate copula. 
However, multivariate copulas need to be used with careful consideration in a spatial
setting. For example, it is common to assume that spatial processes exhibit stronger 
dependence at smaller distances. Thus, copulas such as the multivariate Archimedean
copula that induce an exchangeable dependence structure are inappropriate. Though
spatial vine copula models \citep{graler2014modelling} can resolve this
restriction, their model structure and computation are substantially more complicated
than copula NNMP models.

\end{ex}

\subsection{Mixture Component Specification and Tail Dependence}
\label{sec:tail}

A benefit of building NNMPs from a set of 
base random vectors is that specification of the multivariate
dependence of $Z(\sv)$ given its neighbors is determined mainly by that of 
the base random vectors. In this section, we illustrate this attractive property 
of the model with the establishment of lower bounds for two measures used to assess 
strength of tail dependence.

The main assumption is that the base 
random vector $(U_l,V_l)$ has stochastically increasing positive dependence.
$U_l$ is said to be stochastically increasing in $V_l$, 
if $P\big(U_l > u_l\,|\, V_l = v_l\big)$ increases as $v_l$ increases. %for all $u_l$. 
The definition can be extended to a multivariate random vector $(Z_1,\dots,Z_p)$.
$Z_1$ is said to be stochastically increasing in $(Z_2,\dots, Z_p)$ if 
$P\big(Z_1 > z_1\,|\, Z_2 = z_2,\dots, Z_p = z_p\big) \leq P\big(Z_1 > z_1\,|\, Z_2 = z_2',\dots,
Z_p = z_p'\big)$, for all $(z_2,\dots,z_p)$ and $(z_2',\dots,z_p')$ in the support 
of $(Z_2,\dots,Z_p)$, where $z_j\leq z_j'$, for $j = 2,\dots,p$. 
The conditional density in \eqref{eq:nnmp3} implies that
$$
P\big(Z(\sv) > z\,|\, \bm{Z}_{\tNe(\sv)} = \bz_{\tNe(\sv)}\big) = 
\sum_{l=1}^L w_l(\sv) \, P\big(Z(\sv) > z\,|\, Z(\sv_{(l)}) = z(\sv_{(l)})\big).
$$
Therefore, $Z(\sv)$ is stochastically increasing in $\bm{Z}_{\tNe(\sv)}$ if 
$Z(\sv)$ is stochastically increasing in $Z(\sv_{(l)})$ with respect to
$(U_{\sv,l},V_{\sv,l})$ for all $l$. If the sequence $(U_{\sv,l},V_{\sv,l})$
is built from the vector $(U_l,V_l)$, then the set of base random vectors
determines the stochastically increasing positive dependence of $Z(\sv)$
given its neighbors.

For a bivariate random vector $(U_l,V_l)$, the upper and lower
tail dependence coefficients, denoted as  $\lambda_{\iH,l}$ and $\lambda_{\iL,l}$,
respectively, are $\lambda_{\iH,l} = \lim_{q\rightarrow1^-}
P\big(U_l > F_{U_l}^{-1}(q)\,|\, V_l > F_{V_l}^{-1}(q)\big)$ and
$\lambda_{\iL,l} = \lim_{q\rightarrow0^+}
P\big(U_l \leq F_{U_l}^{-1}(q)\,|\, V_l \leq F_{V_l}^{-1}(q)\big)$.
When $\lambda_{\iH,l} > 0$, we say $U_l$ and $V_l$ have upper tail dependence. 
When $\lambda_{\iH,l} = 0$, $U_l$ and $V_l$ are said to be asymptotically 
independent in the upper tail. Lower tail dependence and asymptotically
independence in the lower tail are similarly defined using $\lambda_{\iL,l}$.
Let $F_{Z(\sv)}$ be the marginal distribution function of $Z(\sv)$.
Analogously, we can define the upper and lower tail dependence coefficients for
$Z(\sv)$ given its nearest neighbors, 
$$
\begin{aligned}
\lambda_{\iH}(\sv) & = \lim_{q\rightarrow1^-}P\big(Z(\sv) > F_{Z(\sv)}^{-1}(q)\mid 
Z(\sv_{(1)}) > F_{Z(\sv_{(1)})}^{-1}(q), \dots, 
Z(\sv_{(L)}) > F_{Z(\sv_{(L)})}^{-1}(q)\big),\\
\lambda_{\iL}(\sv) & = \lim_{q\rightarrow0^+} P\big(Z(\sv) \leq F_{Z(\sv)}^{-1}(q)\mid 
Z(\sv_{(1)}) \leq F_{Z(\sv_{(1)})}^{-1}(q), \dots, 
Z(\sv_{(L)}) \leq F_{Z(\sv_{(L)})}^{-1}(q)\big).
\end{aligned}
$$
The following proposition provides lower bounds for the tail dependence coefficients.

\begin{prop}\label{prop:tail1}
Consider an NNMP for which the component density $f_{\sv,l}$ is specified by 
the conditional density of $U_{\sv,l}$ given $V_{\sv,l}$, where the random 
vector $(U_{\sv,l},V_{\sv,l})$ follows a bivariate distribution with marginal
distribution functions $F_{U_{\sv,l}}$ and $F_{V_{\sv,l}}$, for $l = 1,\dots, L$. 
%For each $\sv$, 
The spatial dependence of random vector $(U_{\sv,l},V_{\sv,l})$ is 
built from the base vector $(U_l,V_l)$, which has a bivariate distribution such 
that $U_l$ is stochastically increasing in $V_l$, for $l = 1,\dots,L$.
Then, for every $\sv$, the lower bound for the upper tail 
dependence coefficient $\lambda_{\iH}(\sv)$ is
$\sum_{l=1}^Lw_l(\sv)\lim_{q\rightarrow1^-}
P\big(Z(\sv) > F_{U_{\sv,l}}^{-1}(q)\,|\,Z(\sv_{(l)}) = F_{V_{\sv,l}}^{-1}(q)\big)$, and
the lower bound for the lower tail dependence coefficient $\lambda_{\iL}(\sv)$ is
$\sum_{l=1}^Lw_l(\sv)\lim_{q\rightarrow0^+}
P\big(Z(\sv) \leq F_{U_{\sv,l}}^{-1}(q)\,|\, 
Z(\sv_{(l)}) = F_{V_{\sv,l}}^{-1}(q)\big)$.
\end{prop}

Proposition \ref{prop:tail1} establishes that the lower and upper tail dependence 
coefficients are bounded below by a convex combination of, respectively, the limits 
of the conditional distribution functions and the conditional survival 
functions. These are fully determined by the dependence structure of the bivariate 
distribution for $(U_l,V_l)$. The result is best illustrated with an example.

\begin{ex}\label{ex:lomax-nnmp}
\normalfont
Consider a Lomax NNMP for which the bivariate distributions of the base random
vectors correspond to a bivariate Lomax distribution \citep{arnold1999conditional},
resulting in conditional density, $p(z(\sv)\,|\,\bz_{\tNe(\sv)}) =$
$\sum_{l=1}^L w_l(\sv) \, \mathrm{Lo}(z(\sv)\,|\, z(\sv_{(l)}) + 
\phi_l, \alpha_l(\sv))$,
where $\mathrm{Lo}(x\,|\, \phi,\alpha) = \alpha\phi^{-1}(1 + x\phi^{-1})^{-(\alpha+1)}$
denotes the Lomax density, a shifted version of the Pareto Type I density.
A small value of $\alpha$ indicates a heavy 
tail. The component conditional survival function of the Lomax NNMP, expressed 
in terms of the quantile $q$, is $\left\{1 +
F_{U_{\sv,l}}^{-1}(q)/\big(F_{V_{\sv,l}}^{-1}(q) +
\phi_l\big)\right\}^{-\alpha_l(\sv)}$ which converges to $2^{-\alpha_l(\sv)}$ as
$q\rightarrow1^-$. Therefore, the lower bound for $\lambda_{\iH}(\sv)$ 
is $\sum_{l=1}^Lw_l(\sv) \, 2^{-\alpha_l(\sv)}$. 
As $\alpha_l(\sv)\rightarrow0$ for all $l$, the lower bound for
$\lambda_{\iH}(\sv)$ tends to one, and hence $\lambda_{\iH}(\sv)$ tends to one,
since $\lambda_{\iH}(\sv)\leq 1$. As $\alpha_l(\sv)\rightarrow\infty$ for all $l$,
the lower bound tends to zero.
\end{ex}

Proposition \ref{prop:tail1} holds for the general framework. 
If the distribution of $(U_l,V_l)$
with $F_{U_l} = F_{V_l}$ has first order partial derivatives and 
exchangeable dependence, namely $(U_l,V_l)$ and $(V_l,U_l)$
have the same joint distribution,
the lower bounds of the tail dependence coefficients 
depend on the component tail dependence coefficients.
The result is summarized in the following corollary.

\begin{coro}\label{coro:oftail1}
Suppose that the base random vector $(U_l,V_l)$ in Proposition \ref{prop:tail1}
is exchangeable, and its bivariate distribution with marginals $F_{U_l} = F_{V_l}$
has first order partial derivatives, for all $l$. Then the upper and lower tail dependence 
coefficients $\lambda_{\iH}(\sv)$ and $\lambda_{\iL}(\sv)$ are 
bounded below by $\sum_{l=1}^Lw_l(\sv)\lambda_{\iH,l}(\sv)/2$
and $\sum_{l=1}^Lw_l(\sv)\lambda_{\iL,l}(\sv)/2$, 
where $\lambda_{\iH,l}(\sv)$ and $\lambda_{\iL,l}(\sv)$ are the tail dependence coefficients
with respect to $(U_{\sv,l},V_{\sv,l})$.
\end{coro}

Under Corollary \ref{coro:oftail1}, if the bivariate distribution of $(U_l,V_l)$ is symmetric,
for instance, an elliptically symmetric distribution, the upper and lower tail dependence 
coefficients coincide, and can simply be denoted as $\lambda(\sv)$. Then, we have
that $\lambda(\sv)\geq\sum_{l=1}^L w_l(\sv)\lambda_l(\sv)/2$, where $\lambda_l(\sv)$
is the tail dependence coefficient with respect to $(U_{\sv,l},V_{\sv,l})$.

Tail dependence can also be quantified using the boundary of the conditional distribution 
function, as proposed in \cite{hua2014strength} for a bivariate random 
vector. In particular, the upper tail dependence of $(U_l,V_l)$ is said to have 
some strength if 
$F_{U_l|V_l}\big(F_{U_l}^{-1}(q)\,|\, F_{V_l}^{-1}(1)\big)$ is positive at $q = 1$.
Likewise, a non-zero $F_{U_l|V_l}\big(F_{U_l}^{-1}(q)\,|\, F_{V_l}^{-1}(0)\big)$
at $q=0$ indicates some strength of dependence in the lower tails.
The functions $F_{U_l|V_l}\big(\cdot\mid F_{V_l}^{-1}(0)\big)$ and 
$F_{U_l|V_l}\big(\cdot\mid F_{V_l}^{-1}(1)\big)$ are referred to as the boundary 
conditional distribution functions.

We use $F_{1|2}\big(\cdot\,|\, F_{\bZ_{\tNe(\sv)}}^{-1}(q)\big)$ for simpler 
notation for the conditional distribution function of $Z(\sv)$, 
$F\big(\cdot\,|\, Z(\sv_{(1)}) = F_{Z(\sv_{(1)})}^{-1}(q),\dots,Z(\sv_{(L)}) = F_{Z(\sv_{(L)})}^{-1}(q)\big)$.
Then $F_{1|2}\big(\cdot\,|\, F_{\bZ_{\tNe(\sv)}}^{-1}(0)\big)$ and 
$F_{1|2}\big(\cdot\,|\, F_{\bZ_{\tNe(\sv)}}^{-1}(1)\big)$ are the boundary conditional 
distribution functions for the NNMP model. The upper tail dependence is said to be
i) strongest if $F_{1|2}\big(F_{Z(\sv)}^{-1}(q)\,|\, F_{\bZ_{\tNe(\sv)}}^{-1}(1)\big)$ 
equals $0$ for $0\leq q < 1$ and has a mass of 1 at $q = 1$;
ii) intermediate if $F_{1|2}\big(F_{Z(\sv)}^{-1}(q)\,|\, F_{\bZ_{\tNe(\sv)}}^{-1}(1)\big)$ 
has positive but not unit mass at $q = 1$;
iii) weakest if $F_{1|2}\big(F_{Z(\sv)}^{-1}(q)\,|\, F_{\bZ_{\tNe(\sv)}}^{-1}(1)\big)$ has 
no mass at $q = 1$. The strength of lower tail dependence is defined likewise using
$F_{1|2}\big(F_{Z(\sv)}^{-1}(q)\,|\, F_{\bZ_{\tNe(\sv)}}^{-1}(0)\big)$.
The following result gives lower bounds for the boundary conditional 
distribution functions.

\begin{prop}\label{prop:tail2}
Consider an NNMP for which the component density $f_{\sv,l}$ is specified by 
the conditional density of $U_{\sv,l}$ given $V_{\sv,l}$.
The spatial dependence of random vector $(U_{\sv,l},V_{\sv,l})$ is 
built from the base vector $(U_l,V_l)$, which has a bivariate distribution such 
that $U_l$ is stochastically increasing in $V_l$, for $l = 1,\dots,L$.
Let $\lambda_{\iL,l}(\sv)$ and  $\lambda_{\iH,l}(\sv)$
be the lower and upper tail dependence coefficients corresponding to 
$(U_{\sv,l},V_{\sv,l})$. If for a given $\sv$, there exists 
$\lambda_{\iL,l}(\sv) > 0$ for some $l$, then the conditional distribution function
$F_{1|2}\big(F_{Z(\sv)}^{-1}(q)\,|\, F_{\bZ_{\tNe(\sv)}}^{-1}(0)\big)$ has strictly positive 
mass $p_0(\sv)$ at $q = 0$ with 
$p_0(\sv)\geq \sum_{l=1}^Lw_l(\sv)\lambda_{\iL,l}(\sv)$. 
Similarly, if for a given $\sv$, there exists $\lambda_{\iH,l}(\sv) > 0$ for some $l$, 
then the conditional distribution function
$F_{1|2}\big(F_{Z(\sv)}^{-1}(q)\,|\, F_{\bZ_{\tNe(\sv)}}^{-1}(1)\big)$ 
has strictly positive mass $p_1(\sv)$ at $q = 1$ with 
$p_1(\sv)\geq \sum_{l=1}^Lw_l(\sv)\lambda_{\iH,l}(\sv)$.
\end{prop}

Proposition \ref{prop:tail2} complements Proposition \ref{prop:tail1} to 
assess strength of tail dependence. It readily applies
for bivariate distributions, especially for copulas which yield explicit 
expressions for the tail dependence coefficients. In particular, the spatially
varying Gumbel copula $C_{\sv,l}$ in Example \ref{ex:gumbel} has upper tail
dependence coefficient $2 - 2^{1/\eta_l(\sv)} > 0$ for $\eta_l(\sv) > 1$, 
so the tail dependence of a Gumbel copula NNMP model
has some strength if $\eta_l(\sv)>1$ for some $l$. In fact, applying the 
result in \cite{hua2014strength}, with a Gumbel copula, 
$F_{1|2}\big(F_{Z(\sv)}^{-1}(q)\,|\, F_{\bZ_{\tNe(\sv)}}^{-1}(1)\big)$ degenerates 
at $q = 1$, implying strongest tail dependence.

\section{Bayesian Hierarchical Model and Inference}
\label{sec:inference}

\subsection{Hierarchical Model Formulation}
\label{sec:hier}

We introduce the general approach for NNMP Bayesian implementation,
treating the observed spatial responses as an NNMP realization.
The inferential framework can be easily extended to incorporate model components
that may be needed in practical settings, such as covariates and additional error terms.
We illustrate the extensions with the real data analysis in Section \ref{sec:app}
and in the supplementary material,
and provide further discussion in Section \ref{sec:sum}.

Our approach for inference is based on a likelihood conditional on the first $L$ elements 
of the realization $\bz_{\BS} = (z(\bs_1),\dots,z(\bs_n))^\top$ over the reference set 
$\BS\subset\D$. Following a commonly used approach for mixture models fitting, we use data
augmentation to facilitate inference. For each $z(\bs_i)$, $i = L+1,\dots,n$, we 
introduce a configuration variable $\ell_i$, taking values in $\{1,\dots,L\}$,
such that $P\big(\ell_i\,|\,\bm w(\bs_i)\big) = \sum_{l=1}^Lw_l(\bs_i)\delta_l(\ell_i)$,
where $\bm w(\bs_i) = (w_1(\bs_i),\dots, w_L(\bs_i))^\top$, and $\delta_l(\ell_i) = 1$
if $\ell_i = l$ and $0$ otherwise. Conditional on the configuration
variables and the vector $(z(\bs_1),\dots,z(\bs_L))^\top$, the augmented model is
\begin{equation}\label{eq:hier}
\begin{aligned}
z(\bs_i) \mid z(\bs_{(i,\ell_i)}),\ell_i,\btheta \,\stackrel{ind.}{\sim}
f_{\bs_i,\ell_i}(z(\bs_i)\,|\,z(\bs_{(i,\ell_i)}),\btheta),\;\;\;
\ell_i\mid \bm w(\bs_i)\, \stackrel{ind.}{\sim}\sum_{l=1}^{L}w_l(\bs_i)\delta_l(\ell_i),\\
\end{aligned}
\end{equation}
where $\btheta$ collects the parameters of the densities $f_{\bs_i,l}$.

A key component of the Bayesian model formulation is the prior model for 
the weights. Weights are allowed to vary in space, adjusting to the
neighbor structure of different reference locations. We describe the construction 
for weights corresponding to a point in the reference set. For non-reference points, 
weights are defined analogously.
Consider a collection of spatially dependent distribution functions
$\{G_{\bs_i}: \bs_i\in\BS\}$ supported on $(0,1)$.
For each $\bs_i$, the weights are defined as the increments of $G_{\bs_i}$ with 
cutoff points $r_{\bs_i,0}\,,\dots,r_{\bs_i,L}$. More specifically, 
\begin{equation}\label{eq:weights}
w_l(\bs_i) = \int \mathbbm{1}_{(r_{\bs_i,l-1},\,r_{\bs_i, l})}(t) \, 
dG_{\bs_i}(t), \;\;\;l = 1,\dots, L,
\end{equation}
where $\mathbbm{1}_A$ denotes the indicator function for set $A$.
The cutoff points 
$0 = r_{\bs_i,0} < r_{\bs_i,1} < \dots < r_{\bs_i,L} = 1$ are such that, for $l = 1,\dots, L$, 
$r_{\bs_i, l}-r_{\bs_i, l-1} = k'(\bs_i,\bs_{(il)}\,|\,\bzeta)/\sum_{l=1}^{L}k'(\bs_i,\bs_{(il)}\,|\,\bzeta)$,
where $k': \D\times\D \rightarrow [0,1]$ is a bounded kernel function with parameters $\bzeta$. 
The kernel and its associated parameters affect the smoothness of the 
resulting random field. We take $G_{\bs_i}$ as a logit Gaussian distribution, denoted as
$G_{\bs_i}(\cdot\,|\,\mu(\bs_i),\kappa^2)$,
such that the 
corresponding Gaussian distribution has mean $\mu(\bs_i)$ and variance 
$\kappa^2$. 
The spatial dependence across the weights is introduced 
through the mean $\mu(\bs_i) = \gamma_0 + \gamma_1s_{i1} + \gamma_2s_{i2}$, 
where $\bs_i = (s_{i1},s_{i2})$.
Given the cutoff points and $\kappa^2$, a small value of $\mu(\bs_i)$
favors large weights for the near neighbors of $\bs_i$. 
A simpler version of the model in \eqref{eq:weights} is obtained by 
letting $G_{\bs_i}$ be the uniform distribution on $(0,1)$. Then the weights become
$ k'(\bs_i,\bs_{(il)}\,|\,\bzeta)/\sum_{l=1}^{L}k'(\bs_i,\bs_{(il)}\,|\,\bzeta)$.
We notice that \cite{cadonna2019bayesian} use a set of fixed, uniform cutoff 
points on $[0,1]$, i.e., $r_{\bs_i,l}-r_{\bs_i,l-1} = 1/L$, 
for spectral density  estimation, with a collection of logit Gaussian 
distributions indexed by frequency.

The full Bayesian model is completed with prior specification for parameters 
$\btheta,\bzeta,\bga = (\gamma_0,\gamma_1,\gamma_2)^\top$, and $\kappa^2$.
The priors for $\btheta$ and $\bzeta$ depend on the choices of the densities 
$f_{\bs_i,l}$ and the cutoff point kernel $k'$, respectively.
For parameters $\bga$ and $\kappa^2$, we specify 
$N(\bga\,|\,\bmu_{\gamma},\bV_{\gamma})$ and
$\mathrm{IG}(\kappa^2\,|\, u_{\kappa^2},v_{\kappa^2})$ priors, respectively,
where $\mathrm{IG}$ denotes the inverse gamma distribution.

Finally, we note that an NNMP model requires selection of the neighborhood size 
$L$. This can be done using standard model comparison metrics, scoring 
rules, or information criteria; for example, \cite{datta2016hierarchical} used
root mean square predictive error and \cite{guinness2018permutation} used 
Kullback–Leibler divergence.
In general, a larger $L$ increases computational cost. \cite{datta2016hierarchical}
conclude that a moderate value $L$ $(\leq 20)$ typically suffices for the 
nearest-neighbor Gaussian process models.
\cite{Peruzzi2020mesh} point out that a smaller $L$ corresponds to a 
larger Kullback–Leibler divergence of $\tilde{p}(\bz_{\BS})$ from $p(\bz_{\BS})$, 
regardless of the distributional assumption of the density. Moreover, it is possible 
that information from the farthest neighbors is also important \citep{stein2004approximating}.
Therefore, for large non-Gaussian data sets with complex dependence, one may seek a larger 
$L$ to obtain a better approximation to the full model. 
Our model for the weights allows taking a relatively large neighbor set with less 
computational demand. We assign small probabilities a priori to distant neighbors.
The contribution of each neighbor will be induced by the mixing, with important neighbors
being assigned large weights a posteriori.

\subsection{Estimation and Prediction}\label{sec:est}

We implement a Markov chain Monte Carlo sampler to simulate from the posterior distribution of 
the model parameters.
To allow for efficient simulation of parameters $\bga$ and $\kappa^2$, we associate
each $y(\bs_i)$ with a latent Gaussian variable $t_i$ with mean $\mu(\bs_i)$ and variance $\kappa^2$.
There is a one-to-one correspondence between the configuration variables $\ell_i$ and 
latent variables $t_i$: $\ell_i = l$ if and only if $t_i\in(r_{\bs_i,l-1}^*,r_{\bs_i,l}^*)$
where $r^*_{\bs_i,l} = \log(r_{\bs_i,l}/(1-r_{\bs_i,l}))$, for $l=1,\dots,L$.
The posterior distribution of the model parameters, based on the new augmented model, is 
$$
\begin{aligned}
p(\btheta, &\bzeta,\bga,\kappa^2, \{t_i\}_{i=L+1}^n\,|\,\bz_{\BS}) 
\propto \; \pi_{\btheta}(\btheta) \times \pi_{\bzeta}(\bzeta)\times
N(\bga\,|\,\bmu_{\gamma},\bV_{\gamma}) \times \mathrm{IG}(\kappa^2\,|\, u_{\kappa^2},v_{\kappa^2})\\
& \times N(\bm t\,|\,\bD\bga,\,\kappa^2\mathbf{I}_{n-L}) 
\times \prod_{i=L+1}^n\sum_{l=1}^L
f_{\bs_i,l}(z(\bs_i)\,|\,z(\bs_{(il)}),\btheta) \, 
\mathbbm{1}_{(r^*_{\bs_i,l-1},r^*_{\bs_i,l})}(t_i),
\end{aligned}
$$
where $\pi_{\btheta}$ and $\pi_{\bzeta}$ are the priors for $\btheta$ and $\bzeta$, respectively,
$\mathbf{I}_{n-L}$ is an $(n-L)\times (n-L)$ identity matrix,
the vector $\bm{t} = (t_{L+1},\dots,t_n)^\top$, and the matrix $\bD$ is $(n-L)\times 3$ such that
the $i$th row is $(1,s_{L+i,1},s_{L+i,2})$.

The posterior full conditional distribution of $\btheta$ depends on the form of 
$f_{\bs_i,l}$. Details for the models implemented in Section \ref{sec:data_examples}
are given in the supplementary material.
To update $\bzeta$, we first marginalize out the latent variables $t_i$ 
from the joint posterior distribution. We then update $\bzeta$ using a random walk
Metropolis step with target density
$\pi_{\bzeta}(\bzeta)\prod_{i=L+1}^n\{G_{\bs_i}(r_{\bs_i,\ell_i}\,|\,\mu(\bs_i),\kappa^2) -G_{\bs_i}(r_{\bs_i,\ell_i-1}\,|\,\mu(\bs_i),\kappa^2)\}$.
The posterior full conditional distribution of the latent variable $t_i$ is
$\sum_{l=1}^{i_L} q_l(\bs_i) 
\mathrm{TN}(t_i \,|\, \mu(\bs_i),\kappa^2; r^*_{\bs_i,l-1}, r^*_{\bs_i,l})$,
where $q_l(\bs_i) \propto w_l(\bs_i)f_{\bs_i,l}(z(\bs_i)\,|\,z(\bs_{(il)}),\btheta)$
and $w_l(\bs_i) = G_{\bs_i}(r_{\bs_i,l}\,|\,\mu(\bs_i),\kappa^2)
-G_{\bs_i}(r_{\bs_i,l-1}\,|\,\mu(\bs_i),\kappa^2)$,
for $l=1,...,L$. Hence, each $t_i$ can be updated by sampling from the 
$l$-th truncated Gaussian with probability proportional to $q_l(\bs_i)$.
The posterior full conditional distribution of $\bga$ is 
$N(\bga\,|\,\bmu_{\gamma}^*, \bV_{\gamma}^*)$, where 
$\bV_{\gamma}^* = (\bV_{\gamma}^{-1} + \kappa^{-2}\bD^\top\bD)^{-1}$
and $\bmu_{\gamma}^* = \bV_{\gamma}^*(\bV_{\gamma}^{-1}\bmu_{\gamma} + 
\kappa^{-2}\bD^\top\bm{t})$.
The posterior full conditional of $\kappa^2$ is 
$\mathrm{IG}(\kappa^2\,|\, u_{\kappa^2}+(n-L)/2, \, v_{\kappa^2}+
\sum_{i=L+1}^n(t_i-\mu(\bs_i))^2/2)$.

Turning to the prediction, let $\sv_0\in\D$. 
We obtain posterior predictive samples of $z(\sv_0)$ as follows. If 
$\sv_0\notin\BS$, for each posterior sample of the parameters,
we first compute the cutoff points $r_{\sv_0,l}$, such that
$r_{\sv_0,l} - r_{\sv_0,l-1} = k'(\sv_0,\sv_{(0l)}\,|\,\bzeta)/\sum_{l=1}^Lk'(\sv_0,\sv_{(0l)}\,|\,\bzeta)$,
and obtain the weights $w_l(\sv_0) = G_{\sv_0}(r_{\sv_0,l}\,|\,\,\mu(\sv_0),\kappa^2)-G_{\sv_0}(r_{\sv_0,l-1}\,|\,\mu(\sv_0),\kappa^2)$,
for $l = 1,\dots, L$. We then predict $z(\sv_0)$ using \eqref{eq:nnmp2}.
If $\sv_0\equiv\bs_i\in\BS$, we generate $z(\sv_0)$ similarly, 
but using samples for the weights collected from the posterior simulation, 
and applying \eqref{eq:nnmp1} instead of \eqref{eq:nnmp2} to generate $z(\sv_0)$.

\section{Data Illustrations}
\label{sec:data_examples}

\subsection{Simulation Study}
\label{sec:sim}

We have conducted several simulation experiments to study the inferential benefits 
of the NNMP modeling framework. Here, we present one synthetic data example. 
Three additional simulation examples are detailed in the supplementary material, 
the first demonstrates that Gaussian NNMP models can effectively approximate 
Gaussian random fields, the second illustrates the ability of skew-Gaussian NNMPs 
to handle data with different levels of skewness, and the third explores a copula 
NNMP model with beta marginals for bounded spatial data.

In this section, we demonstrate the use of copulas to construct NNMPs for tail dependence 
modeling. Our focus is on illustrating the flexibility of copula NNMPs 
for modeling complex dependence structures, and not specifically on extreme value modeling.

To simulate data, we created a regular grid of $200\times200$ resolution on a unit 
square domain $\D$, and generated over this grid a realization from random field
$y(\sv) = F^{-1}\big(T_{\nu}(\omega(\sv))\big),$\\$\sv\in\D$;
see Fig. 1(a). Here, $\omega(\sv)$ is a standard Student-t process with tail parameter 
$\nu$ and scale matrix specified by an exponential correlation function with range 
parameter $\phi_w$, and the distribution functions $F$ and $T_{\nu}$ correspond to 
a gamma distribution, $\mathrm{Ga}(2,2)$ with mean $1$, and a standard Student-t 
distribution with tail parameter $\nu$, respectively.
For a given pair of locations in $\D$ with correlation $\rho_0 = \exp(-d_0/\phi_w)$,
the corresponding tail dependence coefficient of the random field is 
$\chi_{\nu} = 2T_{\nu+1}\big(-\sqrt{(1+\nu)(1-\rho_0)/(1+\rho_0)}\big)$. 
We took $\phi_w = 1/12$, and chose $\nu = 10$ so that the synthetic data exhibits moderate tail 
dependence at close distance, and the dependence decreases rapidly as the distance $d_0$ 
becomes larger. In particular, for $\rho_0 =$ $0.05, 0.5, 0.95$, we obtain 
$\chi_{10} =$ $0.01, 0.08, 0.61$, respectively.

\begin{figure}[t]
    \centering
    \captionsetup[subfigure]{justification=centering}
    \begin{subfigure}[b]{0.32\textwidth}
         \centering
         \includegraphics[width=\textwidth]{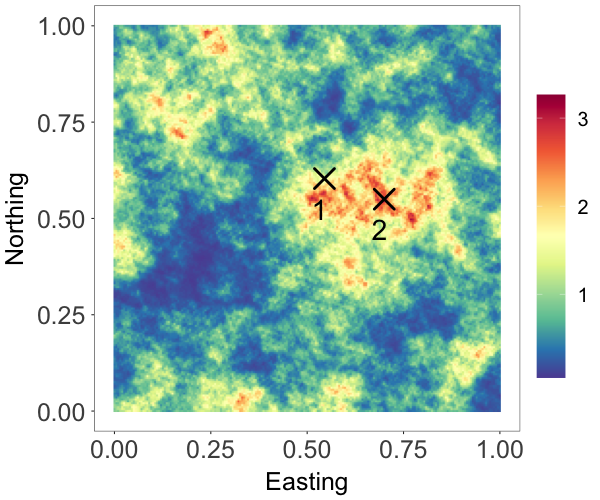}
         \caption{True $y(\sv)$}
     \end{subfigure}
     \hfill
     \begin{subfigure}[b]{0.32\textwidth}
         \centering
         \includegraphics[width=\textwidth]{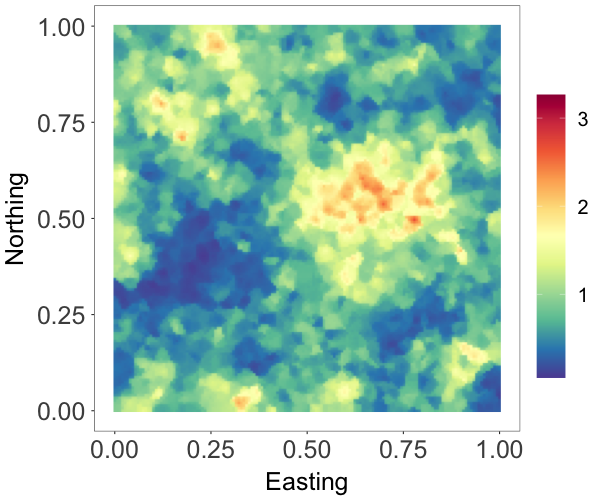}
         \caption{Gaussian copula NNMP}
     \end{subfigure}
     \hfill
     \begin{subfigure}[b]{0.32\textwidth}
         \centering
         \includegraphics[width=\textwidth]{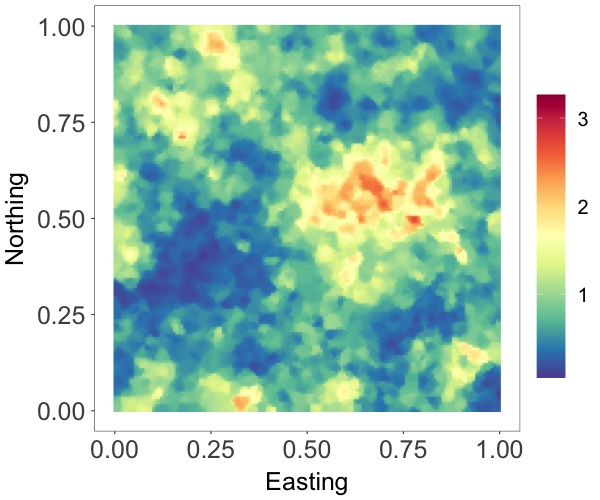}
         \caption{Gumbel copula NNMP}
     \end{subfigure}\\
     \bigskip
    \begin{subfigure}[b]{0.32\textwidth}
         \centering
         \includegraphics[width=\textwidth]{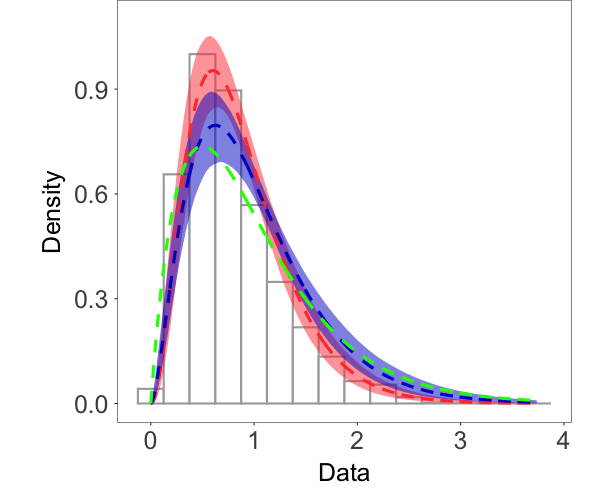}
         \caption{Estimated marginals}
     \end{subfigure}
     \hfill
     \begin{subfigure}[b]{0.32\textwidth}
         \centering
         \includegraphics[width=\textwidth]{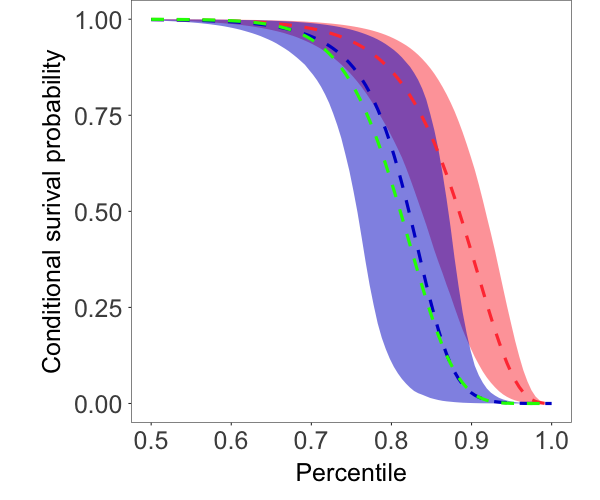}
         \caption{Estimated probability (site 1)}
     \end{subfigure}
     \hfill
     \begin{subfigure}[b]{0.32\textwidth}
         \centering
         \includegraphics[width=\textwidth]{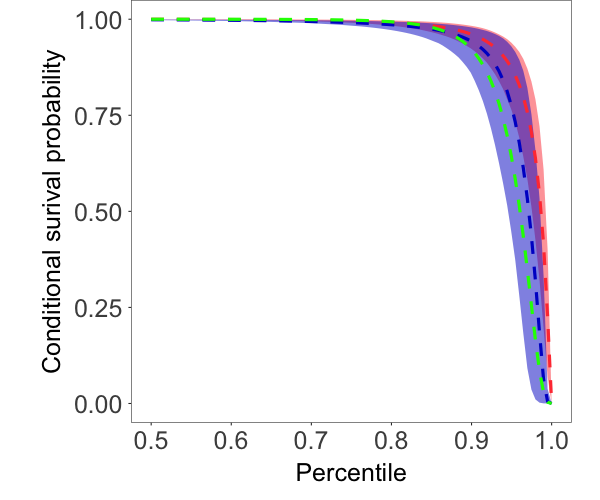}
         \caption{Estimated probability (site 2)}
     \end{subfigure}        
    \caption{
    Synthetic data example. Top panels are interpolated 
    surfaces of the true field and posterior median estimates from both models.
    Bottom panels are estimated marginal densities and conditional survival probabilities
    from the two models. The green dashed lines correspond to the true model. The 
    red (blue) dashed lines and shaded regions are the posterior mean and 95\% credible interval estimates
    from the Gaussian (Gumbel) copula NNMP models.
    }
    \label{fig:copula-nnmp-itpl}
\end{figure}

We applied two copula NNMP models.
The models are of the form in \eqref{eq:copula-nnmp} with stationary 
gamma marginal $\mathrm{Ga}(a,b)$ with mean $a/b$. 
In the first model, the component copula density $c_{\sv,l}$ 
corresponds to a bivariate Gaussian copula, which is known to be 
unsuitable for tail dependence modeling. The spatially varying
correlation parameter of the copula was specified by an exponential correlation 
function with range parameter $\phi_1$.
In the second model, we consider a spatially varying Gumbel copula as in Example \ref{ex:gumbel}.
The spatially varying parameter of the copula density is defined
with the link function $\eta_l(\sv) \equiv \eta_l(||\sv-\sv_{(l)}||) = \min\{(1-\exp(-||\sv-\sv_{(l)}||/\phi_2))^{-1}, 50\}$, 
where the upper bound $50$ ensures numerical stability.
When $\eta_l(d_0) = 50$, 
$\exp(-d_0/\phi_2) = 0.98$. With this link function, we assume that given $\phi_2$, the strength of 
the tail dependence with respect to the $l$th component of the Gumbel model stays the 
same for any distance smaller than $d_0$ between two locations. 
For the cutoff point kernels, we specified an exponential correlation
function with range parameters $\zeta_1$ and $\zeta_2$, respectively, for each model.
The Bayesian model is completed with a 
$\mathrm{IG}(3, 1/3)$ prior for $\phi_1$ and $\phi_2$, 
a $\mathrm{Ga}(1,1)$ prior for $a$ and $b$,
a $\mathrm{IG}(3,0.2)$ prior for $\zeta_1$ and $\zeta_2$, and
$N(\bga\,|\,(-1.5,0,0)^\top,\,2\mathbf{I}_3)$ and $\mathrm{IG}(\kappa^2\,|\,3, 1)$ priors.

We randomly selected 2000 locations as the reference set 
with a random ordering for model fitting. For the purposes of illustration,
we chose neighbor size $L = 10$. Results are based on posterior samples collected
every 10 iterations from a Markov chain of 30000 iterations, with the first 10000
samples being discarded. Implementation details for both models are provided in
the supplementary material. The computing time was around $18$ minutes.

Fig. \ref{fig:copula-nnmp-itpl} shows the random fields, marginal densities, 
and conditional survival probabilities estimated by the two models.
From Fig. \ref{fig:copula-nnmp-itpl}(a)-\ref{fig:copula-nnmp-itpl}(c), 
we see that, comparing with the true field, the posterior median estimate by
the Gumbel copula model seems to recover the large values better than the Gaussian 
copula model. Besides, as shown in Fig. \ref{fig:copula-nnmp-itpl}(d), 
the Gumbel copula model provides a more accurate estimate of
the marginal distribution, especially in the tails.
We computed the conditional survival probabilities at two different unobserved sites
marked in Fig. \ref{fig:copula-nnmp-itpl}(a).
In particular, Site 1/2 is surrounded by reference set observations with 
moderate/large values. The Gumbel copula model provides much more accurate estimates 
of the conditional survival probabilities, 
indicating that the model captures better the tail dependence structure in the data.
Overall, this example demonstrates that the Gumbel copula NNMP model is a useful 
option for modeling spatial processes with tail dependence.

\subsection{Mediterranean Sea Surface Temperature Data Analysis}
\label{sec:app}

The study of Ocean's dynamics is crucial for understanding climate variability.
One of the most valuable sources of information regarding the evolution of the state of the ocean 
is provided by the centuries-long record of temperature observations from the surface of 
the oceans. The record of sea surface temperatures consists of data collected over time at 
irregularly scattered locations.
In this section, we examine the sea surface temperature from the 
Mediterranean Sea area during December 2003.

\begin{figure}[t!]
    \centering
    \captionsetup[subfigure]{justification=centering}
     \begin{subfigure}[b]{0.48\textwidth}
         \centering
         \includegraphics[width=\textwidth]{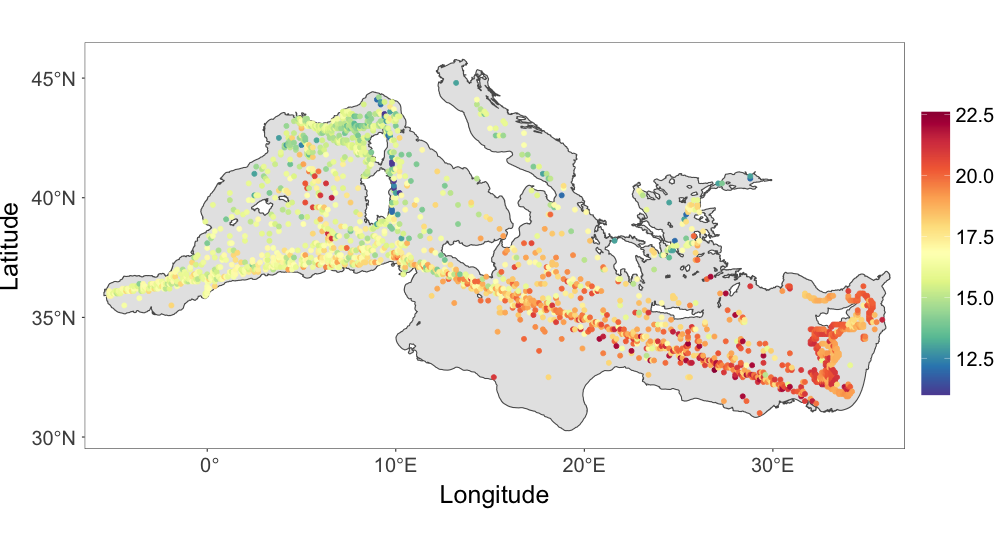}
         \caption{Sea surface temperature observations}
    \end{subfigure}         
    \hfill
    \begin{subfigure}[b]{0.48\textwidth}
         \centering
         \includegraphics[width=\textwidth]{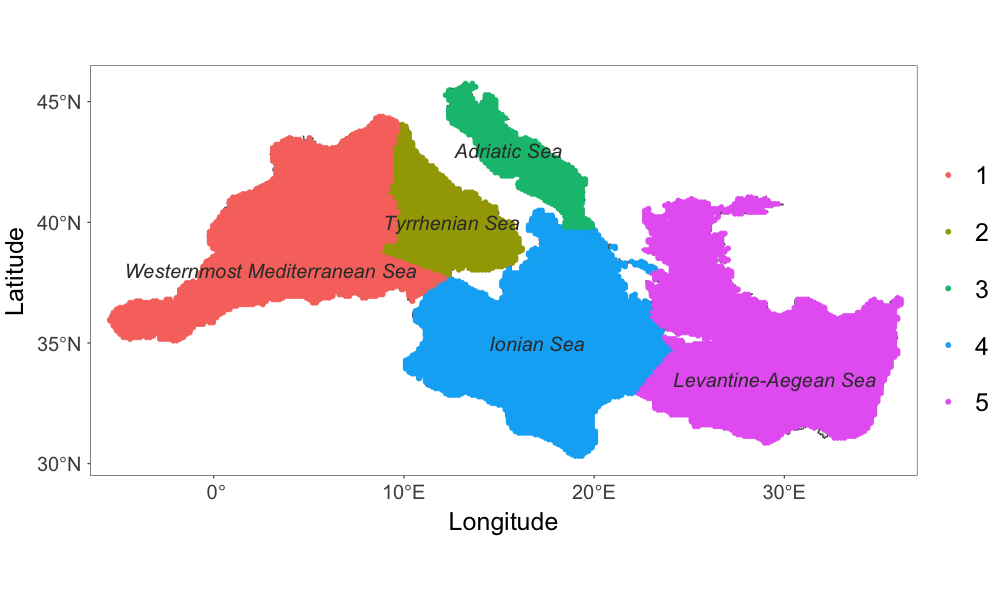}
         \caption{Mediterranean Sea partitions}
     \end{subfigure}    
    \begin{subfigure}[b]{0.48\textwidth}
         \centering
         \includegraphics[width=\textwidth]{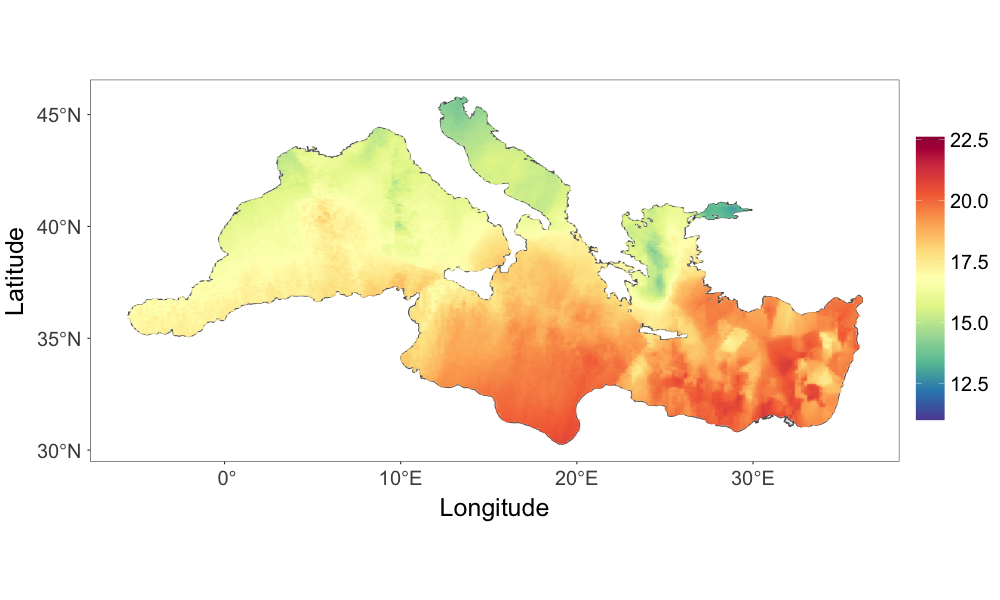}
         \caption{$50\%$ predicted sea surface temperature}
     \end{subfigure}
     \hfill
     \begin{subfigure}[b]{0.48\textwidth}
         \centering
         \includegraphics[width=\textwidth]{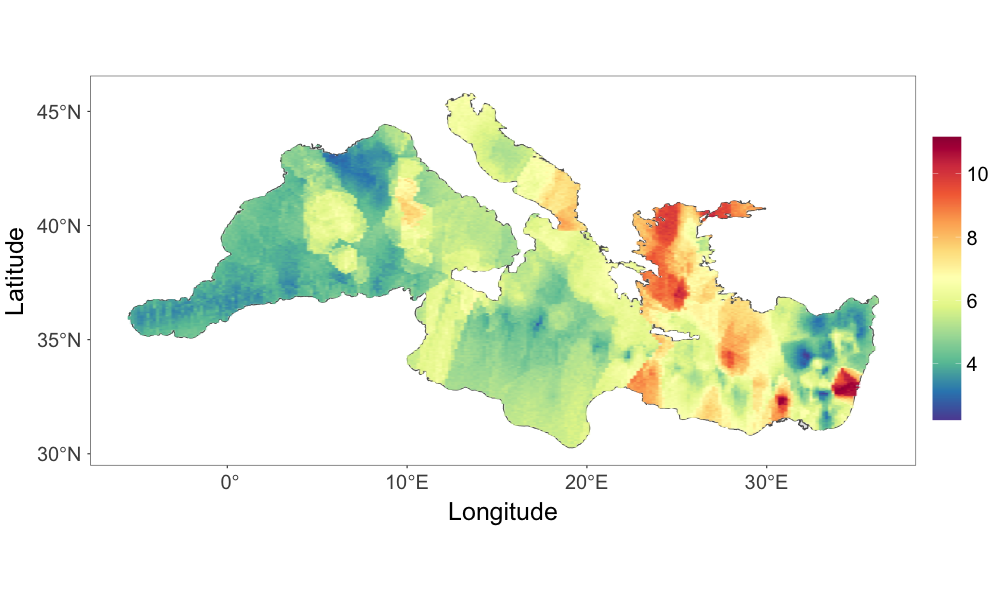}
         \caption{$95\%$ credible interval width}
     \end{subfigure}\\
     \bigskip
    \begin{subfigure}[b]{0.96\textwidth}
         \centering
         \includegraphics[width=\textwidth]{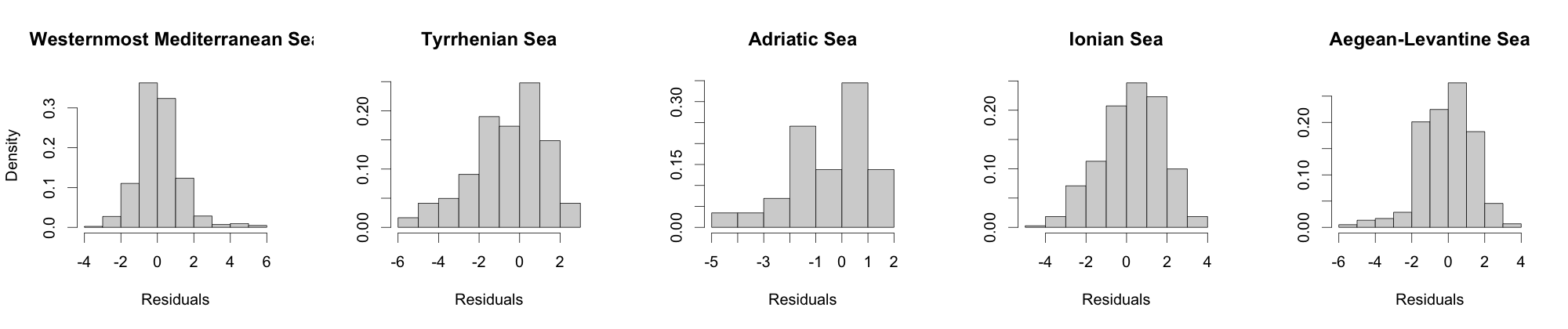}
         \caption{Histograms of residuals}
    \end{subfigure}       
    \caption{
    Mediterranean Sea surface temperature data analysis.
    Panel (a) shows observations.
    Panels (b) and (e) are partitions according to Mediterranean sub-basins and 
    histograms of the residuals obtained from a non-spatial linear model.
    Panels (c) and (d) are posterior median and 95\% credible interval estimates
    of the sea surface temperature from the extended skew-Gaussian NNMP model.
    }
    \label{fig:sst_par}
\end{figure}

It is well known that the Mediterranean Sea area produces very heterogeneous temperature 
fields. A goal of the spatial analysis of sea surface temperature in the area is to generate 
a spatially continuous field that accounts for the complexity of the surrounding coastlines 
as well as the non-linear dynamics of the circulation system. An additional source of 
complexity comes from the data collection process. Historically, the observations are 
collected from different types of devices: buckets launched from navigating vessels, 
readings from the water intake of ships’ engine rooms, moored buoys, 
and drifting buoys \citep{kirsner2020multi}. The source of some observations is known, 
but not all the data are labelled. A thorough case study will be needed to include all 
this information in order to account for possible heterogeneities due to the
different measuring devices. That is beyond the scope of this paper. 
We will focus on demonstrating the ability of the proposed framework to model non-Gaussian 
spatial processes that, hopefully, capture the complexities of the physical process and the 
data collection protocol.
We notice that in the original record several sites had multiple observations. 
In those cases we took the median of the observations, resulting in a total of 1966
observations. The data are shown in Fig. \ref{fig:sst_par}(a).

We first examine the Gaussianity assumption for the data. We compare the Gaussian NNMP 
and the nearest-neighbor Gaussian process models over a subset of the region 
where the ocean dynamics are known to be complex and the observations are 
heterogeneous. The model comparison is detailed in the supplementary material 
and the results support the NNMP model.

In light of the evidence \citep{pisano2020new} that sea surface temperature spatial 
patterns are different over Mediterranean sub-basins, shown in Fig. \ref{fig:sst_par}(b), 
which are characterized by different dynamics and high variability of surface currents 
\citep{bouzaiene2020analysis}, we further investigate the sea surface temperature over 
those sub-basins. We fitted a non-spatial linear model to all data, including 
longitude and latitude as covariates, and obtained residuals from the linear model. 
Fig. \ref{fig:sst_par}(e) shows that the histograms of the residuals are asymmetric 
over the sub-basins, indicating skewness in the marginal distribution, with 
levels of skewness that vary across sub-basins.

The exploratory data analysis suggests the need for a 
spatial model that can capture skewness.
We thus analyze the full data set with an extension of the skew-Gaussian NNMP model.
The new model has two features that extend the skew-Gaussian NNMP: 
(i) it incorporates a fixed effect through
the location parameter of the mixture component; (ii) it allows 
the skewness parameter $\la$ to vary in space. 
More specifically, the spatially varying
conditional density $f_{\sv,l}$ of the model builds from a Gaussian random vector with
mean $\left(\bx(\sv)^\top\bbeta + \la(\sv)z_0(\sv),\,\bx(\sv_{(l)})^\top\bbeta + 
\la(\sv_{(l)})z_0(\sv)\right)^\top$
and covariance matrix $\sigma^2\left(\begin{smallmatrix}1 & \rho_l(\sv)\\ 
\rho_l(\sv) & 1\end{smallmatrix}\right)$, where $\bx(\sv) = (1,v_1,v_2)^\top$ and
$z_0(\sv)\sim\mathrm{TN}(0,1;0,\infty)$, for all $\sv$
and for all $l$, and $(v_1,v_2)$ are longitude and latitude.
The conditional density $p(y(\sv)\,|\,\by_{\tNe(\sv)})$
of the extended model is
\begin{equation}\label{eq:ext-SGNNMP}
    \sum_{l=1}^{L} w_l(\sv) \, \int_0^{\infty}N(y(\sv)\,|\,\mu_l(\sv),\sigma_l^2(\sv))
    \mathrm{TN}(z_0(\sv)\,|\,\mu_{0l}(\sv_{(l)}),\sigma_{0l}^2(\sv_{(l)});0,\infty)dz_0(\sv),
\end{equation}
with parameters $\mu_l(\sv) = \bx(\sv)^\top\bbeta + \la(\sv)z_0(\sv) + 
\rho_l(\sv)\{y(\sv_{(l)}) - \bx(\sv_{(l)})^\top\bbeta - \la(\sv_{(l)})z_0(\sv)\}$,
$\,\sigma_l^2(\sv) = \sigma^2\{1-(\rho_l(\sv))^2\}$,
$\,\mu_{0l}(\sv_{(l)}) = \{y(\sv_{(l)})- \bx(\sv_{(l)})^\top\bbeta\}\la(\sv_{(l)})/\{\sigma^2+(\la(\sv_{(l)}))^2\}$, 
and $\sigma_{0l}^2(\sv_{(l)}) = \sigma^2/\{\sigma^2+(\la(\sv_{(l)}))^2\}$.
After marginalizing out $z_0(\sv)$, 
we obtain that the marginal distribution of $Y(\sv)$ is 
$\mathrm{SN}\big(\bx(\sv)^\top\bbeta, (\la(\sv))^2+\sigma^2, \la(\sv)/\sigma\big)$,
based on the result of Proposition \ref{prop:stationary}.
We model the spatially varying $\lambda(\sv)$ via a partitioning approach. 
In particular, we partition the Mediterranean Sea $\D$ according to the sub-basins,
that is, $\D = \cup_{k=1}^KP_k$, $P_i\cap P_j=\emptyset$ for $i\neq j$, where $K = 5$. 
For all $\sv\in P_k$, we take $\la(\sv) = \la_k$, for $k = 1,\dots, K$. The 
partitions $P_1,\dots, P_K$ correspond to the sub-basins: Westernmost Mediterranean Sea, 
Tyrrhenian Sea, Adriatic Sea, Ionian Sea, and Levantine-Aegean Sea.

We applied the extended skew-Gaussian NNMP model \eqref{eq:ext-SGNNMP} using the whole data set
as the reference set, with $L$ chosen to be $10, 15$ or $20$. 
The regression parameters $\bbeta = (\beta_0,\beta_1,\beta_2)^\top$ were assigned mean-zero, 
dispersed normal priors. For the skewness parameters $\bm\la = (\la_1,\dots, \la_5)$,
each element received a $N(0,5)$ prior.
We used the same prior specification for other parameters as in the first simulation experiment.
Posterior inference was based on thinned samples retaining every 4th iteration, from a total of 
30000 samples with a burn-in of 10000 samples. The computing time was around 14, 16, 
and 20 minutes, respectively, for each of the $L$ values.

We focus on the estimation of regression and skewness parameters $\bbeta$ and $\bm\la$.
We report the estimates for $L = 15$; they were similar for $L = 10$ or $20$.
The posterior mean and $95\%$ credible interval estimates of $\beta_0$, $\beta_1$,
and $\beta_2$ were $30.51\,(28.88, 32.16)$, $0.12\,(0.09, 0.15)$, and 
$-0.37\,(-0.42, -0.33)$, indicating that there was an increasing trend in 
sea surface temperature as longitude increased and latitude decreased. 
The corresponding posterior estimates of $\bm\la$ were 
$-0.38\,(-0.94, 0.14)$, $-1.37\,(-2.10, -0.71)$, $-2.44\,(-4.03, -1.14)$, 
$-1.60\,(-2.54, -0.86)$, and  $-2.69\,(-3.95, -1.82)$.
These estimates suggest different levels of left skewness over the sub-basins 
except for the Westernmost Mediterranean Sea.

Fig. \ref{fig:sst_par}(c) provides the temperature posterior 
median estimate over a dense grid of locations on the Mediterranean Sea.
Compared to Fig. \ref{fig:sst_par}(a), the estimate generally resembles the observed 
pattern. The prediction was quite smooth even for areas with few observations.
The 95\% credible interval width of the prediction over the gridded locations, 
as shown in Fig. \ref{fig:sst_par}(d), demonstrates that the model describes the uncertainty 
in accordance with the observed data structure; the uncertainty is higher in areas 
where there are less observations or the observations are volatile.

\section{Discussion}
\label{sec:sum}

We have introduced a class of geostatistical models for non-Gaussian
processes, and demonstrated its flexibility for modeling complex dependence by 
specification of a collection of bivariate distributions indexed in space. 
The scope of the methodology has been illustrated through synthetic spatial 
data examples corresponding to distributions of different nature and support, 
and with the analysis of sea surface temperature measurements from the 
Mediterranean Sea.

To incorporate covariates, the NNMP can be embedded in an additive or multiplicative 
model. The former is illustrated in the supplementary material 
with a spatially varying regression model. Under an additive model, the posterior
simulation algorithm requires extra care as it involves sequential updating of the
elements in $\bz_{\BS}$. This may induce slow convergence behavior. An alternative
strategy for covariate inclusion is to model the weights or some parameter(s) of the
spatially varying conditional density as a function of covariates. For example, in
Section \ref{sec:app}, we modeled the location 
parameter of the skew-Gaussian marginal as a linear function of the covariates. 
Posterior simulation under this approach is easily developed by modifying the update 
of the relevant parameters discussed in Section \ref{sec:est} to that of the 
regression coefficients.

The NNMP model structure not only bypasses all the potential issues from 
large matrix operations, but it also enhances modeling power. Kernel functions, such as 
wave covariance functions, that are impractical for Gaussian process-based models 
due to numerical instability from matrix inversion, can be used as link functions for 
the spatially varying parameter of the NNMP. One limitation of the NNMP's computation, 
similar to mixture models, is that the posterior simulation algorithm may experience slow 
convergence issues. Further development is needed on efficient algorithms for fast 
computation, especially when dealing with massive, complex data sets.

We have focused in this article on a modeling framework for continuous data. 
The proposed approach can be naturally extended to modeling discrete spatial data.
Modeling options for geostatistical count data in the existing literature involve
either spatial generalized linear mixed models or spatial 
copula models \citep{madsen2009maximum}. However, owing to their structures, both 
models have limitations with respect to the distributional assumption for the spatial 
random effects, as well as in computational efficiency. The extension to discrete 
NNMP models has the potential to provide both inferential and computational benefits 
in modeling large discrete data sets.

It is also interesting to explore the opportunities for the analysis of spatial
extremes using the NNMP framework. We developed guidelines in Section \ref{sec:tail}
to choose NNMP mixture components based on strength of tail dependence. The results
highlight the ability of the NNMP model structure to capture local tail dependence 
at different levels, controlled by the mixture component bivariate distributions,
for instance, with a class of bivariate extreme-value copulas. Moreover, using NNMPs for
spatial extreme value modeling allows for efficient implementation of
inference which is typically a challenge with existing approaches
\citep{davison2012statistical}.

Other research directions include extensions to multivariate and spatio-temporal settings. 
The former extension requires families of high-dimensional multivariate distributions to 
construct an NNMP. Effective strategies will be needed to define the spatially varying 
multivariate distributions that balance flexibility and scalability. 
When it comes to a joint model over time and space, 
there is large scope for exploring the integration of the time component into the model,
including extending the NNMP weights or the NNMP mixture components.

\section*{Supplementary Material}
The supplementary material includes proofs of the propositions, additional
data examples, and implementation details.

\bibliographystyle{jasa3}
\bibliography{ref}

% \newpage
\clearpage\pagebreak\newpage
%%%%%%%%%% Merge with supplemental materials %%%%%%%%%%

\spacingset{1.2} 

\begin{center}
\Large\bf Supplementary Material for ``Nearest-Neighbor Mixture Models for Non-Gaussian Spatial Processes"
\end{center}
% \section*{\hfil \LARGE\bf Supplementary Material\hfil}
%%%%%%%%%% Prefix a "S" to all equations, figures, tables and reset the counter %%%%%%%%%%
\setcounter{section}{0}
\setcounter{equation}{0}
\setcounter{figure}{0}
\setcounter{table}{0}
% \setcounter{page}{1}
% \makeatletter
% \renewcommand{\theequation}{S\arabic{equation}}
% \renewcommand{\thefigure}{S\arabic{figure}}
% \renewcommand{\bibnumfmt}[1]{[S#1]}
% \renewcommand{\citenumfont}[1]{S#1}
%%%%%%%%%% Prefix a "S" to all equations, figures, tables and reset the counter %%%%%%%%%%

\renewcommand{\thesection}{\Alph{section}}  

\vspace{50pt}

\section{Proofs}

\begin{proof}[\textbf{Proof of Proposition 1}]
We consider a univariate spatial process $\{Z(\sv), \sv\in\D\}$, where
$Z(\sv)$ takes values in $\mathcal{X}\subseteq\R$, and $\D\subset\R^p, p\geq 1$. 
Let $\BS\subset\D$ be a reference set. Without loss of generality, we consider the 
continuous case, i.e., $Z(\sv)$ has a continuous distribution for which its density exists,
for all $\sv\in\D$. To verify the proposition, we partition the domain $\D$ into the reference 
set $\BS$ and the nonreference set $\U$. 

Given any $\sv\in\D$, consider a bivariate random vector indexed at $\sv$,
denoted as $(U_{\sv,l},V_{\sv,l})$ taking values in $\mathcal{X}\times\mathcal{X}$. 
We denote $f_{\sv,l}$ as the conditional density of $U_{\sv,l}$ given $V_{\sv,l}$, 
and $f_{U_{\sv,l}}, f_{V_{\sv,l}}$ as the marginal densities of $U_{\sv,l}$, $V_{\sv,l}$,
respectively. Using the proposition assumption that $f_Z = f_{U_{\sv,l}} = f_{V_{\sv,l}}$, 
we have that
\begin{equation}\label{eq:prop1-proof-eq1}
\int_{\mathcal{X}}f_{\sv,l}(u\mid v)f_Z(v)dv = \int_{\mathcal{X}}f_{\sv,l}(u\mid v)f_{V_{\sv,l}}(v)dv 
= f_{U_{\sv,l}}(u) = f_Z(u),
\end{equation}
for every $\sv\in\D$ and for all $l$.

We first prove the result for the reference set $\BS$.
By the model assumption, locations in $\BS$ are 
ordered. In this regard, using the proposition assumptions, we can show that $Z(\bs)\sim f_Z$
for all $\bs\in\BS$ by applying Proposition 1 in \citep{zheng2021construction}.

Turning to the nonreference set $\U$. 
Let $g_{\su}(z(\su))$ be the marginal density of $Z(\su)$ for every $\su\in\U$.
Denote by $\tilde{p}(\bz_{\tNe(\su)})$ the joint density for 
the random vector $\bz_{\tNe(\su)}$ where $\tNe(\su) = \{\su_{(1)},\dots, \su_{(L)}\}\subset\BS$, 
so every element of $\bm{Z}_{\tNe}$ has marginal density $f_Z$. 
Then, the marginal density for $Z(\su)$ is given by:
$$
\begin{aligned}
g_{\su}(z(\su)) & = \int_{\mathcal{X}^L} p(z(\su)\mid \bz_{\tNe(\su)})\tilde{p}(\bz_{\tNe(\su)})
\prod_{\{\bs_i\in\tNe(\su)\}}d(z(\bs_i))\\
& = \sum_{l=1}^Lw_l(\sv)\int_{\mathcal{X}^L}f_{\sv,l}(z(\su)\mid z(\su_{(l)}))
\tilde{p}(\bz_{\tNe(\su)})\prod_{\{\bs_i\in\tNe(\su), \bs_i\neq \su_{(l)}\}}d(z(\bs_i))\\
& = \sum_{l=1}^Lw_l(\sv)\int_{\mathcal{X}} f_{\sv,l}(z(\su)\mid z(\su_{(l)}))
f_Z(z(\su_{(l)}))d(z(\su_{(l)}))\\
& = f_Z(z(\su)),
\end{aligned}
$$
where the second-to-last equality holds by the result that $Z(\bs)\sim f_Z$
for all $\bs\in\BS$ and $\tNe(\su)\subset\BS$ for every $\su\in\U$.
The last equality follows from \eqref{eq:prop1-proof-eq1}.

\end{proof}

\begin{proof}[\textbf{Proof of Proposition 2}]

We verify the proposition by partitioning the domain $\D$ into the reference set $\BS$ and the nonreference
set $\U$. We first prove by induction the result for the joint distribution $\tilde{p}(\bz_{\BS})$ over $\BS$. 
Then to complete the proof, it suffices to show that for every location 
$\su\in\U$, the joint density $\tilde{p}(\bz_{\U_1})$ is a mixture of multivariate Gaussian distributions, 
where $\U_1 = \BS\cup\{\su\}$.

Without loss of generality, we assume $\mu = 0$ for the stationary Gaussian NNMP, 
i.e., the Gaussian NNMP has invariant marginal $f_Z(z) = N(z\,|\, 0, \sigma^2)$. 
The conditional density for the reference set is 
$p(z(\bs_i)|\bz_{\tNe(\bs_i)}) = \sum_{l=1}^{i_L}w_l(\bs_i)
N(z(\bs_i)|\rho_l(\bs_i)z(\bs_{(il)}), \sigma^2(1-(\rho_l(\bs_i))^2))$,
where for, $i = 2,\dots,L$, $i_L = i - 1$, and for $i > L$, $i_L = L$.
For each $i$, we denote as $\{w_{i,l_i}\}_{l_i=1}^{i_L}$ the vector of mixture weights, 
as $\{\rho_{i,l_i}\}_{l_i=1}^{i_L}$ the vector of the correlation coefficients, 
and as $\{z_{i,l_i}\}_{l_i=1}^{i_L}$ the vector of the nearest neighbors of $z_i\equiv z(\bs_i)$,
for $i\geq 2$, where $w_{i,l_i}\equiv w_{l_i}(\bs_i)$, $\rho_{i,l_i}\equiv 
\rho_{l_i}(\bs_i)$, $z_{i,l_i}\equiv z(\bs_{(i,l_i)})$. 
Let $z_1 \equiv z(\bs_1)$. We denote by 
$\bz_{1:k} = (z(\bs_1),\dots,z(\bs_k))$ the realization of $Z(\bs)$ over 
locations $(\bs_1,\dots,\bs_k)^\top$ for $k\geq 2$, and use $\bz_{1:k}^{-z_j}$ to 
denote the random vector $\bz_{1:k}$ with element $z_j$ removed, $1\leq j\leq k$.
In the following, for a vector $\bm{a}=(a_1,\dots,a_m)^\top$, we have that
$\bm{a}c = (a_1c,\dots,a_mc)^\top$, where $c$ is a scalar.

Take $Z_1 \sim N(z_1\,|\,0,\sigma^2)$. When $i = 2$, $i_L = 1$ and $w_{2,1} = 1$.
The joint density of $\bz_{1:2}$ is
$\tilde{p}(\bz_{1:2}) = N(z_2|\rho_{2,1}z_1), \sigma^2(1-\rho_{2,1}^2))
N(z_1|0,\sigma^2) = N(\bz_{1:2}|\bm0, \sigma^2\bm\Omega_{2,1})$,
where $\bm\Omega_{2,1} = \left(\begin{smallmatrix}1 & \rho_{2,1}\\\rho_{2,1} & 1\end{smallmatrix}\right)$.
The joint density of $\bz_{1:3}$ is
$$
\begin{aligned}
\tilde{p}(\bz_{1:3}) & = p_3(z_3|\bz_{1:2})\tilde{p}(\bz_{1:2})\\
& = \sum_{l_3=1}^2w_{3,l_3}N(z_3|\rho_{3,l_3}z_{3,l_3}, \sigma^2(1-\rho_{3,l_3}^2))
N(\bz_{1:2}|\bm0,\sigma^2\bm\Omega_{2,1})\\
& = \sum_{l_3=1}^2w_{3,l_3}
N(z_3|\rho_{3,l_3}z_{3,l_3}, \sigma^2(1-\rho_{3,l_3}^2))
N(\bz_{1:2}^{-z_{3,l_3}}|\rho_{2,l_2}z_{3,l_3},\sigma^2(1-\rho_{2,1}^2))
N(z_{3,l_3}|0,\sigma^2)\\
& = \sum_{l_3=1}^2w_{3,l_3}
N((z_3,\bz_{1:2}^{-z_{3,l_3}})^\top|\bm m_{3,l_3}z_{3,l_3}, 
\bm V_{3,l_3})N(z_{3,l_3}|0,\sigma^2)\\
\end{aligned}
$$
where 
$\bm m_{3,l_3} = (\rho_{3,l_3}, \rho_{2,1})^\top$, and
$\bm V_{3,l_3} = \left(\begin{smallmatrix}\sigma^2(1-\rho_{3,l_3}^2) & 0\\ 
0 & \sigma^2(1-\rho_{2,1}^2)\end{smallmatrix}\right)$.
The last equality follows from the fact that a product of conditionally independent
Gaussian densities is a Gaussian density.
By the properties of the Gaussian distribution and the property of the model that 
has a stationary marginal $N(0,\sigma^2)$, for each $l_3$, we have that
$$
N(\tz_{1:3, l_3}|\bm0, \sigma^2\bm R_{3,l_3}) = 
N((z_3,\bz_{1:2}^{-z_{3,l_3}})^\top|\bm m_{3,l_3}z_{3,l_3},\bm V_{3,l_3})N(z_{3,l_3}|0,\sigma^2),
$$
where $\tz_{1:3,l_3} = (z_3,\bz_{1:2}^{-z_{3,l_3}},z_{3,l_3})^\top$, 
with the following partition relevant to the vector $\tz_{1:3,l_3}$, 
$$
\tz_{1:3,l_3} = \begin{pmatrix}
(z_3,\bz_{1:2}^{-z_{3,l_3}})^\top\\ z_{3,l_3}
\end{pmatrix},\;
E(\tz_{1:3,l_3}) = \begin{pmatrix}\bm0\\ 0\end{pmatrix},\;
\bm R_{3,l_3} = \begin{pmatrix}
\bm R_{3,l_3}^{(11)} & \bm R_{3,l_3}^{(12)}\\
\bm R_{3,l_3}^{(21)} & \bm R_{3,l_3}^{(22)}
\end{pmatrix},
$$
where $\bm R_{3,l_3}^{(22)} = 1$ corresponds to $z_{3,l_3}$. It follows that
\begin{equation}\label{eq:solve}
\begin{aligned}
\bm m_{3, l_3}z_{3,l_3} & = E((Z_3,\tilde{\bm{Z}}_{1:2}^{-Z_{3,l_3}})\,|\, 
Z_{3,l_3} = z_{3,l_3}) = 
\bm R_{3,l_3}^{(12)}z_{3,l_3},\;\;\\
\bm V_{3,l_3} & = \sigma^2(\bm R_{3,l_3}^{(11)} - \bm R_{3,l_3}^{(12)}\bm R_{3,l_3}^{(21)}).
\end{aligned}
\end{equation}
From \eqref{eq:solve}, we obtain $\bm m_{3, l_3} = \bm R_{3,l_3}^{(12)}$
and $\bm R_{3,l_3} = \left(\begin{smallmatrix}
1 & \rho_{2,1}\rho_{3,l_3} & \rho_{3,l_3}\\
\rho_{2,1}\rho_{3,l_3} & 1 & \rho_{2,1}\\
\rho_{3,l_3} & \rho_{2,1} & 1
\end{smallmatrix}\right)$
for $l_3 = 1, 2$. Then we reorder $\tz_{1:3,l_3}$ with a matrix $\bm B_{3,l_3}$ such that 
$\bz_{1:3} = \bm B_{3,l_3}\tz_{1:3,l_3}$. It follows that
$\bm\Omega_{3,l_31} =\bm B_{3,l_3}\bm R_{3,l_3}\bm B_{3,l_3}^T$, and the joint density is
$p(\bz_{1:3}) 
= \sum_{l_3=1}^2w_{3,l_3}N(\bz_{1:3}|\bm0,\sigma^2\bm\Omega_{3,l_31})$.

Similarly, the joint density of $\bz_{1:4}$ is given by
$$
\begin{aligned}
\tilde{p}(\bz_{1:4}) & = p_4(z_4|\bz_{1:3})\tilde{p}(\bz_{1:3})\\
& = \sum_{l_4=1}^3w_{4,l_4}N(z_4|\rho_{4,l_4}z_{4,{l_4}}, \sigma^2(1-\rho_{4,l_4}^2))
\sum_{l_3=1}^2w_{3,l_3}N(\bz_{1:3}|\bm0, \sigma^2\bm\Omega_{3,l_31})\\
& = \sum_{l_4=1}^3\sum_{l_3=1}^2w_{4,l_4}w_{3,l_3}
N(z_4|\rho_{4,l_4}z_{4,l_4}, \sigma^2(1-\rho_{4,l_4}^2))\\
&\;\;\;\;\;\;\;\;\;\;\;\;\;\;\;\;\;\;\;\;\;\;
N((\bz_{1:3}^{-z_{4,{l_4}}})^\top\mid\tilde{\bm\Omega}_{3, l_31}^{(12)}z_{4,l_4},
\sigma^2(\tilde{\bm\Omega}_{3, l_31}^{(11)} - 
\tilde{\bm\Omega}_{3, l_31}^{(12)}\tilde{\bm\Omega}_{3, l_31}^{(21)}))N(z_{4,l_4}|0,\sigma^2)\\
& = \sum_{l_4=1}^3\sum_{l_3=1}^2w_{4,l_4}w_{3,l_3}
N((z_4, \bz_{1:3}^{-z_{4,{l_4}}})^\top|\bm m_{4, l_4l_3}z_{4,l_4}, \bm V_{4,l_4l_3})N(z_{4,l_4}|0,\sigma^2),
\end{aligned}
$$
where $\tilde{{\bm\Omega}}_{3,l_31}  = \tilde{\bm B}_{4,l_4}\bm\Omega_{3,l_31}
\tilde{\bm B}_{4,l_4}^\top$, and $\tilde{\bm B}_{4,l_4}$ is a rotation matrix such that
$(\bz_{1:3}^{-z_{4,l_4}},z_{4,l_4})^\top = \tilde{\bm B}_{4,l_4}\bz_{1:3}$.
We partition the matrix $\tilde{{\bm\Omega}}_{3,l_31}$ such that
$\tilde{\bm\Omega}_{3,l_41}^{(11)}$ and $\tilde{\bm\Omega}_{3,l_41}^{(22)}$ 
correspond to $\bz_{1:3}^{-z_{4,l_4}}$ and $z_{4,l_4}$, respectively.
We have that for $l_3 =1,2,\;l_4 = 1,2,3$,
$$
N(\tz_{1:4,l_4}|\bm0, \sigma^2\bm R_{4, l_4l_3})=
N((z_4, \bz_{1:3}^{-z_{4,{l_4}}})^\top|\bm m_{4, l_4l_3}z_{4,l_4}, \bm V_{4,l_4l_3})N(z_{4,l_4}|0,\sigma^2),
$$
and
$$
\begin{aligned}
\tz_{1:4,l_4} & = (z_4,\bz_{1:3}^{-z_{4,l_4}},z_{4,l_4})^\top,\;\;
\bm m_{4,l_4l_3} = (\rho_{4,l_4}, (\tilde{\bm\Omega}_{3, l_31}^{(12)})^\top)^\top,\\
\bm V_{4,l_4l_3} & = \begin{pmatrix}
\sigma^2(1-\rho_{4,l_4}^2) & \bm0^T\\
\bm0 & \sigma^2(\tilde{\bm\Omega}_{3, l_31}^{(11)} - \tilde{\bm\Omega}_{3, l_31}^{(12)}
\tilde{\bm\Omega}_{3, l_31}^{(21)})
\end{pmatrix},\\
\bm R_{4,l_4l_3}^{(12)} & = (\bm R_{4,l_4l_3}^{(21)})^\top = \bm m_{4,l_4l_3},\;\;
\bm R_{4,l_4l_3}^{(11)} = \bm V_{4,l_4l_3}/\sigma^2 + \bm m_{4, l_4l_3}\bm m_{4, l_4l_3}^{T}.
\end{aligned}
$$

\vspace{10pt}

\noindent We reorder $\tz_{1:4,l_4}$ with matrix $\bm B_{4,l_4}$ such that 
$\bz_{1:4} = \bm B_{4,l_4}\tz_{1:4, l_4}$ and 
let $\bm\Omega_{4,l_4l_31} = \bm B_{4,l_4}\bm R_{4,l_4l_3}\bm B_{4,l_3}^T$.
Then we obtain the joint density of $\bz_{1:4}$ as
$\tilde{p}(\bz_{1:4}) = \sum_{l_4=1}^3\sum_{l_3=1}^2w_{4,l_4}w_{3,l_3}
N(\bz_{1:4}|\bm0,\sigma^2\bm\Omega_{4,l_4l_31})$.

Applying the above technique iteratively for $\tilde{p}(\bz_{1:j})$ for $5\leq j\leq k$,
we obtain the joint density $\tilde{p}(\bz_{1:k}) \equiv \tilde{p}(\bz_{\BS})$, for $k\geq 2$,
namely,
$$
\tilde{p}(\bz_{1:k}) = \sum_{l_{k} = 1}^{k_L}\dots\sum_{l_2=1}^{2_L}w_{k,l_{k}}\dots w_{3,l_3}
w_{2,l_2}N(\bz_{1:k}|\bm0, \sigma^2\bm\Omega_{k,l_{k}\dots l_3l_2})
$$
where $k_L:= (k-1)\wedge L$, $w_{2,1} = 1$, and for $k\geq 3$,
$$
\begin{aligned}
    \tilde{\bm\Omega}_{k-1,l_{k-1}\dots l_31} & = \tilde{\bm B}_{k,l_{k-1}}\bm\Omega_{k-1,l_{k-1}\dots l_31}\tilde{\bm B}_{k,l_{k}},\;\;
    \bm m_{k,l_{k}\dots l_1} = (\rho_{k,l_{k}}, 
    (\tilde{\bm\Omega}_{k-1, l_{k-1}\dots l_31}^{(12)})^\top)^\top,\\
    \bm V_{k,l_{k}\dots l_3} & = \begin{pmatrix}
    \sigma^2(1-\rho_{k,l_{k}}^2) & \bm0\\
    \bm0^T & \sigma^2(\tilde{\bm\Omega}_{k-1, l_{k-1}\dots l_31}^{(11)} -
    \tilde{\bm\Omega}_{k-1, l_{k-1}\dots l_31}^{(12)}
    \tilde{\bm\Omega}_{k-1, l_{k-1}\dots l_31}^{(21)})
    \end{pmatrix},\\
    \bm R_{k,l_{k}\dots l_3}^{(12)} & = (\bm R_{k,l_{k}\dots l_3}^{(21)})^\top = \bm m_{k,l_{k}\dots l_3},\;\;
    \bm R_{k,l_{k}\dots l_3}^{(11)} = \bm V_{k,l_{k}\dots l_3}/\sigma^2 + 
    \bm m_{k,l_{k}\dots l_3}\bm m_{k,l_{k}\dots l_3}^\top,\\
    \bm\Omega_{k, l_{k}\dots l_31} & = \bm B_{k,l_{k}}\bm R_{k,l_{k}\dots l_3}\bm B_{k,l_{k}}^T,
\end{aligned}
$$
where 
 $\tilde{\bm B}_{k,l_{k}}$ is the rotation matrix such that 
$(\bz_{1:(k-1)}^{-z_{k,l_k}}, z_{k,l_k})^\top = \tilde{\bm B}_{k,l_{k}}\bz_{1:(k-1)}$,
and $\bm B_{k,l_{k}}$ is the rotation matrix such that the vector $\bz_{1:k} =
\bm B_{k,l_{k}}\tz_{1:k,l_{k}}$, where 
$\tz_{1:k,l_{k}} = (z_k, \bz_{1:(k-1)}^{-z_{k,l_k}}, z_{k,l_k})^\top$.

To complete the proof, what remains to be shown is that the density $\tilde{p}(\bz_{\U_1})$
is a mixture of multivariate Gaussian distributions, where $\U_1 = \BS\cup\{\su\}$.
We have that
$\tilde{p}(\bz_{\U_1}) = \sum_{l=1}^Lw_l(\su)N(z(\su)\mid\rho_l(\su)
z(\su_{(l)}),\sigma^2(1-(\rho_l(\su))^2))\tilde{p}(\bz_{1:k})$,
where $z(\su_{(l)})$ is an element of $\bz_{1:k}$ for $l = 1,\dots,L$. We can express each 
component density $N(\bz_{1:k}\,|\,\bm0, \sigma^2\Omega_{k,l_{k}\dots l_31})$ of the joint 
density $\tilde{p}(\bz_{1:k})$ as the product of a Gaussian density of 
$\bm Z_{1:k}^{-Z(\su_{(l)})}$ conditional on $Z(\su_{(l)}) = z(\su_{(l)})$ and a
Gaussian density of $Z(\su_{(l)})$. Using the approach in deriving the joint density over
$\BS$, we can show that $\tilde{p}(\bz_{\U_1})$ is a mixture of multivariate Gaussian
distributions.

\end{proof}

\begin{proof}[\textbf{Proof of Proposition 3}]
For an NNMP $Z(\sv)$,
the conditional probability that $Z(\sv)$ is greater than $z$ given its neighbors
$\bZ_{\tNe(\sv)} = \bz_{\tNe(\sv)}$, where 
$\bz_{\tNe(\sv)} = (z_{\sv_{(1)}},\dots,z_{\sv_{(L)}})$,  is
$P(Z(\sv) > z\,|\, \bm{Z}_{\tNe(\sv)} = \bz_{\tNe(\sv)}) = 
\sum_{l=1}^Lw_l(\sv)P(Z(\sv) > z\,|\, Z(\sv_{(l)}) = z(\sv_{(l)}))$,
where the conditional probability $P(Z(\sv) > z\,|\, Z(\sv_{(l)}) = z(\sv_{(l)}))$
corresponds to the bivariate random vector $(U_{\sv,l},V_{\sv,l})$.
If $U_l$ is stochastically increasing in $V_l$ for all $l$,
by the assumption that the sequence $(U_{\sv,l},V_{\sv,l})$ is built from 
the base random vectors $(U_l,V_l)$ for all $l$,
we have that $Z(\sv)$ is stochastically increasing in $\bZ_{\tNe(\sv)}$ for every $\sv\in\D$, 
i.e., $P(Z(\sv) > z\,|\,\bZ_{\tNe(\sv)} = \bz_{\tNe(\sv)})\,\leq\,
P(Z(\sv) > z\,|\,\bZ_{\tNe(\sv)} = \bz_{\tNe(\sv)}')$,
for all $\bz_{\tNe(\sv)}$ and $\bz_{\tNe(\sv)}'$ in the support of $\bZ_{\tNe(\sv)}$, 
such that $z_{\sv_{(l)}} \leq z_{\sv_{(l)}}'$ for all $l$.

Let $F_{Z(\sv)}$ and $F_{Z(\sv_{(1)}),\dots,Z(\sv_{(L)})}$ 
be the distribution functions of $Z(\sv)$ and $\bZ_{\tNe(\sv)}$, respectively.
Denote by $S_{Z(\sv_{(1)}),\dots,Z(\sv_{(L)})}(z_1,\dots,z_L) = 
P(Z(\sv_{(1)}) > z_1,\dots, Z(\sv_{(L)}) > z_L)$ the joint survival probability.
Then for every $\sv\in\D$ and $q\in(0,1)$, 
\begin{equation}\label{eq:la_h}
\begin{aligned}
&\;\;\;\;P(Z(\sv) > F_{Z(\sv)}^{-1}(q)\mid 
Z(\sv_{(1)}) > F_{Z(\sv_{(1)})}^{-1}(q), \dots, Z(\sv_{(L)}) > F_{Z(\sv_{(L)})}^{-1}(q))\\
& = \bigg\{\int_{F_{Z(\sv_{(1)})}^{-1}(q)}^{\infty}\dots\int_{F_{Z(\sv_{(L)})}^{-1}(q)}^{\infty}
P(Z(\sv) > F_{Z(\sv)}^{-1}(q)\mid Z(\sv_{(1)}) = z_1, \dots, Z(\sv_{(L)}) = z_L)\\
&\;\;\;\;dF_{Z(\sv_{(1)}),\dots,Z(\sv_{(L)})}(z_1,\dots,z_L)\bigg\}
\big/S_{Z(\sv_{(1)}),\dots,Z(\sv_{(L)})}(F_{Z(\sv_{(1)})}^{-1}(q),\dots,F_{Z(\sv_{(L)})}^{-1}(q))\\
& \geq \bigg\{\int_{F_{Z(\sv_{(1)})}^{-1}(q)}^{\infty}\dots\int_{F_{Z(\sv_{(L)})}^{-1}(q)}^{\infty}
P(Z(\sv) > F_{Z(\sv)}^{-1}(q)\mid 
Z(\sv_{(1)}) = F_{Z(\sv_{(1)})}^{-1}(q), \dots, Z(\sv_{(L)}) = F_{Z(\sv_{(L)})}^{-1}(q))\\
&\;\;\;\;dF_{Z(\sv_{(1)}),\dots,Z(\sv_{(L)})}(z_1,\dots,z_L)\bigg\}
\big/S_{Z(\sv_{(1)}),\dots,Z(\sv_{(L)})}(F_{Z(\sv_{(1)})}^{-1}(q),\dots,F_{Z(\sv_{(L)})}^{-1}(q))\\
& = P(Z(\sv) > F_{Z(\sv)}^{-1}(q)\mid 
Z(\sv_{(1)}) = F_{Z(\sv_{(1)})}^{-1}(q), \dots, Z(\sv_{(L)}) = F_{Z(\sv_{(L)})}^{-1}(q))\\
& = \sum_{l=1}^Lw_l(\sv)P(Z(\sv) > F_{U_{\sv,l}}^{-1}(q)\mid Z(\sv_{(l)}) = F_{V_{\sv,l}}^{-1}(q)),\\
\end{aligned}
\end{equation}
where the inequality follows the assumption of stochastically increasing positive dependence
of $Z(\sv)$ given $\bZ_{\tNe(\sv)}$. 

Taking $q\rightarrow1^-$ on both sides of \eqref{eq:la_h}, we obtain
$\lambda_{\iH}(\sv) \geq \sum_{l=1}^Lw_l(\sv)
\lim_{q\rightarrow1^-}P(Z(\sv) > F_{U_{\sv,l}}^{-1}(q)\mid 
Z(\sv_{(l)}) = F_{V_{\sv,l}}^{-1}(q))$,
where $F_{U_{\sv,l}}$ and $F_{V_{\sv,l}}$ are the marginal distribution functions
of $(U_{\sv,l}, V_{\sv,l})$.
Similarly, we can obtain the lower bound for $\lambda_{\iL}(\sv)$.

\end{proof}

\begin{proof}[\textbf{Proof of Corollary 1}]

We prove the result for $\lambda_{\iL}(\sv)$. 
The result for $\lambda_{\iH}(\sv)$ is obtained in a similar way.
Consider a bivariate cdf $F_{U_l,V_l}$ for random vector $(U_l,V_l)$, 
with marginal cdfs $F_{U_l} = F_{V_l} = F_l$ and 
marginal densities $f_{U_l} = f_{V_l} = f_l$, for all $l$.
The lower tail dependence coefficient is expressed as 
$\lambda_{\iL,l} = \lim_{q\rightarrow0^+}\frac{F_{U_l,V_l}(F_l^{-1}(q), F_l^{-1}(q))}
{F_{l}(F_l^{-1}(q))}$ with $q\in[0,1]$.
If $F_{U_l,V_l}$ has first order partial derivatives,
applying the L'Hopital's rule, we obtain
$$
\begin{aligned}
\lambda_{\iL,l} 
& = \lim_{q\rightarrow0^+}\frac{\partial{F_{U_l,V_l}}/\partial{V_l}(F_l^{-1}(q), F_l^{-1}(q)) + 
\partial{F_{U_l,V_l}}/\partial{U_l}(F_l^{-1}(q), F_l^{-1}(q))}{f_{l}(F_l^{-1}(q))}\\
& = \lim_{q\rightarrow0^+}P(U_l\leq F_l^{-1}(q)\mid V_l = F_l^{-1}(q)) + 
\lim_{q\rightarrow0^+}P(V_l\leq F_l^{-1}(q)\mid U_l = F_l^{-1}(q)).
\end{aligned}
$$
The above is a reproduced result from Theorem 8.57 of \citep{joe2014dependence}.
If $(U_l,V_l)$ is exchangeable,  we have 
$\lambda_{\iL,l} = 2\lim_{q\rightarrow0^+}P(U_l\leq F_l^{-1}(q)\,|\, V_l = F_l^{-1}(q))$.
If the sequences $(U_{\sv,l},V_{\sv,l})$ of an NNMP model are built from 
the base random vectors $(U_l,V_l)$. By our assumption that $F_{U_l}=F_{V_l}$
for all $l$, the marginal cdfs of $(U_{\sv,l},V_{\sv,l})$ extended from
$(U_l,V_l)$ are $F_{\sv,l} = F_{U_{\sv,l}} = F_{V_{\sv,l}}$ for all $\sv$ and all $l$. 
Then we have 
$\lambda_{\iL,l}(\sv) = 2\lim_{q\rightarrow0^+}
P(U_{\sv,l}\leq F_{\sv,l}^{-1}(q)\,|\,V_{\sv,l} = F_{\sv,l}^{-1}(q))$.
Using the result of Proposition 3, we obtain
$\lambda_{\iL}(\sv) \geq 
\sum_{l=1}^Lw_l(\sv)\lambda_{\iL,l}(\sv)/2.
$

\end{proof}

\begin{proof}[\textbf{Proof of Proposition 4}]

By the assumption that $U_l$ is stochastically increasing in $V_l$
and that $(U_{\sv,l},V_{\sv,l})$ is constructed based on $(U_l,V_l)$,
$U_{\sv,l}$  is stochastically increasing in $V_{\sv,l}$ for all $\sv\in\D$ and for all $l$.
Then for $(Z(\sv), Z(\sv_{(l)}))$ with respect to the 
bivariate distribution of $(U_{\sv,l},V_{\sv,l})$ with marginal distributions $F_{U_{\sv,l}}$
and $F_{V_{\sv,l}}$, we have that
$$
\begin{aligned}
&\;\;\;\;P(Z(\sv)\leq F_{U_{\sv,l}}^{-1}(q)\mid Z(\sv_{(l)}) \leq F_{V_{\sv,l}}^{-1}(q))\\
& = \int_{F_{V_{\sv,l}}^{-1}(0)}^{F_{V_{\sv,l}}^{-1}(q)}P(Z(\sv)\leq F_{U_{\sv,l}}^{-1}(q)\mid 
Z(\sv_{(l)}) = z_l)dF_{V_{\sv,l}}(z_l)\big/\int_{F_{V_{\sv,l}}^{-1}(0)}^{F_{V_{\sv,l}}^{-1}(q)}dF_{V_{\sv,l}}\\
& \leq\int_{F_{V_{\sv,l}}^{-1}(0)}^{F_{V_{\sv,l}}^{-1}(q)}P(Z(\sv)\leq F_{U_{\sv,l}}^{-1}(q)\mid Z(\sv_{(l)}) =  F_{V_{\sv,l}}^{-1}(0))dF_{V_{\sv,l}}(z_l)\big/
\int_{F_{V_{\sv,l}}^{-1}(0)}^{F_{V_{\sv,l}}^{-1}(q)}dF_{V_{\sv,l}}\\
& = P(Z(\sv)\leq F_{U_{\sv,l}}^{-1}(q)\mid Z(\sv_{(l)}) = F_{V_{\sv,l}}^{-1}(0)). 
\end{aligned}
$$
It follows that the boundary cdf of the NNMP model 
\begin{equation}\label{eq:boundary}
\begin{aligned}
&\;\;\;\;F_{1|2}(F_{Z(\sv)}^{-1}(q)\mid F_{\bm{Z}_{\tNe(\sv)}}^{-1}(0))\\
& = P(Z(\sv)\leq F_{Z(\sv)}^{-1}(q)\mid 
Z(\sv_{(1)}) = F_{Z(\sv_{(1)})}^{-1}(0),\dots,Z(\sv_{(L)}) = F_{Z(\sv_{(L)})}^{-1}(0))\\
& = \sum_{l=1}^Lw_l(\sv)P(Z(\sv)\leq F_{U_{\sv,l}}^{-1}(q)\mid Z(\sv_{(l)}) = F_{V_{\sv,l}}^{-1}(0))\\
& \geq \sum_{l=1}^Lw_l(\sv)P(Z(\sv)\leq F_{U_{\sv,l}}^{-1}(q)\mid Z(\sv_{(l)}) \leq F_{V_{\sv,l}}^{-1}(q)),
\end{aligned}
\end{equation}
Taking $q\rightarrow0^+$ on both sides of \eqref{eq:boundary}, we obtain
$F_{1|2}(F_{Z(\sv)}^{-1}(0)\,|\,F_{\bm{Z}_{\tNe(\sv)}}^{-1}(0))
\geq\sum_{l=1}^Lw_l(\sv)\lambda_{\iL,l}(\sv)$.
If there exists some $l$ such that $\lambda_{\iL,l}(\sv) > 0$, then
$F_{1|2}(F_{Z(\sv)}^{-1}(q)\,|\, F_{\bm{Z}_{\tNe(\sv)}}^{-1}(0))$ has 
strictly positive mass at $q=0$. Similarly, we can prove the result for 
$F_{1|2}(F_{Z(\sv)}^{-1}(q)\,|\, F_{\bm{Z}_{\tNe(\sv)}}^{-1}(1))$.

\end{proof}

\section{Additional Data Illustrations}

We provide three additional simulation experiments to demonstrate the benefits 
of the proposed NNMP framework. The first simulation experiment demonstrates that the 
Gaussian NNMP (GNNMP) provides a good approximation to Gaussian random fields.
The second example illustrates the ability of the framework to handle skewed data
using a skew-Gaussian NNMP (skew-GNNMP) model. Next, we demonstrate the 
effectiveness of the NNMP model for bounded spatial data. The last part of 
this section examines the non-Gaussian process assumption for the analysis of 
Mediterranean Sea surface temperature data example presented in the main paper, 
using a limited region to compare the GNNMP and the nearest-neighbor Gaussian 
process (NNGP) models.

In each of the experiments, we created a regular grid of $200\times200$ resolution 
on a unit square domain, and generated data on each grid location. 
We then randomly selected a subset of the data as the reference set 
with a random ordering for model fitting. For the purpose of illustration,
we chose neighbor size $L = 10$ for both cases.
Results are based on posterior samples collected every 10 
iterations from a Markov chain of 30000 iterations, with the first 
10000 samples as a burn-in.

\subsection{Additional Simulation Experiment 1}
\label{sec:add1}

We generated data from a spatially varying regression,
$y(\sv) = \x(\sv)^\top\bbeta + z(\sv) + \epsilon(\sv),\sv\in\D$,
where $z(\sv)$ is a standard Gaussian process with 
an exponential correlation function with range parameter $1/12$.
We included an intercept and a covariate drawn from $N(0,1)$ in the model, and
chose $\bbeta = (\beta_0,\beta_1)^\top = (1, 5)^\top$,  and 
$\tau^2 = 0.1$. The setting followed the simulation experiment in
\cite{datta2016hierarchical}.

We applied two models. The first one assumes that $z(\sv)$
follows an NNGP model with variance $\sigma^2_0$ and exponential 
correlation function with range parameter $\phi_0$.
The second one assumes that $z(\sv)$ follows a stationary GNNMP model,
i.e., $\mu_l = 0$ and $\sigma_l^2 = \sigma^2$ for all $l$, 
such that $z(\sv)$ has a stationary marginal $N(0,\sigma^2)$. 
For the GNNMP, we used exponential correlation functions with range parameter $\phi$ and $\zeta$,
respectively, for the correlation with respect to the component density and the cutoff points kernel 
function. For the NNGP model, we implement the latent NNGP algorithm from
the spNNGP package in R \citepsm{spnngp}. We trained both models using 2000 of the 2500
observations, and used the remaining 500 observations for model comparison.

\begin{figure*}[t]
    \centering
    \begin{subfigure}[b]{0.32\textwidth}
         \centering
         \includegraphics[width=\textwidth]{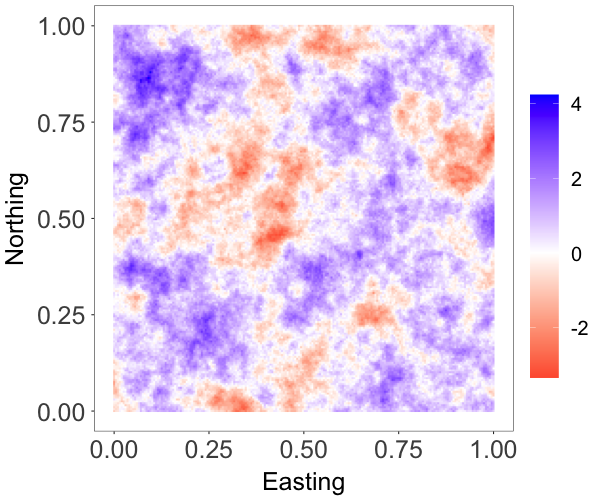}
         \caption{True GP}
     \end{subfigure}
     \hfill
     \begin{subfigure}[b]{0.32\textwidth}
         \centering
         \includegraphics[width=\textwidth]{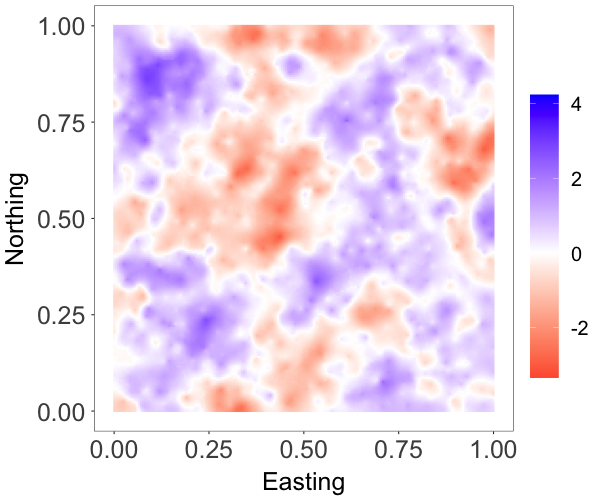}
         \caption{NNGP ($L = 10$)}
     \end{subfigure}
     \hfill
     \begin{subfigure}[b]{0.32\textwidth}
         \centering
         \includegraphics[width=\textwidth]{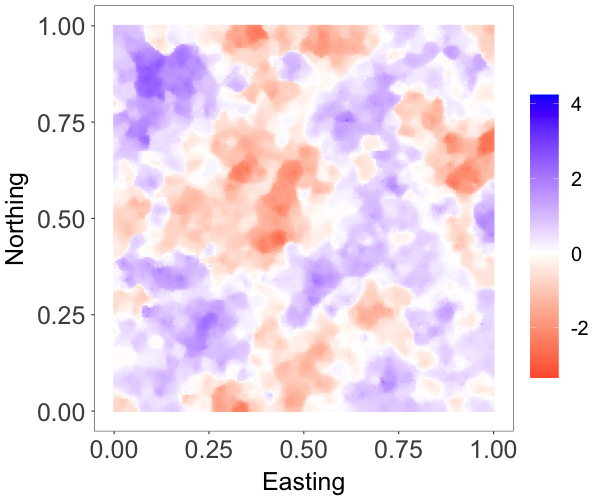}
         \caption{GNNMP ($L = 10$)}
     \end{subfigure}
    \caption{
    Additional simulation experiment 1 data analysis.
    Interpolated surface of the true Gaussian random field
    and posterior median estimates from the NNGP and GNNMP models.}
    \label{fig:gnnmp}
\end{figure*}

For both models, the regression coefficients $\bbeta$ were assigned flat priors.
The variances $\sigma_0^2$ and $\sigma^2$ received the same inverse gamma prior
$\mathrm{IG}(2,1)$,  and $\tau^2$ was assigned $\mathrm{IG}(2,0.1)$. The range parameter
$\phi_0$ of the NNGP received a uniform prior $\mathrm{Unif}(1/30,1/3)$, while the range
parameters $\phi$ and $\zeta$ of the GNNMP received inverse gamma priors 
$\mathrm{IG}(3, 1/3)$ and $\mathrm{IG}(3, 0.2)$, respectively. Regarding 
the logit Gaussian distribution parameters, $\bga$ and $\kappa^2$, we used  
$N(\bga\,|\,(-1.5,0,0)^\top,\,2\mathbf{I}_3)$ and $\mathrm{IG}(3, 1)$ priors, respectively.

The posterior estimates from the two models for the common parameters, $\bbeta$ and $\tau^2$,
were quite close. The posterior mean and $95\%$ credible interval estimates 
of $\beta_0$ and $\beta_1$ were
$1.32\,(1.11, 1.54)$ and $5.01\,(4.99, 5.04)$ from the GNNMP model, and 
$1.25\,(0.83, 1.62)$ and $5.01\,(4.99, 5.04)$ from the NNGP model. 
The corresponding estimates of $\tau^2$ were 
$0.12\,(0.09, 0.15)$ and $0.10\,(0.07, 0.12)$ from the GNNMP and NNMP models, respectively.

\begin{table}[t]
    \captionsetup{font=footnotesize}
    \caption{Additional simulation experiment 1 data analysis. Performance metrics of different models}
    \centering
    \begin{threeparttable}
    \begin{tabular*}{\hsize}{@{\extracolsep{\fill}}lcccccccc}
    % \\[-5pt]
\hline
  & RMSPE & 95\% CI & 95\% CI width & CRPS & PPLC & DIC\\
\hline
GNNMP & 0.57 & 0.96 & 2.51 & 0.32 & 461.40 & 2656.67\\
\hline
NNGP & 0.54 & 0.95 & 2.09 & 0.30 & 382.15 & 2268.79\\
\hline
    \end{tabular*}
    \begin{tablenotes}[para,flushleft]
        \small 
    \end{tablenotes}
    \end{threeparttable}
    \label{tbl:add1}
\end{table}

Table \ref{tbl:add1} shows the performance metrics of the two models.
The performance metrics of the GNNMP model are comparable to those of the NNGP model, 
the model assumptions of which are more well suited to the particular synthetic data example.
The posterior median estimate of the spatial random effects from both models
are shown in Figure \ref{fig:gnnmp}. We can see that the predictive field 
given by the GNNMP looks close to the true field and that predicted by the NNGP.  
On the whole, the GNNMP model provides a reasonably good approximation
to the Gaussian random field.

\subsection{Additional Simulation Experiment 2}
\label{sec:add2}

\begin{figure}[bt]
    \centering
    \captionsetup[subfigure]{justification=centering, font=footnotesize}
    \begin{subfigure}[b]{0.3\textwidth}
         \centering
         \includegraphics[width=\textwidth]{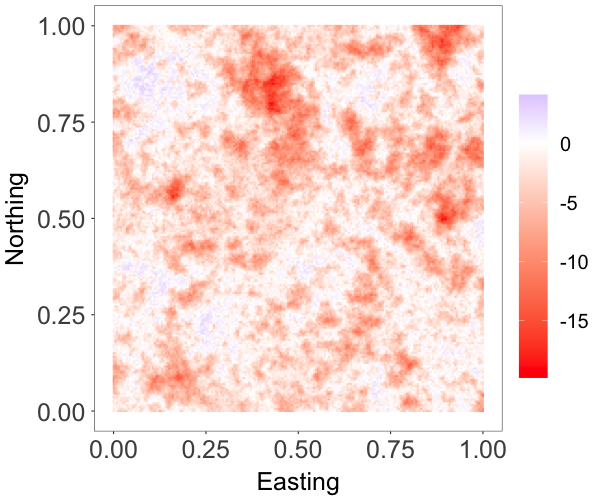}
         \caption{True $y(\sv)$ ($\sigma_1 = -5$)}
     \end{subfigure}
     \hfill
     \begin{subfigure}[b]{0.3\textwidth}
         \centering
         \includegraphics[width=\textwidth]{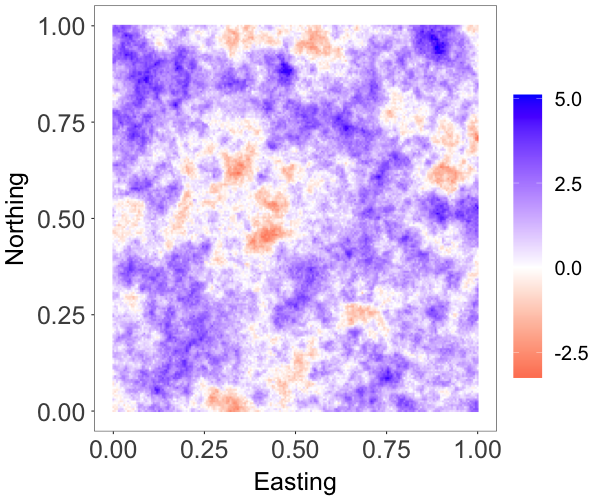}
         \caption{True $y(\sv)$ ($\sigma_1 = 1$)}
     \end{subfigure}
     \hfill
     \begin{subfigure}[b]{0.3\textwidth}
         \centering
         \includegraphics[width=\textwidth]{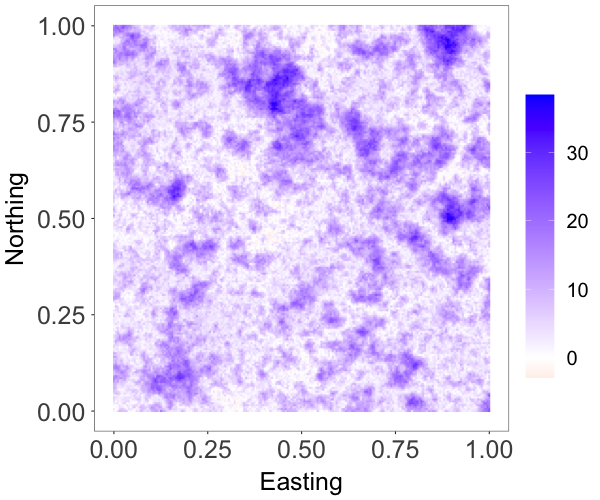}
         \caption{True $y(\sv)$ ($\sigma_1 = 10$)}
     \end{subfigure}\\
     \bigskip
         \begin{subfigure}[b]{0.3\textwidth}
         \centering
         \includegraphics[width=\textwidth]{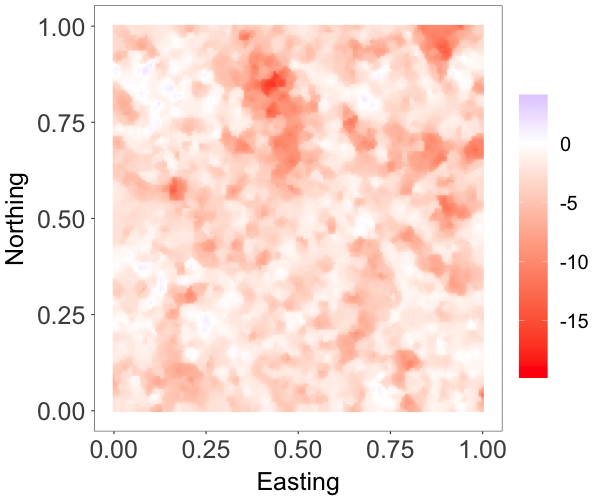}
         \caption{Skew-GNNMP}
     \end{subfigure}
     \hfill
     \begin{subfigure}[b]{0.3\textwidth}
         \centering
         \includegraphics[width=\textwidth]{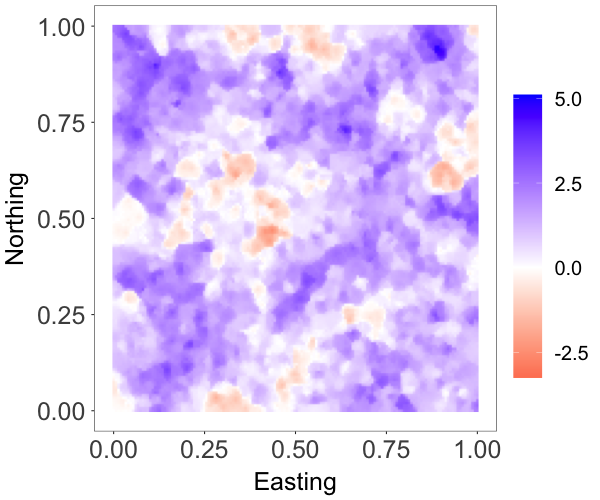}
         \caption{Skew-GNNMP }
     \end{subfigure}
     \hfill
     \begin{subfigure}[b]{0.3\textwidth}
         \centering
         \includegraphics[width=\textwidth]{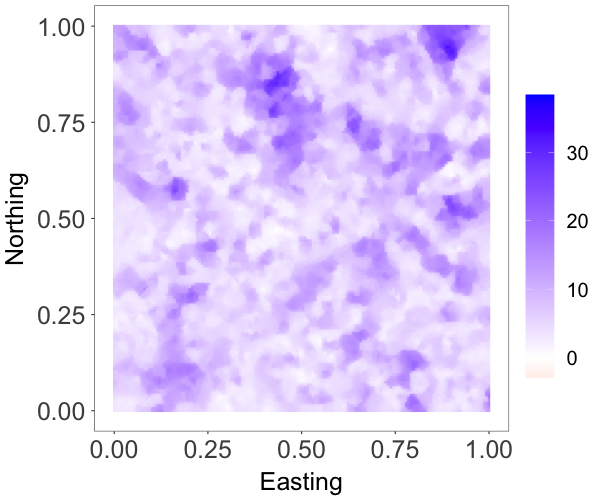}
         \caption{Skew-GNNMP}
     \end{subfigure}
    \caption{
    Additional simulation example 2 data analysis. 
    Top panels are interpolated surfaces of $y(\sv)$ generated by \eqref{eq:sim-sn}.
    Bottom panels are the posterior median estimates from the skew-GNNMP model.
    }
    \label{fig:skew-gnnmp}
\end{figure}

We generated data from the skew-Gaussian process \citep{zhang2010spatial},
\begin{equation}\label{eq:sim-sn}
y(\sv) = \sigma_1 \, |\omega_1(\sv)| + \sigma_2 \, \omega_2(\sv),\;\;\sv\in\D
\end{equation}
where $\omega_1(\sv)$ and $\omega_2(\sv)$ are both standard Gaussian processes with 
correlation matrix specified by an exponential correlation function with range parameter $1/12$. 
The parameter $\sigma_1\in\R$ controls the skewness, whereas $\sigma_2 > 0$ is a scale parameter.
The model has a stationary skew-Gaussian marginal
density $\mathrm{SN}(0, \sigma_1^2+\sigma_2^2, \sigma_1/\sigma_2)$.
We took $\sigma_2 = 1$, and generated data with $\sigma_1 = -5$, $1$ and $10$,
resulting in three different random fields that are, respectively, 
moderately negative-skewed, slightly positive-skewed, and strongly positive-skewed,
as shown in Figure \ref{fig:skew-gnnmp}(a)-\ref{fig:skew-gnnmp}(c).

We applied the stationary skew-GNNMP model.
The model is obtained as a special case of the skew-GNNMP model discussed 
in Section 2.3 of the main paper,
taking  $\la_l = \la$, $\mu_l = 0$, and $\sigma_l^2 = \sigma^2$,  
for all $l$. Here, $\lambda\in\R$ controls the skewness, such that 
a large positive (negative) value of $\lambda$ indicates strong positive (negative) skewness.
If $\la = 0$, the skew-GNNMP model reduces to the GNNMP model.
After marginalizing out $z_0$, we obtain a stationary skew-Gaussian marginal density 
$\mathrm{SN}(0, \la^2+\sigma^2, \la/\sigma)$.
We completed the full Bayesian specification for the model, by assigning priors 
$N(\lambda\,|\,0, 5)$, $\mathrm{IG}(\sigma^2\,|\,2,1)$, $\mathrm{IG}(\phi\,|\,3, 1/3)$, 
$\mathrm{IG}(\zeta\,|\,3, 0.2)$,
$N(\bga\,|\,(-1.5,0,0)^\top,\,2\mathbf{I}_3)$, and $\mathrm{IG}(\kappa^2\,|\,3, 1)$,
where $\zeta$ is the range parameter of the exponential correlation function 
specified for the cutoff point kernel.

\begin{figure*}[t]
    \centering
    \captionsetup[subfigure]{justification=centering, font=footnotesize}
    \begin{subfigure}[b]{0.3\textwidth}
         \centering
         \includegraphics[width=\textwidth]{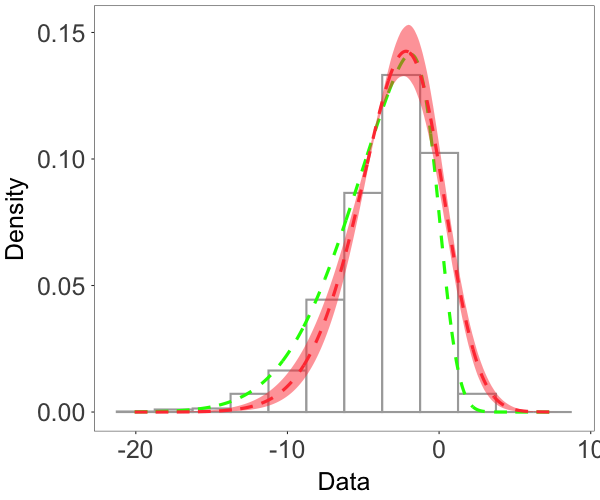}
         \caption{Estimated marginal ($\sigma_1 = -5$)}
     \end{subfigure}
     \hfill
     \begin{subfigure}[b]{0.3\textwidth}
         \centering
         \includegraphics[width=\textwidth]{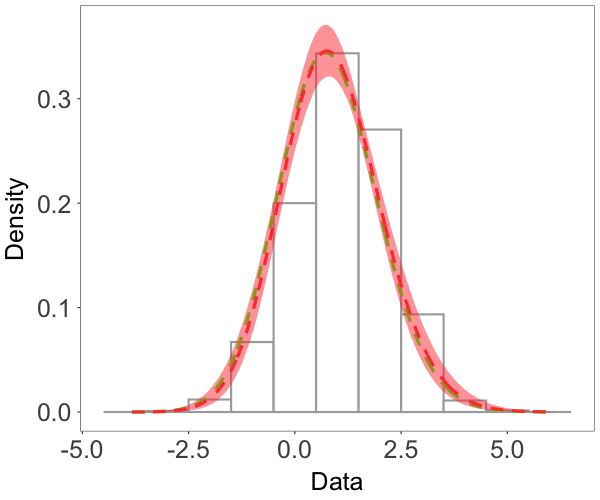}
         \caption{Estimated marginal ($\sigma_1 = 1$)}
     \end{subfigure}
     \hfill
     \begin{subfigure}[b]{0.3\textwidth}
         \centering
         \includegraphics[width=\textwidth]{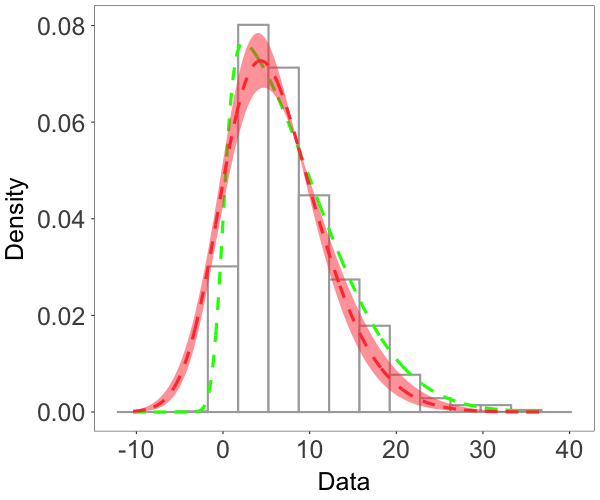}
         \caption{Estimated marginal ($\sigma_1 = 10$)}
     \end{subfigure}    
\caption{
Additional simulation example 2 data analysis.
Green lines are true marginal densities. Posterior means (dashed lines) 
and 95\% credible intervals (shaded regions) of the estimated marginal densities.}
    \label{fig:skew-gnnmp-hist}
\end{figure*}

We focus on the model performance on capturing skewness . 
The posterior mean and 95\% credible interval of 
$\lambda$ for the three scenarios 
were $-3.65\,(-4.10, -3.27)$, $1.09\,(0.91, 1.28)$ and $7.69\,(6.88, 8.68)$, respectively, 
indicating the model's ability to estimate different levels of skewness. 
The bottom row of Figure \ref{fig:skew-gnnmp} shows that the posterior median estimates 
of the surfaces capture well features of the true surfaces, even when the level of skewness
is small, thus demonstrating that the model is also able to recover near-Gaussian features.
Figure \ref{fig:skew-gnnmp-hist} plots the posterior mean and pointwise 95\%
credible interval for the marginal density, overlaid on the histogram of the 
simulated data for each of the three cases. These estimates demonstrate the adaptability 
of the skew-GNNMP model in capturing skewed random fields with different levels of skewness.

\subsection{Additional Simulation Experiment 3}
\label{sec:add3}

Many spatial processes are measured over a compact interval. As an example,
data on proportions are common in ecological applications. 
In this experiment, we demonstrate the effectiveness of the NNMP model for 
directly modeling bounded spatial data. We generated data using the model
$y(\sv) = F^{-1}\big(\Phi(\omega(\sv))\big)$,
where the cdf $F$ corresponds to a beta 
distribution, denoted as $\mathrm{Beta}(a_0,b_0)$,
and $\omega(\sv)$ is a standard Gaussian process with exponential correlation function 
with range parameter $0.1$. We set $a_0 = 3$, $b_0 = 6$.

\begin{figure}[t]
    \centering
    \captionsetup[subfigure]{justification=centering, font=footnotesize}
    \begin{subfigure}[b]{0.32\textwidth}
         \centering
         \includegraphics[width=\textwidth]{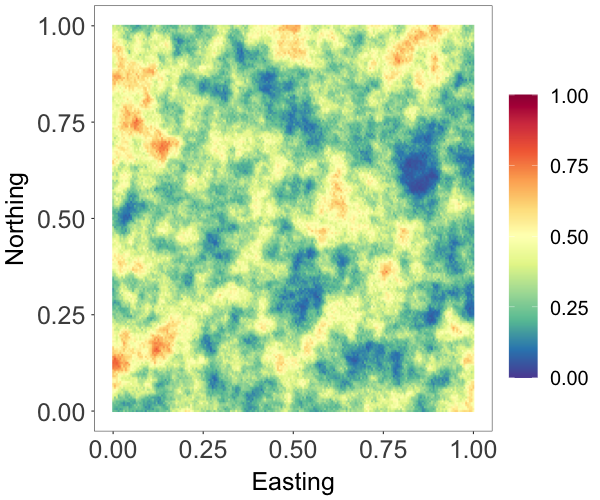}
         \caption{True $y(\sv)$}
     \end{subfigure}
     \hfill
     \begin{subfigure}[b]{0.32\textwidth}
         \centering
         \includegraphics[width=\textwidth]{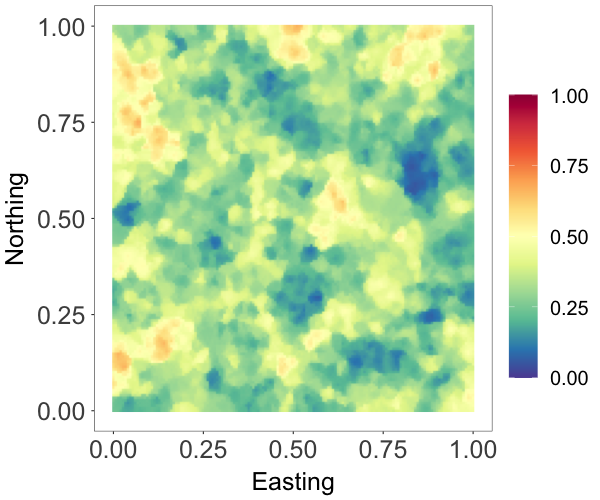}
         \caption{Beta NNMP}
     \end{subfigure}
     \hfill
     \begin{subfigure}[b]{0.32\textwidth}
         \centering
         \includegraphics[width=\textwidth]{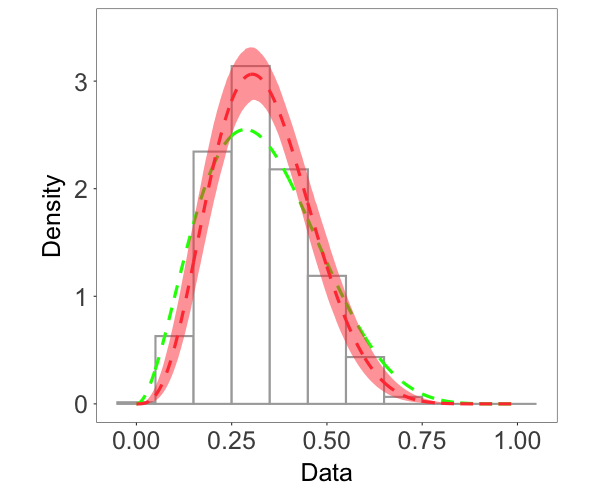}
         \caption{Estimated marginals}
     \end{subfigure} \\
    \caption{
    Additional simulation example 3 data analysis.
    Panels (a) and (b) are interpolated surfaces of the true field and
    posterior median estimate from the beta NNMP model, respectively.
    In Panel (c), the green dotted line corresponds to the true marginal.
    The red dash line and shaded region are the posterior mean and 
    pointwise $95\%$ credible interval for the estimated marginal.
    }
    \label{fig:sim-beta}
\end{figure}

We applied a Gaussian copula NNMP model with stationary marginal $\mathrm{Beta}(a, b)$, 
with the same spatial Gaussian copula and prior specification used in the 
second experiment. We used 2000 observations to train the model.
Figure \ref{fig:sim-beta}(b) shows the estimated random field which captures well
the main features of the true field in Figure \ref{fig:sim-beta}(a). 
The posterior mean and pointwise $95\%$ credible interval of the estimated marginal 
density in Figure \ref{fig:sim-beta}(c) overlay on the data histogram. These show that 
the beta NNMP estimation and prediction provide good approximation to the true field.

It is worth mentioning that implementing the beta NNMP model is simpler
than fitting existing models for data corresponding to proportions. 
For example, a spatial Gaussian copula model, that corresponds to the data generating 
process of this experiment, involves computations for large matrices. 
Alternatively, if a multivariate non-Gaussian copula is used, the 
resulting likelihood  can be intractable and require certain approximations. 
Another model that is commonly used in this setting is defined analogously to a spatial 
generalized linear mixed model. The spatial element in the model is introduced through
the transformed mean of the observations. A sample-based approach to fit such a model
requires sampling a large number of highly correlated latent variables. We conducted 
an experiment to demonstrate the effectiveness of the beta NNMP 
to approximate random fields simulated by the link function approach. 
We generated data as follows.
$$
\begin{aligned}
y(\sv)\,|\,\mu(\sv),\psi & \sim \mathrm{Beta}(\mu(\sv)\psi, 
(1-\mu(\sv))\psi),\\
\mathrm{logit}(\mu(\sv)) & = \mu_0 + \sigma_0\omega(\sv).
\end{aligned}
$$
The above model is analogous to a spatial generalized linear mixed model where the mean 
$\mu(\sv)$ of the beta distribution is modeled via a logit link function,
and $\omega(\sv)$ is a standard Gaussian process with exponential correlation function 
with range parameter $0.1$. We set $\psi=20$, $\mu_0 = -0.5$ and $\sigma_0 = 0.8$.

Since our purpose is primarily demonstrative, 
we applied a Gaussian copula NNMP model with a stationary beta marginal 
$\mathrm{Beta}(a, b)$, referred to as the beta NNMP model.
The correlation parameter of the Gaussian copula was specified by an exponential 
correlation function with range parameter $\phi$.
We specified an exponential correlation
function for the random cutoff points kernel function with range parameter $\zeta$. 
The Bayesian model is fully specified with a 
$\mathrm{IG}(3, 1/3)$ prior for $\phi$, 
a $\mathrm{Ga}(1,1)$ prior for $a$ and $b$,
a $\mathrm{IG}(3,0.2)$ prior for $\zeta$,
$N(\bga\,|\,(-1.5,0,0)^\top,\,2\mathbf{I}_3)$ and $\mathrm{IG}(\kappa^2\,|\,3, 1)$.

\begin{figure*}[t]
    \centering
    \captionsetup[subfigure]{justification=centering, font=footnotesize}
    \begin{subfigure}[b]{0.32\textwidth}
         \centering
         \includegraphics[width=\textwidth]{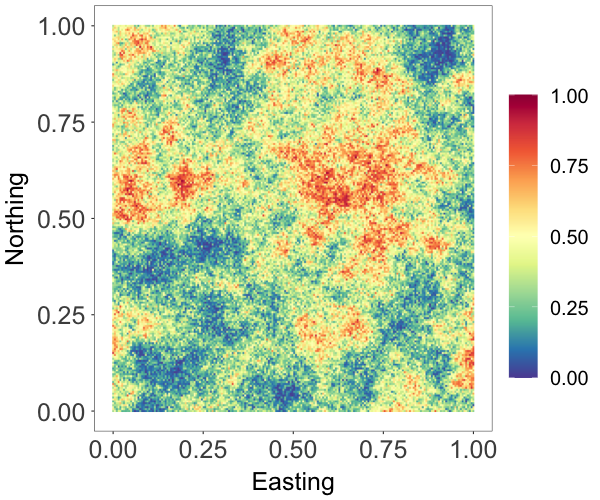}
         \caption{True $y(\sv)$}
     \end{subfigure}
    %  \hfill
    \hspace{15pt}
     \begin{subfigure}[b]{0.32\textwidth}
         \centering
         \includegraphics[width=\textwidth]{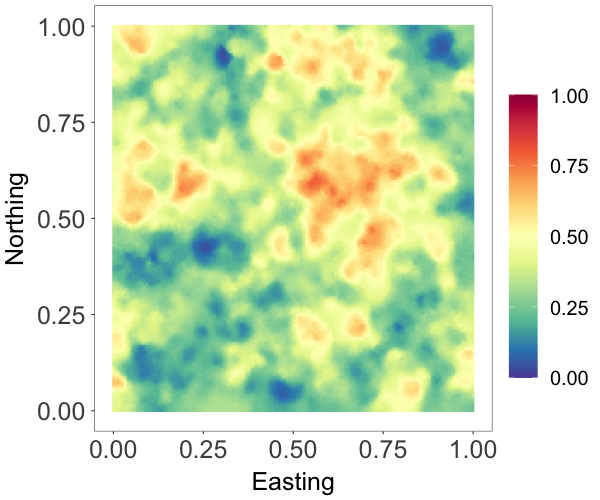}
         \caption{Beta NNMP}
     \end{subfigure}
    %  \hfill
    \caption{
    Additional simulation experiment 3 data analysis.
    Interpolated surfaces of the true field and 
    posterior median estimate from the beta NNMP model.
    }
    \label{fig:sim-beta2}
\end{figure*}

We trained the model using 2000 observations.
Figure \ref{fig:sim-beta2}(a)-(b) shows the interpolated surface of the true field and 
the predictive field given by the beta NNMP model. Although the beta NNMP's 
stationary marginal distribution assumption does not align with the true model, 
we can see that the predictive filed was able to capture the main feature of 
the true field. 
Moreover, it is worth mentioning that the MCMC algorithm for
the beta NNMP to fit the data set took around 18 minutes with 30000 iterations. 
This is substantially faster than the MCMC algorithm for fitting the true model
which involves sampling a large number of highly correlated latent variables.

\subsection{Mediterranean Sea Surface Temperature Regional Analysis}
\label{sec:regional}

\begin{figure}[t]
    \centering
    \captionsetup[subfigure]{justification=centering, font=footnotesize}
    \begin{subfigure}[b]{0.32\textwidth}
         \centering
         \includegraphics[width=\textwidth]{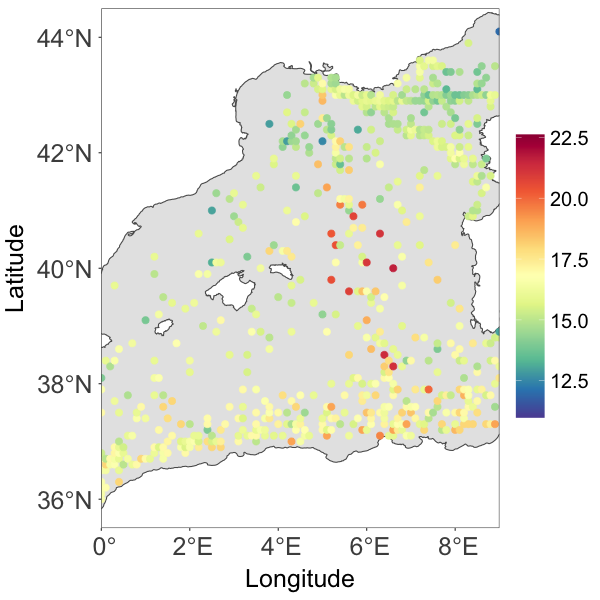}
         \caption{Regional SST}
     \end{subfigure}
     \hfill
     \begin{subfigure}[b]{0.32\textwidth}
         \centering
         \includegraphics[width=\textwidth]{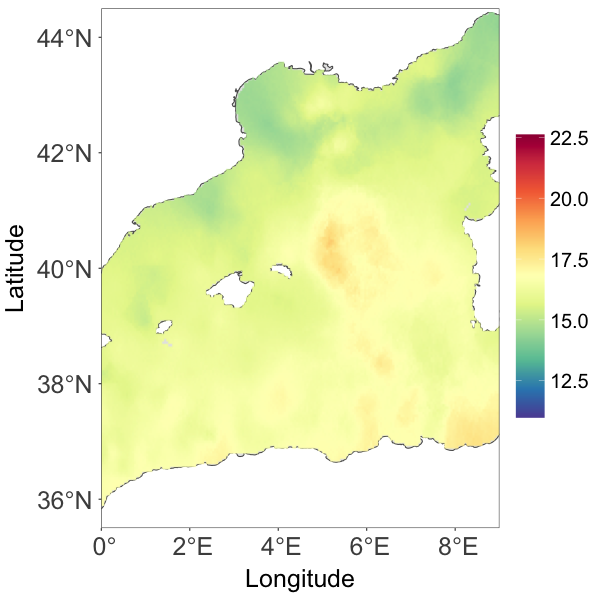}
         \caption{Predicted SST (GNNMP)}
     \end{subfigure}
     \hfill
     \begin{subfigure}[b]{0.32\textwidth}
         \centering
         \includegraphics[width=\textwidth]{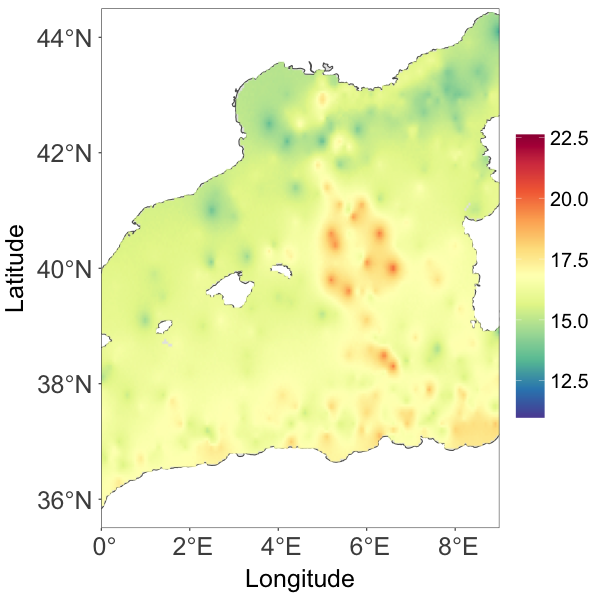}
         \caption{Predicted SST (NNGP)}
     \end{subfigure} \\
    \caption{
    SST data analysis.
    Panel (a) shows the observations at the selected region. 
    Panels (b) - (d) plot posterior median estimates of the SST by different models.
    }
    \label{fig:subset_sst_est}
\end{figure}

In this section, we examine the non-Gaussian process assumption for the Mediterranean 
Sea surface temperature data.
We focus on sea surface temperature (SST) over an area near the Gulf of Lion, along the 
islands near the shores of Spain, France, Monaco and Italy, between 0 - 9 E. longitude 
and 33.5 - 44.5 N. latitude. The SST observations in the region, as shown in Figure
\ref{fig:subset_sst_est}(a), are very heterogeneous, implying that the short range 
variability can be non-Gaussian. We compare the GNNMP with the NNGP in a spatially 
varying regression model, demonstrating the benefit of using a non-Gaussian process to 
explain the SST variability. In particular, the GNNMP has the same Gaussian marginals 
as the NNGP, but its finite-dimensional distribution is a mixture of Gaussian distributions.

We consider the following spatially varying regression model,
\begin{equation}\label{eq:spAdd}
y(\sv) = \x(\sv)^\top\bbeta + z(\sv) + \epsilon(\sv),\;\;\sv\in\D,
\end{equation}
where $y(\sv)$ is the SST observation,
$\x(\sv) = (1, v_1, v_2)^\top$ includes longitude $v_1$ and latitude $v_2$ to account for the 
long range variability in SST with regression parameters 
$\bbeta = (\beta_0, \beta_1, \beta_2)^\top$,
$z(\sv)$ is a spatial process, and $\epsilon(\sv)\stackrel{i.i.d.}{\sim}N(0,\tau^2)$ 
represents the micro-scale variability and/or the measurement error.

We model $z(\sv)$ with the GNNMP defined in \eqref{eq:sm-GNNMP} 
with $\mu_l = 0$ and $\sigma_l^2 = \sigma^2$, for all $l$. 
For comparison, we also applied an NNGP model for $z(\sv)$ with variance $\sigma^2_0$ and 
exponential correlation function with range parameter $\phi_0$.
For the GNNMP, we used exponential correlation functions with range parameter $\phi$ and $\zeta$,
respectively, for the correlation with respect to the component density, and the cutoff 
point kernel. For both models, the regression coefficients $\bbeta$ were assigned flat priors.
The variances $\sigma_0^2$ and $\sigma^2$ received the same inverse gamma prior 
$\mathrm{IG}(2,1)$,  and $\tau^2$ was assigned $\mathrm{IG}(2,0.1)$. The range parameter
$\phi_0$ of the NNGP received a uniform prior $\mathrm{Unif}(1/30,1/3)$, while the range
parameters $\phi$ and $\zeta$ of the GNNMP received inverse gamma priors 
$\mathrm{IG}(3, 1/3)$ and $\mathrm{IG}(3, 0.2)$, respectively. 
Regarding the logit Gaussian distribution parameters, $\bga$ and $\kappa^2$, we used 
$N((-1.5,0,0)^\top,\,2\mathbf{I}_3)$ and $\mathrm{IG}(3, 1)$ priors, respectively.

We took around 10\% of the data in the region, as the 
held-out data for model comparison, and used the remaining 580 observations to train models.
We compare models based on RMSPE,
95\% posterior credible interval coverage rate (95\% CI coverage),
deviance information criterion (DIC; \citealtsm{spiegelhalter2002bayesian}),
PPLC \citepsm{gelfand1998model}, and continuous ranked probability score
(CRPS; \citealtsm{gneiting2007strictly}).
To effectively compare the GNNMP and NNGP models, we first trained the NNGP model 
with neighbor sizes $L$ from 5 to 20, and selected the optimal neighbor size, $L = 11$ 
that corresponds to the smallest RMSPE. We then trained the GNNMP model with the same $L$.
In all cases, we ran the MCMC with 120000 iterations, discarding the first 20000 samples,
and collected samples every 20 iterations. 
The computing time was around 12 and 9 minutes for the GNNMP and NNGP 
models, respectively.

\begin{table}[t]
    \captionsetup{font=footnotesize}
    \caption{SST data analysis. Performance metrics of different models}
    \centering
    \begin{threeparttable}
    \begin{tabular*}{\hsize}{@{\extracolsep{\fill}}lcccccccc}
\hline
  & RMSPE & 95\% CI coverage & 95\% CI width & CRPS & PPLC & DIC\\
\hline
GNNMP & 0.92 & 0.97 & 4.09 & 0.50 & 135.58 & 539.53\\
\hline
NNGP & 1.00 & 0.97 & 4.30 & 0.56 & 526.81 & 1083.61\\
\hline
    \end{tabular*}
    \begin{tablenotes}[para,flushleft]
        \small 
    \end{tablenotes}
    \end{threeparttable}
    \label{tbl:sst}
\end{table}

We report the results for both models. The posterior mean and $95\%$ credible 
interval estimates of the regression intercept from the GNNMP was 
higher than the NNGP. They were $28.55\,(22.81, 35.34)$ and 
$27.18\,(23.53, 30.94)$, respectively. 
The corresponding posterior estimates of the coefficients for longitude and latitude 
given by the GNNMP and the NNGP were 
$0.09\,(-0.03, 0.24);$
$ -0.32\,(-0.50, -0.17)$ and $0.09\,(0.01, 0.17); -0.29\,(-0.39, -0.19)$,
respectively. Both models indicated that there was a trend of SST decreasing in the latitude 
at the selected region. For the error variance $\tau^2$,
the GNNMP provided a smaller estimate $0.12\,(0.02, 0.38)$,
compared to $0.45\,(0.02, 0.92)$ from the NNGP.

Regarding the model performance metrics in Table \ref{tbl:sst},
both the PPLC and DIC suggest that the GNNMP had a better goodness-of-fit than the NNGP. 
For out-of-sample prediction, the GNNMP produced smaller RMSPE and CRPS than the NNGP.
Both models gave the same 95\% CI coverage, while the one from GNNMP had a narrower width.
Figure \ref{fig:subset_sst_est}(b)-\ref{fig:subset_sst_est}(c) 
show the posterior median estimates of the temperature field from both models. 
We can see that both models yield estimates that resemble the pattern in the observations.
The predictive surface produced by the NNGP depicts some very localized, unrealistic features. 
These are not present in the results from the GNNMP.

\section{Implementation Details}

We provide model implementation details for the data examples. In particular, 
Section \ref{sec:gnnmp} corresponds to the GNNMP model applied in Sections \ref{sec:add1}
and \ref{sec:regional}. Section 
\ref{sec:sgnnmp} discusses the stationary skew-GNNMP model for the data example 
in Section \ref{sec:add2}, and the extended skew-GNNMP model for the Mediterranean 
Sea surface data analysis of the main paper. Section \ref{sec:copnnmp} introduces the 
Gaussian and Gumbel copula NNMP models implemented for the data example in Section
\ref{sec:add3}, and the simulation study of the main paper.

\subsection{GNNMP Models}
\label{sec:gnnmp}

We consider the spatially varying regression model, 
$y(\sv) = \bx(\sv)^\top\bbeta + z(\sv) + \epsilon(\sv),\sv\in\D$,
where $\epsilon(\sv)\stackrel{i.i.d.}{\sim}N(0,\tau^2)$,
and the spatial random effect $z(\sv)$ follows a stationary GNNMP model.
The associated conditional density of the model is
\begin{equation}\label{eq:sm-GNNMP}
p(z(\sv)\,|\,\bz_{\tNe(\sv)}) = \sum_{l=1}^{L}w_l(\sv)N(z(\sv)\,|\,
\rho_l(\sv)z(\sv_{(l)}), \sigma^2(1-(\rho_l(\sv))^2)),
\end{equation}
where $\rho_l(\sv) \equiv \rho_l(||\sv-\sv_{(l)}||) = \exp(-||\sv - \sv_{(l)}||/\phi)$.
For the weights, we consider an exponential correlation function with 
range parameter $\zeta$ for the kernel function that defines the random cutoff points. 
The Bayesian model is completed with priors 
$N(\bbeta\,|\,\bmu_{\bbeta},\bV_{\bbeta})$, 
$\mathrm{IG}(\sigma^2\,|\,u_{\sigma^2},v_{\sigma^2})$,
$\mathrm{IG}(\tau^2\,|\,u_{\tau^2},v_{\tau^2})$,
$\mathrm{IG}(\phi\,|\,u_{\phi},v_{\phi})$,
$\mathrm{IG}(\zeta\,|\,u_{\zeta},v_{\zeta})$,
$N(\bga\,|\,\bmu_{\bga},\bV_{\bga})$, 
$\mathrm{IG}(\kappa^2\,|\,u_{\kappa^2},v_{\kappa^2})$.

Let $y(\bs_i)$, $i = 1,\dots, n$, be the observations over reference set 
$\BS = (\bs_1,\dots,\bs_n)$. We introduce the MCMC sampler. 
It involves sampling the latent variables
$z(\bs_i)$, but it is easily developed based on the algorithm described in the main paper.
For each $z(\bs_i)$, $i = 3,\dots,n$, we 
introduce a configuration variable $\ell_i$, taking values in $\{1,\dots,i_L\}$ where
$i_L = (i-1)\wedge L$,
such that $\mathrm{Pr}(\ell_i\,|\,\bm w(\bs_i)) = \sum_{l=1}^{i_L}w_l(\bs_i)\delta_l(\ell_i)$,
where $\bm w(\bs_i) = (w_1(\bs_i),\dots, w_{i_L}(\bs_i))^\top$ and $\delta_l(\ell_i) = 1$
if $\ell_i = l$ and $0$ otherwise.
To allow for efficient simulation of parameters $\bga = (\ga_0,\ga_1,\ga_2)^\top$ and $\kappa^2$ for the weights, 
we associate each $z(\bs_i)$ with a latent Gaussian variable 
with mean $\mu(\bs_i) = \ga_0+s_{i1}\ga_1 + s_{i2}\ga_2$ and variance $\kappa^2$, 
where $\bs_i = (s_{i1},s_{i2})$, for $i = 3,\dots, n$.
There is a one-to-one correspondence between the configuration variables $\ell_i$ and 
latent variables $t_i$: $\ell_i = l$ if and only if $t_i\in(r_{\bs_i,l-1}^*,r_{\bs_i,l}^*)$
where $r^*_{\bs_i,l} = \log(r_{\bs_i,l}/(1-r_{\bs_i,l}))$, for $l=1,\dots,i_L$.
The posterior distribution of the model parameters is given by 
$$
\begin{aligned}
& N(\bbeta\,|\,\bmu_{\bbeta},\bV_{\bbeta}) \times
\mathrm{IG}(\tau^2\,|\,u_{\tau^2},v_{\tau^2}) \times
\mathrm{IG}(\sigma^2\,|\,u_{\sigma^2},v_{\sigma^2}) 
\times \mathrm{IG}(\phi\,|\,u_{\phi},v_{\phi})\times
\mathrm{IG}(\zeta\,|\,u_{\zeta},v_{\zeta})\\
& \times N(\bga\,|\,\bmu_{\bm\gamma},\bV_{\bm\gamma}) \times \mathrm{IG}(\kappa^2\,|\, u_{\kappa^2},v_{\kappa^2})
\times \prod_{i=1}^nN(y(\bs_i)\,|\,\x(\bs_i)^\top\bbeta + z(\bs_i),\tau^2)\\
&\times N(\bm t\,|\,\bD\bga,\,\kappa^2\mathbf{I}_{n-2})
\times N(z(\bs_1)\,|\,0,\sigma^2) \times N(z(\bs_2)\mid \rho_1(\bs_2)z(\bs_1),\sigma^2(1-(\rho_1(\bs_2))^2))\\
&\times \prod_{i=3}^n\sum_{l=1}^{i_L}
N(z(\bs)\mid \rho_l(\bs_i)z(\bs_{(il)}), \sigma^2(1-(\rho_l(\bs_i))^2))\mathbbm{1}_{(r^*_{\bs_i,l-1},r^*_{\bs_i,l})}(t_i),\\
\end{aligned}
$$
where the vector $\bm{t} = (t_3,\dots,t_n)^\top$, and the matrix $\bD$ is $(n-2)\times 3$ 
such that the $i$th row is $(1,s_{2+i,1},s_{2+i,2})$.

We describe the MCMC sampler to simulate from the posterior distribution of 
model parameters $(\bbeta, \bga, \sigma^2,\phi, \zeta, \tau^2,\kappa^2)$ and 
latent variables $\{t_i\}_{i=3}^n, \{z(\bs_i)\}_{i=1}^n$.
Denote by $\y_{\BS} = (y(\bs_1),\dots,y(\bs_n))^\top$,
$\bz_{\BS} = (z(\bs_1),\dots,z(\bs_n))^\top$, and let
$\X$ be the covariate matrix with the $i$th row being $\x(\bs_i)^\top$.
The posterior full conditional distribution for $\bbeta$ is
$N(\bbeta\,|\,\bm{\mu}_{\beta}^*,\bV_{\bbeta}^*)$ where 
$\bV_{\bbeta}^*=(\bV_{\bbeta}^{-1} + \tau^{-2}\X^\top\X)^{-1}$
and $\bm{\mu}_{\bbeta}^* = \bV_{\bbeta}^*(\bV_{\bbeta}^{-1}\bm{\mu}_{\bbeta} + 
\tau^{-2}\X^\top(\by_{\BS} - \bz_{\BS}))$.
An inverse gamma prior for $\tau^2$ yields an 
$\mathrm{IG}(\tau^2\,|\, u_{\tau^2}+n/2, v_{\tau^2}+\sum_{i=1}^ne_i^2/2)$
posterior full conditional,
where $e_i = y(\bs_i) - \x(\bs_i)^\top\bbeta - z(\bs_i)$.

To update the parameter $\zeta$, we first marginalize out
the latent variables $t_i$ from the joint posterior distribution.
The posterior full conditional distribution of $\zeta$ is proportional to 
$\mathrm{IG}(\zeta\,|\,u_{\zeta},v_{\zeta})
\prod_{i=3}^n\{G_{\bs_i}(r_{\bs_i,\ell_i}\,|\,\mu(\bs_i),\kappa^2)
-G_{\bs_i}(r_{\bs_i,\ell_i-1}\,|\,\mu(\bs_i),\kappa^2)\}$.
We update $\zeta$ on its log scale using a random walk Metropolis step.
The posterior full conditional distribution of $t_i$ is
$\sum_{l=1}^{i_L} q_l(\bs_i) \mathrm{TN}(t_i \,|\, \mu(\bs_i),
\kappa^2; r^*_{\bs_i,l-1}, r^*_{\bs_i,l})$,
where $q_l(\bs_i) \propto w_l(\bs_i)f_{\bs_i,l}(z(\bs_i)\,|\,z(\bs_{(il)}),\btheta)$
and $w_l(\bs_i) = G_{\bs_i}(r_{\bs_i,l}\,|\,\mu(\bs_i),\kappa^2)
-G_{\bs_i}(r_{\bs_i,l-1}\,|\,\mu(\bs_i),\kappa^2)$,
for $l=1,...,L$. Hence, each $t_i$ can be readily updated by sampling from the 
$l$-th truncated Gaussian with probability proportional to $q_l(\bs_i)$.
The posterior full conditional distribution of $\bga$ is 
$N(\bga\,|\,\bmu_{\bm\gamma}^*, \bV_{\bm\gamma}^*)$ where 
$\bV_{\bm\gamma}^* = (\bV_{\bm\gamma}^{-1} + \kappa^{-2}\bD^\top\bD)^{-1}$
and $\bmu_{\bm\gamma}^* = \bV_{\bm\gamma}^*(\bV_{\bm\gamma}^{-1}\bmu_{\bm\gamma} + 
\kappa^{-2}\bD^\top\bm{t})$.
The posterior full conditional distribution of $\kappa^2$ is 
$\mathrm{IG}(\kappa^2\,|\, u_{\kappa^2}+(n-2)/2, v_{\kappa^2}+
\sum_{i=3}^n(t_i-\mu(\bs_i))^2/2)$.

The posterior full conditional distribution of $\sigma^2$ is 
$\mathrm{IG}(\sigma^2\,|\, u_{\sigma^2}+n/2, v_{\sigma^2} +
\sum_{i=1}^n(z(\bs_i)-\rho_{\ell_i}(\bs_i)z(\bs_{(i,\ell_i)}))^2/\{2(1-(\rho_{\ell_i}(\bs_i))^2)\})$. 
The posterior full conditional distribution of $\phi$ is proportional to
$\mathrm{IG}(\phi\,|\, u_{\phi},v_{\phi})\prod_{i=2}^n
N(z(\bs_i)\mid \rho_{\ell_i}(\bs_i)z(\bs_{(i,\ell_i)}), 
\sigma_l^2(1-(\rho_{\ell_i}(\bs_i))^2))$. 
We update $\phi$ on its log scale with a random walk Metropolis step.
Denote by $\bm A_j^{(i)} = \{j: z(\bs_{(j,\ell_j)}) = z(\bs_i)\}$,
and assume $\rho_l(\bs_1) = 0, z(\bs_{(1l)}) = 0$ for every $l$.
The posterior full conditional of the latent spatial 
random effects $z(\bs_i)$ is 
$N(z(\bs_i)\,|\, \tilde{\sigma}_i^2\tilde{\mu}_i, \tilde{\sigma}_i^2)$
where $\tilde{\sigma}_i^2 = \big(\tau^{-2} + 
\sigma^{-2}(1-(\rho_{\ell_i}(\bs_i))^2)^{-1} + 
\sum_{j:j\in \bm A_j^{(i)}}\tilde{s}_{ij}^{-2}\big)^{-1}$ and 
$\tilde{\mu}_i = \tau^{-2}(y(\bs_i)-x(\bs_i)^\top\bbeta) + 
\sigma^{-2}(1-(\rho_{\ell_i}(\bs_i))^2)^{-1}\rho_{\ell_i}(\bs_i)z(\bs_{(i,\ell_i)}) + 
\sum_{j:j\in\bm A_j^{(i)}}z(\bs_j)(\rho_{\ell_j}(\bs_j))^{-1}\tilde{s}_{ij}^{-2}$ with
$\tilde{s}_{ij}^2 = \sigma^2(1-(\rho_{\ell_j}(\bs_j))^2)/(\rho_{\ell_j}(\bs_j))^2$,
for $i = 1,\dots, n$.

\subsection{Skew-GNNMP Models}
\label{sec:sgnnmp}

\subsubsection{Bivariate Skew-Gaussian Distribution}

Exploiting the location mixture representation of the skew-Gaussian distribution 
\citepsm{azzalini1996multivariate} for bivariate random vector $\bm Z = (U,V)$, we can write
\begin{equation}\label{eq:latent-bi-sn}
f(\bz\mid z_0) \sim N\left(\begin{pmatrix}\xi_u+\la_u z_0\\ \xi_v+\la_v z_0\end{pmatrix},
\sigma^2\begin{pmatrix}1 & \rho\\ \rho & 1\end{pmatrix}\right),\;\;
z_0\sim N(z_0\mid 0, 1)I(z_0\geq 0).
\end{equation}
It follows that, conditional on $Z_0 = z_0$, the marginal densities of $\bm Z$ are
$N(u\,|\,\xi_u+\la_u z_0, \sigma^2)$ and $N(v\,|\,\xi_v+\la_v z_0, \sigma^2)$,
respectively. Then the conditional density of $Z_0$ given $V = v$ is
$p(z_0\,|\,v) \propto N(z_0\,|\,(v-\xi_v)/\la_v, \sigma^2/\la_v^2)N(z_0\,|\, 0,1)I(z_0\geq 0)$.
Therefore, the conditional density $p(z_0\,|\,v)$ is a Gaussian distribution with
mean $\mu_0(v) = (v-\xi_v)\la_v/(\sigma^2+\la_v^2)$
and variance $\tau_0^2(v) = \sigma^2/(\sigma^2+\la_v^2)$,
truncated at $[0,\infty)$, denoted as $\mathrm{TN}_0(z_0\,|\,\mu_0(v),\tau_{0}^2(v))$.
Then the conditional distribution of $U$ given $V$ can be written as
\begin{equation}\label{eq:sn-cond-dens1}
f_{U|V}(u\,|\, v) = \int_0^{\infty}N(u\,|\, \mu_u + \rho(v-\mu_v), \sigma^2(1-\rho^2))
\mathrm{TN}(z_0\,|\, \mu_{0}(v), \tau_{0}^2(v))dz_0,
\end{equation}
where $\mu_u = \xi_u+\la_uz_0$, $\mu_v = \xi_v + \la_vz_0$.

Let $\bxi = (\xi_u, \xi_v)^\top$ and $\bla = (\la_u,\la_v)$. 
After marginalizing out $z_0$, the joint density of $\bm Z$ is given by
$f(\bz) = 2N(\bz\,|\, \bxi, \bm\Sigma)\,\Phi((1-\bla^\top\bm\Sigma^{-1}\bla)^{-1/2}
\bla^\top\bm\Sigma^{-1}(\y-\bxi))$,
where $\bm\Sigma = \sigma^2\bm R + \bla\bla^\top$,
$\bm R = \left(\begin{smallmatrix} 1 & \rho \\ \rho & 1\end{smallmatrix}\right)$,
and $\Phi$ is the cdf of a standard Gaussian distribution.
The marginal density of $U$ is
$f_U(u) = 2N(u\,|\,\xi_u, \omega_u^2)\,\Phi(\alpha_u(u-\xi_u)/\omega_u)$,
where $\omega_u^2 = \la_u^2+\sigma^2$ and $\alpha_u = \la_u/\sigma$.
We denote $f_U(u)$ as $\mathrm{SN}(u\,|\,\xi_u,\omega_u^2,\alpha_u)$.
Similarly, the marginal density of $V$ is $f_V(v) = \mathrm{SN}(\xi_v,\omega_v^2, \alpha_v)$.
It follows that the conditional density of $U$ given $V = v$ is
\begin{equation}\label{eq:sn-cond-dens2}
    \begin{aligned}
f_{U|V}(u\,|\, v) 
= N(u \,|\, \xi_u +\ga(v - \xi_v),\tilde{\omega}^2)\,\Phi(\alpha_1(u-\xi_u) + \alpha_2(v-\xi_v))/\Phi(\alpha_v(v-\xi_v)/\omega_v),
\end{aligned}
\end{equation}
where $\ga = (\rho\sigma^2+\la_u\la_v)/(\sigma^2+\la_v^2)$,\;\;
$\tilde{\omega}^2 = \sigma^2+\la_u^2 - (\rho\sigma^2+\la_u\la_v)^2/(\sigma^2+\la_v^2)$,
$\alpha_1 = (\la_u - \rho\la_v) / m$, $\alpha_2 = (\la_v - \rho\la_u) / m$,
and $m = \sqrt{(1-\rho)s^2}\sqrt{(1-\rho)s^2 + \la_u^2+\la_v^2 - 2\rho\la_u\la_v}$.

In the special case where $\xi_u = \xi_v = 0$ and $\la_u = \la_v = \la$, 
let $\omega^2 = \la^2 + \sigma^2$ and $\alpha = \la/\sigma$. 
The joint density of $\bm Z$ can be written as
$f(\bz) = 2N(\bz\mid\bm 0,\bm\Sigma)\,\Phi(\la(1-\la^2\bm1_2^\top\bm\Sigma^{-1}\bm1_2)^{-1/2}
\bm1_2^\top\bm\Sigma^{-1}\bz)$,
where the marginal density of $\bm{Z}$ is $\mathrm{SN}(x\,|\,0,\omega^2,\alpha)$.
The conditional density of $U$ given $V = v$ is then given by
\begin{equation}\label{eq:sn-cond-dens3}
f_{U|V}(u\,|\,v) = N(u\,|\,\trho v,\omega^2(1-\trho^2))\,\Phi(\alpha'(u+v)/\omega')/
\Phi(\alpha v/\omega),
\end{equation}
where $\trho = (\rho\sigma^2+\lambda^2)/(\sigma^2+\lambda^2)$, $\alpha' = \lambda/s$,
$\omega'^2 = s^2 + 2\lambda^2$, and $s^2 = (1+\rho)\sigma^2$.

\subsubsection{Stationary Skew-GNNMP Models}

We take a set of base random vectors $(U_l,V_l) \equiv (U,V)$ for all $l$,
where $(U,V)$ is a bivariate skew-Gaussian vector with distribution given by \eqref{eq:latent-bi-sn},
and take $\xi_u=\xi_v = 0$, $\la_u = \la_v = \la$.
We then extend $(U_l,V_l)$ to $(U_{\sv,l},V_{\sv,l})$ by extending 
$\rho$ to $\rho_l(\sv)$ using an exponential correlation function
such that $\rho_l(\sv) \equiv \rho_l(||\sv-\sv_{(l)}||) = 
\exp(-||\sv-\sv_{(l)}||/\phi)$, for $l = 1,\dots,L$. 
Using the resulting bivariate distribution for $(U_{\sv,l},V_{\sv,l})$,
we define the spatially varying density $f_{\sv,l}=f_{U_{\sv,l}|V_{\sv,l}}$ 
based on the formulation in \eqref{eq:sn-cond-dens1}.
The resulting associated conditional density of the stationary skew-GNNMP is
\begin{equation}\label{eq:sta-sgnnmp}
    p(y(\sv)\,|\,\by_{\tNe(\sv)}) = 
    \sum_{l=1}^{L} w_l(\sv) \int_0^{\infty}N(y(\sv)\,|\,\mu_l(\sv),\sigma_l^2(\sv))
    \mathrm{TN}(z_0(\sv)\,|\,\mu_{0}(\sv_{(l)}),\sigma_{0}^2)dz_0(\sv),
\end{equation}
where $\mu_l(\sv) = (1-\rho_l(\sv))\la z_0(\sv) + \rho_l(\sv)y(\sv_{(l)})$, 
$\sigma^2(\sv) = \sigma^2(1-(\rho_l(\sv))^2)$,
$\mu_{0}(\sv_{(l)}) = y(\sv_{(l)})\la/(\sigma^2+\la^2)$, 
and $\sigma_{0}^2 = \sigma^2/(\sigma^2+\la^2)$.

The component conditional density in \eqref{eq:sta-sgnnmp} is sampled via 
a latent variable $z_0(\sv)$. We marginalize out $z_0(\sv)$ to facilitate computation. 
Based on \eqref{eq:sn-cond-dens3}, we obtain the associated conditional 
density of the skew-GNNMP as
\begin{equation}\label{eq:sta-sgnnmp2}
p(y(\sv)\mid\y_{\tNe(\sv)}) = \sum_{l=1}^Lw_l(\sv)\,b_l(\sv)N(y(\sv)\mid \tilde{\rho}_l(\sv)y(\sv_{(l)}),
\omega^2(1-(\tilde{\rho}_l(\sv))^2)),
\end{equation}
where $b_l(\sv) = \Phi(\alpha'_l(\sv)(y(\sv)+y(\sv_{(l)}))/
\omega'_l(\sv))/\Phi(\alpha y(\sv_{(l)})/\omega)$,
$\trho_l(\sv) = (\rho_l(\sv)\sigma^2+\lambda^2)/(\sigma^2+\lambda^2)$,
$\alpha'_l(\sv) = \lambda/\sqrt{(1+\rho_l(\sv))\sigma^2}$,
$\omega'^2_l(\sv) = (1+\rho_l(\sv))\sigma^2 + 2\lambda^2$,
$\alpha = \la/\sigma$, and $\omega^2 = \sigma^2 + \la^2$.
For the weights $w_l(\sv)$, we use an exponential correlation
function with range parameter $\zeta$ for the kernel functions of the random cutoff points.
The Bayesian model is completed with prior specifications for 
$\lambda,\sigma^2,\phi,\zeta,\bga,\kappa^2$.
In particular, we consider priors $N(\lambda\,|\,\mu_{\lambda},\sigma^2_{\lambda})$,
$\mathrm{IG}(\sigma^2\,|\,u_{\sigma^2},v_{\sigma^2})$,
$\mathrm{IG}(\phi\,|\,u_{\phi},v_{\phi})$,
$\mathrm{IG}(\zeta\,|\,u_{\zeta},v_{\zeta})$,
$N(\bga\,|\,\bmu_{\bm\gamma},\bm{V}_{\bm\gamma})$ and
$\mathrm{IG}(\kappa^2\,|\,u_{\kappa^2},v_{\kappa^2})$.

Given observations $y(\bs_i)$, $i = 1,\dots, n$, over reference set $\BS = (\bs_1,\dots,\bs_n)$,
we perform Bayesian inference based a likelihood conditional on 
the first $L$ observations. The posterior distribution of the model parameters, 
given the conditional likelihood, is given by
$$
\begin{aligned}
& N(\lambda\,|\,\mu_{\lambda},\sigma^2_{\lambda})\times \mathrm{IG}(\sigma^2\,|\,u_{\sigma^2},v_{\sigma^2}) 
\times \mathrm{IG}(\phi\,|\,u_{\phi},v_{\phi})\times
\mathrm{IG}(\zeta\,|\,u_{\zeta},v_{\zeta})\\
&\times N(\bga\,|\,\bmu_{\bm\gamma},\bV_{\bm\gamma}) \times \mathrm{IG}(\kappa^2\,|\, u_{\kappa^2},v_{\kappa^2})
\times N(\bm t\,|\,\bD\bga,\,\kappa^2\mathbf{I}_{n-L})\\
& \times \prod_{i=L+1}^n\sum_{l=1}^{L}
b_l(\bs_i)N(y(\bs_i)\mid \tilde{\rho}_l(\bs_i)y(\bs_{(il)}),\omega^2(1-(\tilde{\rho}_l(\bs_i))^2))
\mathbbm{1}_{(r^*_{\bs_i,l-1},r^*_{\bs_i,l})}(t_i),\\
\end{aligned}
$$
where the vector $\bm{t} = (t_{L+1},\dots,t_n)^\top$, and the matrix $\bD$ is $(n-L)\times 3$ such that
the $i$th row is $(1,s_{L+i,1},s_{L+i,2})$.

The MCMC sampler to obtain samples from
the joint posterior distribution is described in the main paper. 
We present the posterior updates of $\lambda,\sigma^2$
and $\phi$. Note that the configuration variables $\ell_i$ are such
that $\ell_i = l$ if $t_i\in(r_{\bs_i,l-1}^*,r_{\bs_i,l}^*)$ for $i\geq L+1$.
Denote by $f_{\bs_i,l} = b_l(\bs_i)
N(y(\bs_i)\,|\,\trho_l(\bs_i)y(\bs_{(il)}),\omega^2(1-(\trho_l(\bs_i))^2))$.
We use a random walk Metropolis step to update $\lambda$ with 
target density $N(\lambda\,|\,\mu_{\lambda},\sigma^2_{\lambda})\prod_{i=L+1}^nf_{\bs_i,\ell_i}$.
The posterior full conditional distributions of $\sigma^2$ and $\phi$ are proportional to
$\mathrm{IG}(\sigma^2\,|\,u_{\sigma^2},v_{\sigma^2})\prod_{i=L+1}^nf_{\bs_i,\ell_i}$, and 
$\mathrm{IG}(\phi\,|\,u_{\phi},v_{\phi})\prod_{i=L+1}^nf_{\bs_i,\ell_i}$, respectively.
For each parameter, we update it on its log scale with a random walk Metropolis step.

\subsubsection{Extended Skew-GNNMP Models}

Again, we take a set of base random vectors $(U_l,V_l)\equiv (U,V)$ for all $l$,
where $(U,V)$ is a bivariate skew-Gaussian vector with distribution given by \eqref{eq:latent-bi-sn}.
We extend $(U_l,V_l)$ to $(U_{\sv,l},V_{\sv,l})$ by extending $\rho$ to $\rho_l(\sv)$
using an exponential correlation function such that $\rho_l(\sv) \equiv \rho_l(||\sv-\sv_{(l)}||) = 
\exp(-||\sv-\sv_{(l)}||/\phi)$, and extending
$\xi_u = \x(\sv)^\top\bbeta$, $\xi_v = \x(\sv_{(l)})^\top\bbeta$, $\la_u$ to $\la(\sv)$,
and $\la_v$ to $\la(\sv_{(l)})$, for $l = 1,\dots,L$. 
Using the resulting bivariate distribution for $(U_{\sv,l},V_{\sv,l})$,
we define the spatially varying density $f_{\sv,l}=f_{U_{\sv,l}|V_{\sv,l}}$
based on the formulation in \eqref{eq:sn-cond-dens1}. 
The resulting associated conditional density of the extended skew-GNNMP is
\begin{equation}\label{eq:ext-sgnnmp}
    p(y(\sv)\,|\,\by_{\tNe(\sv)}) = 
    \sum_{l=1}^{L} w_l(\sv)\int_0^{\infty}N(y(\sv)\,|\,\mu_l(\sv),\sigma_l^2(\sv))
    \mathrm{TN}(z_0(\sv)\,|\,\mu_{0l}(\sv_{(l)}),\sigma_{0l}^2(\sv_{(l)}))dz_0(\sv),
\end{equation}
where $\mu_l(\sv) = \x(\sv)^\top\bbeta + \la(\sv)z_0(\sv) + \rho_l(\sv)(y(\sv_{(l)}) - 
\x(\sv_{(l)})^\top\bbeta - \la(\sv_{(l)})z_0(\sv))$,
$\,\sigma_l^2(\sv) = \sigma^2(1-(\rho_l(\sv))^2)$,
$\,\mu_{0l}(\sv_{(l)}) = (y(\sv_{(l)})- \x(\sv_{(l)})^\top\bbeta)\la(\sv_{(l)})/(\sigma^2+(\la(\sv_{(l)}))^2)$, 
and $\sigma_{0l}^2(\sv_{(l)}) = \sigma^2/(\sigma^2+(\la(\sv_{(l)}))^2)$.
After marginalizing out $z_0(\sv)$, the conditional density \eqref{eq:ext-sgnnmp} based on formulation
\eqref{eq:sn-cond-dens2} can be written as
\begin{equation}\label{eq:ext-sgnnmp2}
p(y(\sv)\mid\y_{\tNe(\sv)}) = \sum_{l=1}^Lw_l(\sv)\,\tilde{b}_l(\sv)N(y(\sv)\mid \tilde{\mu}_l(\sv),
\tilde{\omega}^2_l(\sv)),
\end{equation}
where $\tilde{\mu}_l(\sv) = \x(\sv)^\top\bbeta + \tilde{\gamma}_l(\sv)(y(\sv_{(l)}) - \x(\sv_{(l)})^\top\bbeta)$,
$\tilde\gamma_l(\sv) = (\rho_{l}(\sv)\sigma^2+\la(\sv)\la(\sv_{(l)}))/\omega^2(\sv_{(l)})$, 
$\tilde{\omega}^2_l(\sv) = \omega(\sv)^2 - (\rho_{l}(\sv)\sigma^2+\la(\sv)\la(\sv_{(l)}))^2/\omega^2(\sv_{(l)})$,
$s_l^2(\sv) = (1 + \rho_{l}(\sv))\sigma^2$, $\alpha(\sv) = \la(\sv)/\sigma$, 
$(\omega(\sv))^2 = \la(\sv)^2 + \sigma^2$, and
$$
\begin{aligned}
\tilde{b}_l(\sv) & = 
\frac{\Phi(\alpha_1(\sv,\sv_{(l)})(y(\sv)-\x(\sv)^\top\bbeta) + 
\alpha_2(\sv,\sv_{(l)})(y(\sv_{(l)})-\x(\sv_{(l)})^\top\bbeta))}
{\Phi(\alpha(\sv_{(l)})(y(\sv_{(l)})-\x(\sv_{(l)})^\top\bbeta)/\omega(\sv_{(l)}))},\\
\alpha_1(\sv,\sv_{(l)}) & = (\la(\sv)-\rho_{l}(\sv)\la(\sv_{(l)})) / m(\sv),\;\;
\alpha_2(\sv,\sv_{(l)}) = (\la(\sv_{(l)})-\rho_{l}(\sv)\la(\sv)) / m(\sv),\\
m(\sv) & = \sqrt{(1-\rho_{l}(\sv))s_l^2(\sv)}\sqrt{(1-\rho_{l}(\sv))s^2_l(\sv) + (\la(\sv))^2 + 
(\la(\sv_{(l)}))^2 - 2\rho_{l}(\sv)\la(\sv)\la(\sv_{(l)})}.\\
\end{aligned}
$$

We model the spatially varying $\lambda(\sv)$ via a partitioning approach. 
In particular, we partition the domain $\D$ such that 
$\D = \cup_{k=1}^KP_k$, $P_i\cap P_j=\emptyset$ for $i\neq j$.
For all $\sv\in P_k$, we take $\la(\sv) = \la_k$, for $k = 1,\dots, K$. 
For the weights $w_l(\sv)$, we use an exponential correlation
function with range parameter $\zeta$ for the kernel function of the random cutoff points.
The Bayesian model is completed with prior specifications for 
$\bbeta,\bla = (\la_1,\dots,\la_K),\sigma^2,\phi,\zeta,\bga,\kappa^2$.
We assign a $N(\bbeta,\,|\,\bmu_{\bbeta},\bV_{\bbeta})$ to the regression parameter $\bbeta$ and 
$N(\la\,|\,\mu_{\la k},\sigma^2_{\la k})$ to $\la_k$, for $k = 1,\dots,K$.
For other parameters, we take $\mathrm{IG}(\sigma^2\,|\,u_{\sigma^2},v_{\sigma^2})$,
$\mathrm{IG}(\phi\,|\,u_{\phi},v_{\phi})$,
$\mathrm{IG}(\zeta\,|\,u_{\zeta},v_{\zeta})$,
$N(\bga\,|\,\bmu_{\bm\gamma},\bm{V}_{\gamma})$ and
$\mathrm{IG}(\kappa^2\,|\,u_{\kappa^2},v_{\kappa^2})$.

Given observations $y(\bs_i)$, $i = 1,\dots, n$, over reference set $\BS = (\bs_1,\dots,\bs_n)$,
we perform Bayesian inference based a likelihood conditional on 
the first $L$ observations.  The posterior distribution of the model parameters is
$$
\begin{aligned}
& N(\bbeta\,|\,\bmu_{\bbeta},\bV_{\bbeta})
\prod_{k=1}^KN(\lambda_k\,|\,\mu_{\lambda k},\sigma^2_{\lambda k})\times \mathrm{IG}(\sigma^2\,|\,u_{\sigma^2},v_{\sigma^2}) 
\times \mathrm{IG}(\phi\,|\,u_{\phi},v_{\phi})\times
\mathrm{IG}(\zeta\,|\,u_{\zeta},v_{\zeta})\\
&\times N(\bga\,|\,\bmu_{\bm\gamma},\bV_{\bm\gamma}) \times \mathrm{IG}(\kappa^2\,|\, u_{\kappa^2},v_{\kappa^2})
\times N(\bm t\,|\,\bD\bga,\,\kappa^2\mathbf{I}_{n-L})\\
& \times \prod_{i=L+1}^n\sum_{l=1}^{L}
\tilde{b}_l(\bs_i)N(y(\bs_i)\mid \tilde{\mu}_l(\bs_i),\tilde{\omega}^2_l(\bs_i))
\mathbbm{1}_{(r^*_{\bs_i,l-1},r^*_{\bs_i,l})}(t_i),\\
\end{aligned}
$$
where the vector $\bm{t} = (t_{L+1},\dots,t_n)^\top$, and the matrix $\bD$ is $(n-L)\times 3$ such that
the $i$th row is $(1,s_{L+i,1},s_{L+i,2})$.

The MCMC sampler to obtain samples from
the joint posterior distribution is described in the main paper. 
We present the posterior updates of $\bbeta, \bla, \sigma^2$
and $\phi$. Note that the configuration variables $\ell_i$ are such
that $\ell_i = l$ if $t_i\in(r_{\bs_i,l-1}^*,r_{\bs_i,l}^*)$ for $i\geq L+1$.
Denote by $f_{\bs_i,l} = \tilde{b}_l(\bs_i)N(y(\bs_i)\,|\,\tilde{\mu}_l(\bs_i),\tilde{\omega}^2_l(\bs_i))$.
We use a random walk Metropolis step to update $\bbeta$ with 
target density $N(\bbeta\,|\,\bmu_{\bbeta},\bV_{\bbeta})\prod_{i=L+1}^nf_{\bs_i,\ell_i}$. 
Let $B_k = \{i: \bs_i\in P_k\}\cup\{i: \bs_{(i\ell_i)}\in P_k\}$. 
The posterior full conditional of $\la_k$ is proportional to
$N(\la_k\,|\,\mu_{\la k},\sigma^2_{\la k})\prod_{i: i\in B_k}f_{\bs_i,\ell_i}$, 
and we use a random walk Metropolis step to sample $\la_k$, for $k = 1,\dots,K$. 
The posterior full conditional distributions of $\sigma^2$ and $\phi$ are proportional to
$\mathrm{IG}(\sigma^2\,|\,u_{\sigma^2},v_{\sigma^2})\prod_{i=L+1}^nf_{\bs_i,\ell_i}$, and 
$\mathrm{IG}(\phi\,|\,u_{\phi},v_{\phi})\prod_{i=L+1}^nf_{\bs_i,\ell_i}$, respectively.
For each parameter, we update it on its log scale with a random walk Metropolis step.

\subsection{Copula NNMP Models}
\label{sec:copnnmp}

\subsubsection{Gaussian and Gumbel Copulas}

We consider a continuous bivariate vector $(X_1,X_2)$ with marginal cdfs 
$F_1$ and $F_2$ such that $F_1(x_1) = t_1$ and $F_2(x_2) = t_2$.
We introduce basic properties of the Gaussian and Gumbel copulas.
For more details we refer to \cite{joe2014dependence}.

\paragraph{Gaussian copula} A Gaussian copula with correlation parameter 
$\rho\in(-1,1)$ for $(X_1,X_2)$ is 
$C(F_1(x_1), F_2(x_2)) = C(t_1,t_2\mid\rho) = \Phi_2(\Phi^{-1}(t_1)+\Phi^{-1}(t_2))$,
where $\Phi_2$ is the cdf of a standard bivariate Gaussian distribution with correlation
$\rho$. The copula is asymptotically independent in both the lower and the upper tails. 
The corresponding copula density is
\begin{equation}\label{eq:gaus_copula}
\frac{1}{\sqrt{1-\rho^2}}\exp\left(\frac{2\rho\Phi^{-1}(t_1)\Phi^{-1}(t_2) - 
\rho^2\{(\Phi^{-1}(t_1))^2 + (\Phi^{-1}(t_2))^2\}}{2(1-\rho^2)}\right).
\end{equation}
Denote by $C_{1|2}(t_1\,|\,t_2)$ the conditional cdf of $T_1$ given $T_2 = t_2$. Then we have
$C_{1|2}(t_1\,|\,t_2) = \frac{\partial C(t_1,t_2)}{\partial t_2} = 
\Phi\left(\frac{\Phi^{-1}(t_1) - \rho\Phi^{-1}(t_2)}{\sqrt{1-\rho^2}}\right)$.
We sample $X_1$, given $X_2 = x_2$, with the following steps.
Given a realization $x_2$ of $X_2$, we compute $t_2 = F_2(x_2)$.
We then generate a random number $z$ from a uniform distribution on $[0,1]$, and compute $t_1 = C_{1|2}^{-1}(z\,|\,t_2)$ where
$C_{1|2}^{-1}(z\,|\,t_2) = \Phi\left(\sqrt{(1-\rho^2)}\Phi^{-1}(z) + \rho\Phi^{-1}(t_2)\right)$ is the inverse of $C_{1|2}(t_1\,|\,t_2)$.
Finally, we obtain $x_1$ from the inverse cdf $F_1^{-1}(t_1)$.

\paragraph{Gumbel copula} 
A Gumbel copula with parameter $\eta\in[1,\infty)$ is 
$C(F_1(x_1), F_2(x_2)) = C(t_1,t_2\,|\, \eta) = \exp(-[(-\log(t_1))^{\eta} + (-\log(t_2))^{\eta}]^{1/\eta})$.
It is asymptotically independent in the lower tail and asymptotically dependent in
the upper tail with tail dependence coefficient $2 - 2^{1/\eta}$.
The corresponding copula density is
\begin{equation}\label{eq:gum_copula}
\exp(-(u_1^{\eta}+u_2^{\eta})^{1/\eta})((u_1^{\eta}+u_2^{\eta})^{1/\eta} + \eta - 1)
(u_1^{\eta}+u_2^{\eta})^{1/\eta-2}(u_1u_2)^{\eta-1}(t_1t_2)^{-1}.
\end{equation}

Let $u_1 = -\log(t_1)$ and $u_2 = -\log(t_2)$. In particular,
the Gumbel copula can be written as
$\overline{C}(u_1,u_2\,|\,\eta) = \exp(-(u_1^{\eta} + u_2^{\eta})^{1/\eta})$,
which is a bivariate exponential survival function, with marginals corresponding
to a unit rate exponential distribution.
The conditional cdf of $T_1$ given $T_2 = t_2$ is
$C_{1|2}(t_1\,|\,t_2) = \overline{C}_{1|2}(u_1\,|\,u_2) = u_2^{-1}\exp(-(u_1^{\eta}+u_2^{\eta})^{1/\eta})
(1 + (u_1/u_2)^{\eta})^{1/\eta-1}$.
The inverse conditional cdf $C_{1|2}^{-1}(\cdot\,|\,t_2)$ does not have a closed form. 
To generate $X_1$ given $X_2 = x_2$, following \cite{joe2014dependence}, 
we first define $y = (u_1^{\eta}+u_2^{\eta})^{1/\eta}$.
Then we have a realization of $X_1$, say 
$x_1 = (y_0^{\eta}-u_2^{\eta})^{1/\eta}$, where $y_0$ is the root of 
$h(y) = y + (\eta-1)\log(y) - (u_2 + (\eta-1)\log(u_2) - \log z) = 0$,
where $y\geq u_2$, and $z$ is a random number generated from a uniform distribution on $[0,1]$.

\subsubsection{Copula NNMP Models and Inference}

We take a set of base random vectors $(U_l,V_l) \equiv (U, V)$ where its bivariate
distribution is specified by a Gaussian copula with correlation parameter $\rho$. 
We extend $(U_l,V_l)$ to $(U_{\sv,l},V_{\sv,l})$
by extending $\rho$ to $\rho_l(\sv) \equiv \rho_l(||\sv-\sv_{(l)}||) = \exp(-||\sv-\sv_{(l)}||/\phi)$,
creating a sequence of spatially varying Gaussian copula $C_{\sv,l}$ for $(U_{\sv,l},V_{\sv,l})$
with marginal cdfs $F_{U_{\sv,l}} = F_{V_{\sv,l}} = F_Y$ for all $\sv$ and all $l$.
The cdf $F_Y$ corresponds to the stationary marginal distribution of the model.
The associated conditional density of the Gaussian copula NNMP is
% \begin{equation}\label{eq:sm-copula-nnmp}
$$
p(y(\sv)\mid\by_{\tNe(\sv)}) = \sum_{l=1}^L w_l(\sv) \,
c_{\sv,l}(y(\sv),y(\sv_{(l)}))f_Y(y(\sv)),
$$
% \end{equation}
where $c_{\sv,l}(y(\sv),y(\sv_{(l)}))$ is the Gaussian copula density obtained 
by replacing $\rho$ in \eqref{eq:gaus_copula} with $\rho_l(\sv)$, 
and $f_Y$ is the density of $F_Y$.

Similarly, we can obtain the Gumbel copula NNMP model using a collection of
spatially varying Gumbel copulas by extending  $\eta$ in \eqref{eq:gum_copula} to
$\eta_l(\sv) \equiv \eta_l(||\sv-\sv_{(l)}||) = 
\min\{(1-\exp(-||\sv-\sv_{(l)}||/\phi))^{-1}, 50\}$,
where the upper bound $50$ ensures numerical stability.

We discuss the inferential approach for the Gaussian copula NNMP; the approach for the Gumbel copula
NNMP model is similar. Assume the stationary marginal density is a gamma density, denoted as 
$f_Y = \mathrm{Ga}(a,b)$, with mean $E(Y) = a/b$.
For the weights $w_l(\sv)$, we use an exponential correlation
function with range parameter $\zeta$ to define the random cutoff points.
The Bayesian model is completed with prior specifications for 
$a, b,\phi,\zeta,\bga,\kappa^2$.
In particular, we consider priors 
$\mathrm{Ga}(u_a,v_a)$, $\mathrm{Ga}(u_b,v_b)$,
$\mathrm{IG}(\phi\,|\,u_{\phi},v_{\phi})$,
$\mathrm{IG}(\zeta\,|\,u_{\zeta},v_{\zeta})$,
$N(\bga\,|\,\bmu_{\bm\gamma},\bm{V}_{\bm\gamma})$ and
$\mathrm{IG}(\kappa^2\,|\,u_{\kappa^2},v_{\kappa^2})$.

Given observations $y(\bs_i), i = 1,\dots, n$, over reference set $\BS = (\bs_1,\dots,\bs_n)$, 
we perform Bayesian inference using a likelihood conditional on $(y(\bs_1),\dots,y(\bs_L))$.
The posterior distribution of the model parameters is
$$
\begin{aligned}
& \mathrm{Ga}(u_a,v_a) \times \mathrm{Ga}(u_b,v_b)
\times \mathrm{IG}(\phi\,|\,u_{\phi},v_{\phi})\times
\mathrm{IG}(\zeta\,|\,u_{\zeta},v_{\zeta})
\times N(\bga\,|\,\bmu_{\bm\gamma},\bV_{\bm\gamma}) \times \mathrm{IG}(\kappa^2\,|\, u_{\kappa^2},v_{\kappa^2})\\
&\times N(\bm t\,|\,\bD\bga,\,\kappa^2\mathbf{I}_{n-L})
\times \prod_{i=L+1}^n\sum_{l=1}^{L}c_{\bs_i,l}(y(\bs_i),y(\bs_{(il)}))f_Y(y(\bs_i))
\mathbbm{1}_{(r^*_{\bs_i,l-1},r^*_{\bs_i,l})}(t_i),\\
\end{aligned}
$$
where the vector $\bm{t} = (t_{L+1},\dots,t_n)^\top$, and the matrix $\bD$ is $(n-L)\times 3$ such that
the $i$th row is $(1,s_{L+i,1},s_{L+i,2})$.
We provide the updates for parameters $(a, b, \phi)$.
Note that the configuration variables $\ell_i$ are such 
that $\ell_i = l$ if $t_i\in(r_{\bs_i,l-1}^*,r_{\bs_i,l}^*)$ for $i\geq L+1$.
Denote by $f_{\bs_i,l} = c_{\bs_i,l}(y(\bs_i),y(\bs_{(il)}))f_Y(y(\bs_i))$.
The posterior full conditional distributions for parameters $a$, $b$ and $\phi$ 
are proportional to 
$\mathrm{IG}(a\,|\,u_a,v_a)\prod_{i=L+1}^nf_{\bs_i,\ell_i}$,
$\mathrm{IG}(b\,|\,u_b,v_b)\prod_{i=L+1}^nf_{\bs_i,\ell_i}$, and
$\mathrm{IG}(\phi\,|\,u_{\phi},v_{\phi})
\prod_{i=L+1}^nc_{\bs_i,l}(y(\bs_i),y(\bs_{(il)}))$,
respectively. Each parameter is updated on its log scale with 
a random walk Metropolis step.

\bibliographystylesm{jasa3}
\bibliographysm{ref}

\end{document}